\documentclass[floatfix,twocolumn,showpacs,preprintnumbers,amsmath,amssymb,pra,superscriptaddress,longbibliography]{revtex4-1}
\usepackage{color}
\usepackage[usenames,dvipsnames,svgnames,table]{xcolor}
\usepackage[colorlinks=true,linkcolor=blue,urlcolor=blue,citecolor=blue]{hyperref}
\usepackage{mathtools}
\usepackage{graphicx}
\usepackage{dcolumn}
\usepackage{array}
\usepackage{lipsum}
\usepackage{bm}
\usepackage{subfigure}
\usepackage{amssymb}
\usepackage{multirow}
\usepackage{tabularx}
\usepackage{amsmath}
\usepackage{braket}
\usepackage{csquotes}
\graphicspath{{plots/}}
 \usepackage{lipsum}
\usepackage{mathrsfs}
\usepackage{MnSymbol}
	
\newcommand{\beq}{\begin{equation}}
\newcommand{\eeq}{\end{equation}}
\newcommand{\bea}{\begin{eqnarray}}
\newcommand{\eea}{\end{eqnarray}}

\begin{document}
\title{Exact series expansion for even frequency moments of the dynamic structure factor}
\author{Panagiotis Tolias}
\email{tolias@kth.se}
\affiliation{Space and Plasma Physics, Royal Institute of Technology (KTH), Stockholm, SE-100 44, Sweden}
\author{Jan Vorberger}
\affiliation{Helmholtz-Zentrum Dresden-Rossendorf (HZDR), D-01328 Dresden, Germany}
\author{Tobias Dornheim}
\affiliation{Center for Advanced Systems Understanding (CASUS), D-02826 G\"orlitz, Germany}
\affiliation{Helmholtz-Zentrum Dresden-Rossendorf (HZDR), D-01328 Dresden, Germany}

\begin{abstract}
An exact series representation of the even frequency moments of the dynamic structure factor is derived. Truncations are proposed that allow to evaluate the explicitly unknown second, fourth and fifth frequency moments for the finite temperature uniform electron gas. Their applicability range in terms of degeneracy parameter and wavenumber is determined by exploiting the non-interacting limit and by comparing with the quasi-exact results of path integral Monte Carlo simulations.  
\end{abstract}
\maketitle

\section{Introduction}

The accurate understanding of interacting quantum many-body systems at finite temperatures constitutes a highly active frontier in condensed matter physics and quantum chemistry\,\cite{stefanucci2013nonequilibrium,bonitz_book,abrikosov_book}. As a prime example, warm dense matter (WDM) is an exotic yet ubiquitous state of matter that emerges in dense astrophysical objects (planetary deep interiors\,\cite{Howard2023,Hernandez2023,Breuer2023}, white dwarf envelopes\,\cite{Kritcher2020,SAUMON20221}, neutron star crusts\,\cite{chamel_lrr_2008,neutronstar_2021}) as well as in the early stages of inertial confinement fusion\,\cite{Hurricane_RevModPhys_2023,hu_ICF}. WDM presents a particular challenge due to the complex interplay between moderate electronic degeneracy, partial ionization, moderate Coulomb interactions and strong thermal excitations\,\cite{Remington_2005,wdm_book,Riley_2021,vorberger2025roadmapwarmdensematter}.

A pre-requisite for the understanding of WDM is the comprehension of the physical properties of the finite temperature uniform electron gas (UEG), which is a rich-in-physics model system where interacting electrons are immersed in a rigid uniform neutralizing background and thus explicit treatment of the ionic component is circumvented\,\cite{Baus_Hansen_OCP,plasma2,IIT,review}. The UEG state points are uniquely specified by the quantum coupling parameter $r_{\mathrm{s}}=d/a_{\mathrm{B}}$ (ratio of the Wigner-Seitz radius over the first Bohr radius) and the degeneracy parameter $\Theta=T/E_{\mathrm{f}}$ (ratio of the thermal energy over the Fermi energy). In WDM conditions, there is an absence of small parameters ($r_{\mathrm{s}}\sim\Theta\sim1$), which clearly indicates that the full interplay of quantum diffraction, exchange effects, Pauli blocking, Coulomb coupling and thermal excitations needs to be considered\,\cite{review,Dornheim_review}. Thus, the warm dense UEG is ideally studied with quantum Monte Carlo (QMC) simulations\,\cite{Dornheim_POP_2017,new_POP}. Despite the notorious fermion sign problem\,\cite{dornheim_sign_problem,Dornheim_2021}, which leads to an exponential increase in the required compute time with important system parameters such as the inverse temperature $1/T$ or system size $N$, a wealth of quasi-exact data are nowadays available for thermodynamic\,\cite{dornheim_prl,dornheim2024directfreeenergycalculation,dornheim2024chemicalpotentialwarmdense}, structural\,\cite{Dornheim_PRL_2020_ESA,dornheim_ML,Dornheim_PRB_ESA_2021,Dornheim_HEDP_2022,Dornheim_JPCL_2024}, dynamic\,\cite{dornheim_dynamic,dynamic_folgepaper,Hamann_PRB_2020,Dornheim_JCP_xi_2023,Dornheim_PRB_2024,Dornheim_EPL_2024} and nonlinear properties\,\cite{Dornheim_PRL_2020,Dornheim_JCP_ITCF_2021,Tolias_2023} of the warm dense UEG. Yet, few exact properties of the finite temperature UEG are known and many of those concern the frequency moments of spectral functions\,\cite{tkachenko_book,quantum_theory}.

The frequency moments of a spectral function $A(\boldsymbol{q},\omega)$ are generally defined by\,\cite{tkachenko_book,quantum_theory}
\begin{equation}
M^{(\alpha)}_{\mathrm{A}}(\boldsymbol{q})=\langle\omega^{\alpha}\rangle_{\mathrm{A}}=\int_{-\infty}^{+\infty}d\omega\;\omega^{\alpha}A(\boldsymbol{q},\omega)\,,\label{momentsgen}
\end{equation}
with $\alpha$ an arbitrary positive or negative integer. Spectral functions of particular interest are the imaginary part of the linear density response $\Im\{\chi(\boldsymbol{q},\omega)\}$, the dynamic structure factor $S(\boldsymbol{q},\omega)$ and the dielectric loss function $-\Im\{\epsilon^{-1}(\boldsymbol{q},\omega)\}/\omega$. The most fundamental is $\Im\{\chi(\boldsymbol{q},\omega)\}$ that serves as the generator of odd frequency moments. In particular, a celebrated expression of linear response theory, which originates from the combination of a high-frequency Kramers-Kronig expansion with a short-time Kubo formula expansion, connects the odd frequency moments of $\Im\{\chi(\boldsymbol{q},\omega)\}$ to the equilibrium expectation values of equal-time commutators\,\cite{quantum_theory,Ichimaru_BookII,SingwiTosi_Review}. This foundational formula ultimately involves iterated commutators with the number of Hamiltonian operator nests coinciding with the frequency moment order\,\cite{quantum_theory}. Hence, the operator algebra calculations become exceedingly cumbersome. As a result, only the first and third frequency moments of $\Im\{\chi(\boldsymbol{q},\omega)\}$ have been calculated explicitly\,\cite{quantum_theory,Ichimaru_BookII,SingwiTosi_Review,Placzek_1952,Puff_1965}. On the other hand, all even frequency moments of $\Im\{\chi(\boldsymbol{q},\omega)\}$ are zero courtesy of its odd frequency parity.

Moreover, linear response theory provides simple correspondence rules that connect the odd frequency moments of $\Im\{\chi(\boldsymbol{q},\omega)\}$ to the odd frequency moments of $S(\boldsymbol{q},\omega)$. The $S(\boldsymbol{q},\omega)$ frequency moments are naturally defined by  
\begin{equation}
M^{(\alpha)}_{\mathrm{S}}(\boldsymbol{q})=\langle\omega^{\alpha}\rangle_{\mathrm{S}}=\int_{-\infty}^{+\infty}d\omega\;\omega^{\alpha}S(\boldsymbol{q},\omega)\,.\label{momentsDSF}
\end{equation}
At arbitrary quantum degeneracy, only four $S(\boldsymbol{q},\omega)$ moments have been calculated explicitly\,\cite{quantum_theory,Dornheim_PRB_2023}: the inverse moment (a consequence of the Kramers-Kronig relations and the fluctuation--dissipation theorem), the zeroth moment (a trivial consequence of the Fourier transform) as well as the first moment and third moment (both emerging from the fundamental nested commutator formula). In the thermodynamic limit, 
\begin{align}
M&^{(-1)}_{\mathrm{S}}(\boldsymbol{q})=-\frac{\hbar}{2n}\chi(\boldsymbol{q})\,,\label{DSFMOMinv}\\
M&^{(0)}_{\mathrm{S}}(\boldsymbol{q})=S(\boldsymbol{q})\,,\label{DSFMOM0}\\
M&^{(1)}_{\mathrm{S}}(\boldsymbol{q})=\frac{\hbar{q}^2}{2m}\,,\label{DSFMOM1}
\end{align}
\begin{align}
M&^{(3)}_{\mathrm{S}}(\boldsymbol{q})=\frac{\hbar{q}^2}{2m}\left\{\left(\frac{\hbar{q}^2}{2m}\right)^2+\frac{{n}q^2}{m}U(\boldsymbol{q})+\frac{2q^2}{m}\langle\hat{K}\rangle_0+\right.\nonumber\\&\left.\frac{1}{m}\int\frac{d^3k}{(2\pi)^3}U(\boldsymbol{k})\left(\frac{\boldsymbol{k}\cdot\boldsymbol{q}}{q}\right)^2\left[S(\boldsymbol{q}-\boldsymbol{k})-S(\boldsymbol{k})\right]\right\},\label{DSFMOM3}
\end{align}
with $n$ the density, $\chi(\boldsymbol{q})=\chi(\boldsymbol{q},\omega=0)$ the static density response function, $S(\boldsymbol{q})$ the static structure factor, $U(\boldsymbol{q})=4\pi{e}^2/q^2$ the Fourier transformed Coulomb pair interaction and $\langle\hat{K}\rangle_0$ the kinetic energy per particle. Thus, even seemingly simple frequency moments such as the second $S(\boldsymbol{q},\omega)$ moment currently lie beyond our reach.

It is emphasized that the previous statements do not apply to the \emph{classical limit} $T/\hbar\omega\to\infty$, where (i) the linear response theory correspondence rule connects the odd frequency moments of $\Im\{\chi(\boldsymbol{q},\omega)\}$ to the even frequency moments of $S(\boldsymbol{q},\omega)$\,\cite{kugler_classical}, (ii) the detailed balance collapses to an even frequency parity of the dynamic structure factor and thus all the odd $S(\boldsymbol{q},\omega)$ frequency moments are zero\,\cite{kugler_bounds,Arkhipov_2017}, (iii) the sixth and eighth $S(\boldsymbol{q},\omega)$ frequency moments have been evaluated explicitly\,\cite{Forster_1968,Bansal_1974,Ichimaru_1975,Ailawadi_1980}
benefiting from the replacement of commutators with Poisson brackets and the use of Yvon's theorem\,\cite{hansen2013theory}.

For completeness, it is pointed out that additional frequency moments are known for the \emph{ground state} $T/\hbar\omega\to0$\,\cite{quantum_theory,Feenberg_Book}. In particular, by restricting the definition of frequency moments in the positive frequency domain and by utilizing the ground state limit of the Lehmann representation, one obtains a fundamental expression that connects the even -- positive domain -- frequency moments of $\Im\{\chi(\boldsymbol{q},\omega)\}$ to the equilibrium expectation values of the product of equal-time operators\,\cite{quantum_theory}. On the basis of the Lehmann representation, a similar foundational formula exists for the even frequency moments of $S(\boldsymbol{q},\omega)$\,\cite{Feenberg_Book}. In fact, the second $S(\boldsymbol{q},\omega)$ moment has been calculated explicitly\,\cite{Huang_1964}; it depends on the so-called kinetic structure factor and it has found applications in theoretical studies of elementary excitations in superfluid helium\,\cite{Hall_1971,Dalfovo_1992,Stringari_1992,Boronat_1995}.

The utility of the frequency moment sum rules is exemplified within the Feynman ansatz, according to which all the significant contributions to the dynamic structure factor stem from collective excitations\,\cite{Feynman_ansatz}. When combined with the low-order frequency moments of $S(\boldsymbol{q},\omega)$, the Feynman ansatz leads to the famous connection between the excitation spectrum and static structure factor, which can provide an accurate description of the collective modes of classical systems\,\cite{Silvestri_2019,Arkhipov_2017} and quantum systems\,\cite{Kalman_2010} under certain conditions. This naturally leads us to the truncated Hamburger problem; the reconstruction of any spectral function from a finite set of its frequency moments\,\cite{tkachenko_book}. The self-consistent method of moments has proven to be successful in the description of the dynamic properties of non-ideal plasmas. On the basis of Nevanlinna's theorem, dynamic properties are expressed via two sets of polynomials (which solely depend on the frequency moments) and the Nevanlinna parameter function (which is analytic on the upper half-plane, has a positive imaginary part and admits a known integral representation). External structural input and physical assumptions are required in order to approximate the unknown Nevanlinna parameter function. In classical dense plasmas, the framework is formulated with respect to the dynamic structure factor or the loss function and an accurate five moment version has been devised based on a Nevanlinna static approximation combined with the empirical observation that $S(\boldsymbol{q},\omega)$ exhibits broad extrema at $\omega=0$\,\cite{Arkhipov_2017,Arkhipov_2020}. In non-ideal degenerate systems, the framework is more conveniently formulated with respect to the loss function and an accurate nine moment version has been recently devised based on the determination of the sixth and eighth loss function frequency moments from a two-parameter Shannon information entropy maximization procedure combined again with a Nevanlinna static approximation\,\cite{Filinov_PRB_2023,Filinov_RSTA_2023}. Evidently, the knowledge of additional frequency moments would lead to improved variants of the self-consistent method of moments.

Very recently, Dornheim \emph{et al.} introduced an exact framework for the computation of positive $S(\boldsymbol{q},\omega)$ frequency moments from path integral Monte Carlo simulations (PIMC)\,\cite{Dornheim_PRB_2023,Dornheim_MRE_2023}. Finite temperature QMC simulations are typically restricted to the imaginary--time domain and provide access to the imaginary--time density-density correlation function (ITCF) that corresponds one-to-one to the dynamic structure factor in the real frequency domain via the two-sided Laplace transform (with $\tau\in[0,\beta]$ denoting the imaginary time argument)\,\cite{Berne_JCP_1983,cep}
\begin{equation}
F(\boldsymbol{q},\tau)=\int_{-\infty}^{+\infty}d\omega\;{S}(\boldsymbol{q},\omega)e^{-\hbar\omega\tau}\,.\label{LaplaceITCF}
\end{equation}
The framework allows the extraction of $M^{(\alpha)}_{\mathrm{S}}(\boldsymbol{q})$ directly from the ITCF data, thus circumventing the Laplace inversion of the ITCF\,\cite{Dornheim_T_2022,Dornheim_Nature_2025,dornheim_POP_2025,schwalbe2025staticlineardensityresponse}, which is a well-known ill-posed problem with respect to the Monte Carlo error bars\,\cite{epstein2008badtruth,JARRELL1996133,chuna2025dualformulationmaximumentropy}. In particular, the first five moments of the dynamic structure factor were successfully computed for the warm dense UEG\,\cite{review,Dornheim_review}; the first and the third frequency moments were benchmarked against the known sum rules while the second, the fourth and the fifth frequency moments were evaluated for the first time\,\cite{Dornheim_PRB_2023}.

Here, based on exact results of linear density response theory, we derive a novel series that expresses the (unknown) even frequency moments of the dynamic structure factor via the (known) odd frequency moments. For the moderately to weakly degenerate UEG, we propose and validate truncated sums for the second, fourth and fifth frequency moments. The paper is organized as follows: In Sec.~\ref{sec:theory}, we derive the exact series representation for the even frequency moments~(\ref{subsec:theoryseries}), the truncations of the series that facilitate approximations of unknown frequency moments~(\ref{subsec:theorytruncation}), the relation between the frequency moments and the ITCFs~(\ref{subsec:theoryPIMC}) and the frequency moments in the non-interacting limit~(\ref{subsec:theoryFermi}). Sec.\ref{sec:results} is devoted to determining the applicability range of the approximations by utilizing the non-interacting limit (\ref{subsec:resultsFermi}) and by comparing with PIMC simulations (\ref{subsec:resultsPIMC}). The paper is concluded by a summary and outlook in Sec.~\ref{sec:outlook}.

\section{Theory\label{sec:theory}}

\subsection{Exact series representation for the even frequency moments of the dynamic structure factor\label{subsec:theoryseries}}

The starting point is the general definition of the frequency moments of the dynamic structure factor, Eq.(\ref{momentsDSF}). Substituting for the fluctuation--dissipation theorem and exploiting the odd frequency parity of $\Im\{\chi(\boldsymbol{q},\omega)\}$ yields
\begin{align}
M^{(\alpha)}_{\mathrm{S}}(\boldsymbol{q})&=-\frac{\hbar}{\pi{n}}\int_{0}^{+\infty}d\omega\;\omega^{\alpha}\Im\{\chi(\boldsymbol{q},\omega)\}\times\nonumber\\&\quad\left[\frac{1}{1-e^{-\beta\hbar\omega}}+\frac{(-1)^{\alpha+1}}{1-e^{\beta\hbar\omega}}\right]\nonumber\,.
\end{align}
In case of odd frequency moments, $\alpha=2k+1$, this leads to the known correspondence between the frequency moments of $S(\boldsymbol{q},\omega)$ and of $\Im\{\chi(\boldsymbol{q},\omega)\}$ that reads as\,\cite{Dornheim_PRB_2023}
\begin{equation}
M^{(2k+1)}_{\mathrm{S}}(\boldsymbol{q})=-\frac{\hbar}{2\pi{n}}M^{(2k+1)}_{\mathrm{Im}\chi}(\boldsymbol{q})\,.\label{momentcorr}
\end{equation}
In case of even frequency moments, $\alpha=2k$, after some algebraic manipulations, this leads to
\begin{equation*}
M^{(2k)}_{\mathrm{S}}(\boldsymbol{q})=-\frac{\hbar}{2\pi{n}}\int_{-\infty}^{+\infty}d\omega\;\omega^{2k}\coth{\left(\frac{\beta\hbar\omega}{2}\right)}\Im\{\chi(\boldsymbol{q},\omega)\}\,.
\end{equation*}
The series representation of the hyperbolic cotangent involves the Bernoulli numbers\,\cite{GradshteynRyzhik},
\begin{equation*}
\coth{(x)}=\sum_{\ell=0}^{\infty}\frac{2^{2\ell}\mathrm{B}_{2\ell}}{(2\ell)!}x^{2\ell-1}\,,
\end{equation*}
with $B_{2\ell}=\{1,1/6,-1/30,1/42,-1/30,5/66\cdots\}$. Substitution of the above Laurent series and interchange of the series-integral operators lead to
\begin{align*}
M^{(2k)}_{\mathrm{S}}(\boldsymbol{q})&=-\frac{\hbar}{2\pi{n}}\sum_{\ell=0}^{\infty}\frac{2^{2\ell}\mathrm{B}_{2\ell}}{(2\ell)!}\left(\frac{\beta\hbar}{2}\right)^{2\ell-1}\times\nonumber\\&\quad\int_{-\infty}^{+\infty}d\omega\;\omega^{2\ell+2k-1}\Im\{\chi(\boldsymbol{q},\omega)\}\,.
\end{align*}
The series converges for $|x|\leq\pi$, which translates to $|\hbar\omega|\leq2\pi{T}$, thus this is a high temperature approximation. In view of the correspondence between odd $S(\boldsymbol{q},\omega)$ and odd $\Im\{\chi(\boldsymbol{q},\omega)\}$ moments, see Eq.(\ref{momentcorr}), one ends up with
\begin{equation}
M^{(2k)}_{\mathrm{S}}(\boldsymbol{q})=\sum_{\ell=0}^{\infty}\frac{2^{2\ell}\mathrm{B}_{2\ell}}{(2\ell)!}\left(\frac{\beta\hbar}{2}\right)^{2\ell-1}M^{(2\ell+2k-1)}_{\mathrm{S}}(\boldsymbol{q})\,.\label{momentseriesgen}
\end{equation}
This finite temperature expression connects any even $2k$-$S(\boldsymbol{q},\omega)$ moment with the infinite sum of all odd-$S(\boldsymbol{q},\omega)$ moments starting from the $2k-1$ moment. In numerical evaluations, it is preferable to work with the normalized frequency moments $\widetilde{M}^{(\alpha)}_{\mathrm{S}}(\boldsymbol{q})=(\hbar/E_{\mathrm{f}})^{\alpha}M^{(\alpha)}_{\mathrm{S}}(\boldsymbol{q})$. Within this notation, after introducing the degeneracy parameter $\Theta=T/E_{\mathrm{f}}$, the series representation becomes
\begin{equation}
\widetilde{M}^{(2k)}_{\mathrm{S}}(\boldsymbol{q})=2\sum_{\ell=0}^{\infty}\frac{\mathrm{B}_{2\ell}}{(2\ell)!}\frac{1}{\Theta^{2\ell-1}}\widetilde{M}^{(2\ell+2k-1)}_{\mathrm{S}}(\boldsymbol{q})\,.\label{momentseriesgennorm}
\end{equation}

\subsection{Truncated sum approximations for unknown frequency moments of the dynamic structure factor\label{subsec:theorytruncation}}

Setting $k=0$ in Eq.(\ref{momentseriesgen}), keeping the four leading order terms, substituting for the explicit forms of the inverse frequency moment, the zeroth frequency moment and the first frequency moment, see Eqs.(\ref{DSFMOMinv},\ref{DSFMOM0},\ref{DSFMOM1}), a truncated sum is derived for the static structure factor that reads as
\begin{equation}
S(\boldsymbol{q})\simeq-\frac{\chi(\boldsymbol{q})}{n\beta}+\frac{\beta\hbar^2q^2}{12m}-\frac{\beta^3\hbar^3}{360}M_{S}^{(3)}(\boldsymbol{q})+\frac{\beta^5\hbar^5}{15120}M_{S}^{(5)}(\boldsymbol{q})-\cdots\,.\,.\label{momentseries0}
\end{equation}
Since the third frequency moment $M_{S}^{(3)}(\boldsymbol{q})$ is also known, see Eq.(\ref{DSFMOM3}), the truncated sum of Eq.(\ref{momentseries0}) can provide estimates for the fifth frequency moment $M_{S}^{(5)}(\boldsymbol{q})$, whose explicit form remains unknown. 

Setting $k=1$ in Eq.(\ref{momentseriesgen}), keeping the three leading order terms and substituting for the explicit form of the first frequency moment, see Eq.(\ref{DSFMOM1}), a truncated sum is derived for the second frequency moment $M_{S}^{(2)}(\boldsymbol{q})$ that reads as
\begin{equation}
M_{S}^{(2)}(\boldsymbol{q})\simeq\frac{q^2}{m\beta}+\frac{\beta\hbar}{6}M_{S}^{(3)}(\boldsymbol{q})-\frac{\beta^3\hbar^3}{360}M_{S}^{(5)}(\boldsymbol{q})+\cdots\,.\label{momentseries1}
\end{equation}
As a consequence, the truncated sums of Eqs.(\ref{momentseries0},\ref{momentseries1}) can provide estimates for the unknown second frequency moment $M_{S}^{(2)}(\boldsymbol{q})$. 

Setting $k=2$ in Eq.(\ref{momentseriesgen}), keeping the two leading order terms, a truncated sum is derived for the fourth frequency moment $M_{S}^{(4)}(\boldsymbol{q})$ that reads as
\begin{equation}
M_{S}^{(4)}(\boldsymbol{q})\simeq\frac{2}{\beta\hbar}M_{S}^{(3)}(\boldsymbol{q})+\frac{\beta\hbar}{6}M_{S}^{(5)}(\boldsymbol{q})-\cdots\,.\label{momentseries2}
\end{equation}
As a consequence, the truncated sums of Eqs.(\ref{momentseries0},\ref{momentseries2}) can provide estimates for the unknown fourth frequency moment $M_{S}^{(4)}(\boldsymbol{q})$. 

Essentially, the early truncation of the expansion of Eq.(\ref{momentseriesgen}) corresponds to $\beta\hbar\omega\ll1$ or to $T\gg\hbar\omega$, \emph{i.e.} it constitutes a semi-classical approximation. Thus, it is no surprise that the first term of the static structure factor expansion, see Eq.(\ref{momentseries0}), reproduces the known classical result $S(\boldsymbol{q})=-\chi(\boldsymbol{q})/(n\beta)$\,\cite{kugler_classical,Tolias_PoPbrief_2021}. It is emphasized that the aforementioned truncations of the infinite sum constitute the only possible way to estimate the second, fourth and fifth frequency moments without utilizing an approximation for the dynamic structure factor. It is evident that the accuracy of the estimates should strongly depend on value of the degeneracy parameter $\Theta$.

\subsection{Frequency moments of the dynamic structure factor from imaginary--time correlation functions\label{subsec:theoryPIMC}}

We follow Refs.\cite{Dornheim_PRB_2023,Dornheim_MRE_2023} and we repeat the main derivation for the sake of completeness. The starting point is the Laplace transform definition of ITCFs, see Eq.(\ref{LaplaceITCF}). The exponential factor is Taylor expanded and then the series-integral operators are interchanged yielding
\begin{equation}
F(\boldsymbol{q},\tau)=\sum_{k=0}^{\infty}\frac{(-\hbar)^{\alpha}}{\alpha!}\left\{\int_{-\infty}^{+\infty}\omega^{\alpha}S(\boldsymbol{q},\omega)d\omega\right\}\tau^{\alpha}\,.\nonumber
\end{equation}
The frequency moments of the dynamic structure factor, see Eq.(\ref{momentsDSF}), are easily identified, which leads to 
\begin{equation}
F(\boldsymbol{q},\tau)=\sum_{\alpha=0}^{\infty}\frac{(-\hbar)^{\alpha}}{\alpha!}M_{S}^{(\alpha)}(\boldsymbol{q})\tau^{\alpha}\,.\label{ITCFmom1}
\end{equation}
On the other hand, ITCFs can also be Taylor expanded around the imaginary--time origin $\tau=0$,
\begin{equation}
F(\boldsymbol{q},\tau)=\sum_{\alpha=0}^{\infty}\frac{1}{\alpha!}\left.\frac{\partial^{\alpha}F(\boldsymbol{q},\tau)}{\partial\tau^{\alpha}}\right|_{\tau=0}\tau^{\alpha}\,.\label{ITCFmom2}
\end{equation}
As Eqs.(\ref{ITCFmom1},\ref{ITCFmom2}) are valid for any permissible imaginary--time, the equal-power series coefficients should be equal. Thus, one obtains an expression for any $\alpha-S(\boldsymbol{q},\omega)$ moment via the $\alpha-$order derivative of the ITCF at $\tau=0$~\cite{Dornheim_PRB_2023}
\begin{equation}
M_{S}^{(\alpha)}(\boldsymbol{q})=\frac{(-1)^{\alpha}}{\hbar^{\alpha}}\left.\frac{\partial^{\alpha}F(\boldsymbol{q},\tau)}{\partial\tau^{\alpha}}\right|_{\tau=0}\,.\label{ITCFmom3}
\end{equation}

The above exact correspondence was first exploited in Ref.\cite{Dornheim_PRB_2023}, where the first five moments of the $S(\boldsymbol{q},\omega)$ for the warm dense UEG were extracted from high quality PIMC simulation data for the ITCF. In particular, this was achieved through a canonical polynomial representation of the ITCF for each wavenumber
\begin{equation}
F(\boldsymbol{q},\tau)=\sum_{\alpha=0}^{\alpha_{\mathrm{max}}}c_{\alpha}(\boldsymbol{q})\tau^{\alpha}\,,\label{ITCFcanonical}
\end{equation}
which translates to
\begin{equation}
M_{S}^{(\alpha)}(\boldsymbol{q})=\frac{(-1)^{\alpha}}{\hbar^{\alpha}}\alpha!c_{\alpha}(\boldsymbol{q})\,.\label{ITCFmom4}
\end{equation}
This polynomial expansion provided much more accurate results than the numerical evaluation of the $\tau-$derivatives of the ITCF through finite difference approximations on the given PIMC $\tau-$grid\,\cite{Dornheim_PRB_2023}.

\subsection{Frequency moments of the dynamic structure factor in the non-interacting limit\label{subsec:theoryFermi}}

In the non-interacting limit, all the even and odd order frequency moments of the non-interacting dynamic structure factor can be quite easily evaluated. The evaluation can be based on the known imaginary part of the linear density response function (for odd frequency orders only), the known dynamic structure factor (for even and odd orders), the known ITCF (for even and odd orders) or the fundamental iterated commutator expression (for odd orders only). 

The most straightforward calculation is based on the combination of the $\tau=0$ ITCF differentiation property of Eq.(\ref{ITCFmom3}) with the known expression for the non-interacting ITCF\,\cite{Tolias_CtPP2025}
\begin{align}
F_{\mathrm{HF}}(x,t^{\star})&=\frac{3\Theta}{8}\int_{0}^{+\infty}\frac{\cosh{\left[\frac{xy}{\Theta}(t^{\star}-1/2)\right]}}{\sinh{\left(\frac{xy}{2\Theta}\right)}}\times\nonumber\\&\quad\ln{\left\{\frac{1+\exp{\left[\bar{\mu}-\frac{(x-y)^2}{4\Theta}\right]}}{1+\exp{\left[\bar{\mu}-\frac{(x+y)^2}{4\Theta}\right]}}\right\}}dy\,.\label{ITCFnoninteracting}    
\end{align}
In the above, $t^{\star}=T\tau$ and $x=q/q_{\mathrm{F}}$ denote the normalized imaginary--time $t^{\star}$ and the normalized wavenumber $x$, while $\bar{\mu}=\mu/T$ denotes the normalized chemical potential. The interchange of the differential and integral operators, the use of the chain-differentiation rule, the utilization of the normalized frequency moments and the substitution of the derivatives $(\partial/\partial{x})^{2k+1}\cosh{(ax)}=a^{2k+1}\sinh{(ax)}$, $(\partial/\partial{x})^{2k}\cosh{(ax)}=a^{2k}\cosh{(ax)}$ leads to the following numerically convenient expressions for the odd and even moments of the non-interacting ITCF, respectively,
\begin{align}
&\widetilde{M}_{S_{\mathrm{HF}}}^{(2k+1)}(x)=\frac{3\Theta}{8}x^{2k+1}\int_{0}^{+\infty}y^{2k+1}\times\nonumber\\&\quad\ln{\left\{\frac{1+\exp{\left[\bar{\mu}-\frac{(x-y)^2}{4\Theta}\right]}}{1+\exp{\left[\bar{\mu}-\frac{(x+y)^2}{4\Theta}\right]}}\right\}}dy\,,\label{ITCFmomHFodd}\\
&\widetilde{M}_{S_{\mathrm{HF}}}^{(2k)}(x)=\frac{3\Theta}{8}x^{2k}\int_{0}^{+\infty}y^{2k}\coth{\left(\frac{xy}{2\Theta}\right)}\times\nonumber\\&\quad\ln{\left\{\frac{1+\exp{\left[\bar{\mu}-\frac{(x-y)^2}{4\Theta}\right]}}{1+\exp{\left[\bar{\mu}-\frac{(x+y)^2}{4\Theta}\right]}}\right\}}dy\,.\label{ITCFmomHFeven}
\end{align}

As a sanity check, we have analytically and numerically confirmed that, for $k=0$ and $k=1$, the general non-interacting odd moment expression of Eq.(\ref{ITCFmomHFodd}) yields
\begin{align*}
\widetilde{M}_{S_{\mathrm{HF}}}^{(1)}(x)&=x^2\,,\\
\widetilde{M}_{S_{\mathrm{HF}}}^{(3)}(x)&=x^6+6\Theta^{5/2}I_{3/2}(\bar{\mu})x^4\,,
\end{align*}
with $I_{\nu}$ the complete Fermi integral of the $\nu-$order\,\cite{GradshteynRyzhik}, 
\begin{equation}
I_{\nu}(\bar{\mu})=\int_0^{\infty}dz\frac{z^v}{\exp{\left(z-\bar{\mu}\right)}+1}\,,
\end{equation}
which are the f-sum rule and the non-interacting limit of the third moment sum rule, respectively\,\cite{quantum_theory}. 

Notice that the $y^{2k+1}$ pre-factor in the integrand of Eq.(\ref{ITCFmomHFodd}) enables a very simple integration by parts regardless of the value of $k$, which ultimately leads to the emergence of complete Fermi integrals that correspond to $\langle\sum_i{\hat{\boldsymbol{p}}_i}^{2k}\rangle_0$ equilibrium expectation values. On the other hand, the $y^{2k}\coth{(xy/2\Theta)}$ pre-factor in the integrand of Eq.(\ref{ITCFmomHFeven}) does not enable a simple integration by parts, which reflects the lack of a foundational formula that expresses the even frequency moments of the dynamic structure factor through the equilibrium expectation value of equal time operators. 

\begin{figure*}[!h]\centering
\includegraphics[width=1.0\textwidth]{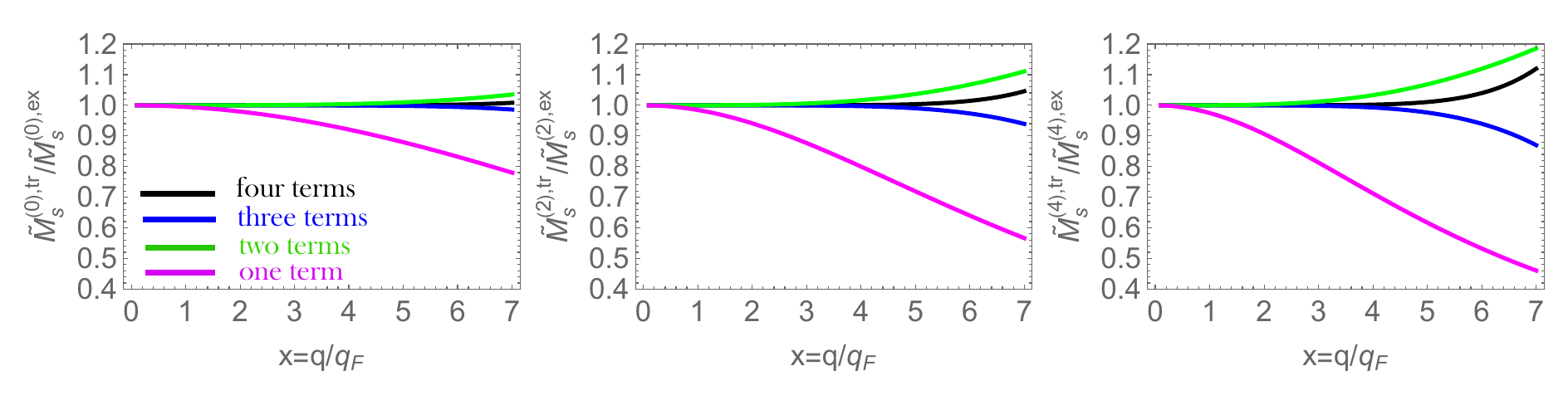}\\\vspace*{-0.25cm}
\includegraphics[width=1.0\textwidth]{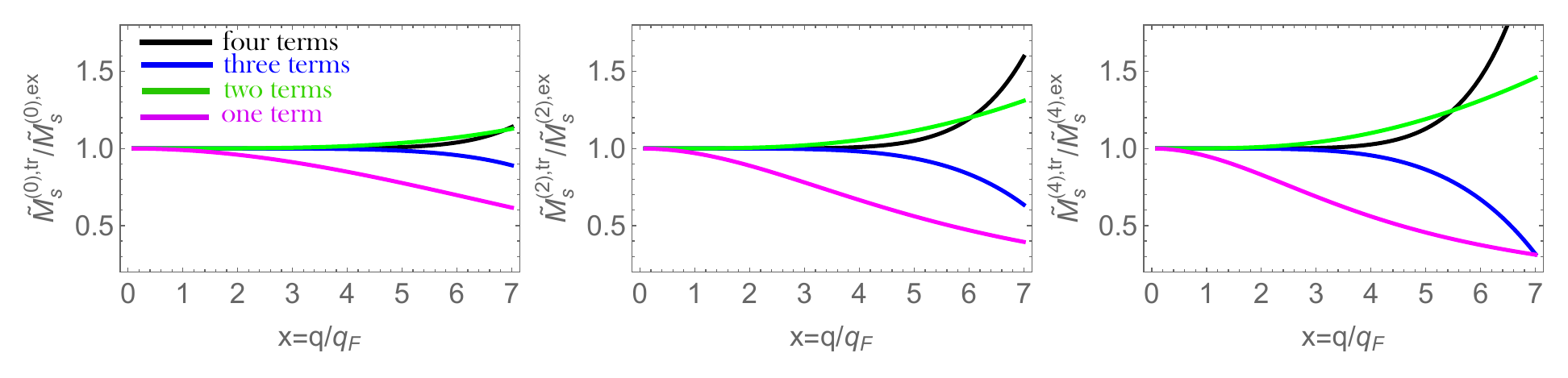}\\\vspace*{-0.25cm}
\includegraphics[width=1.0\textwidth]{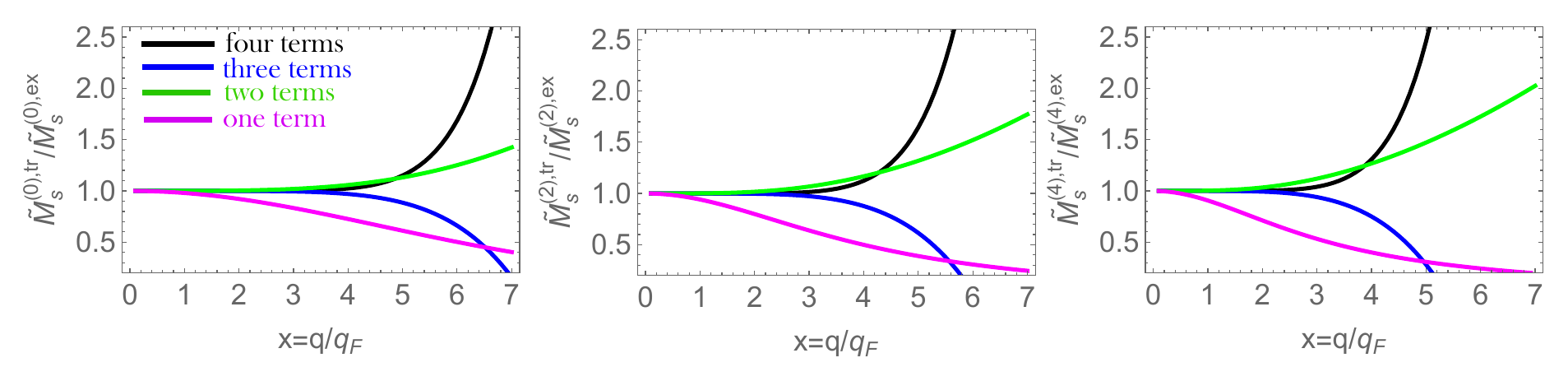}\\\vspace*{-0.25cm}
\includegraphics[width=1.0\textwidth]{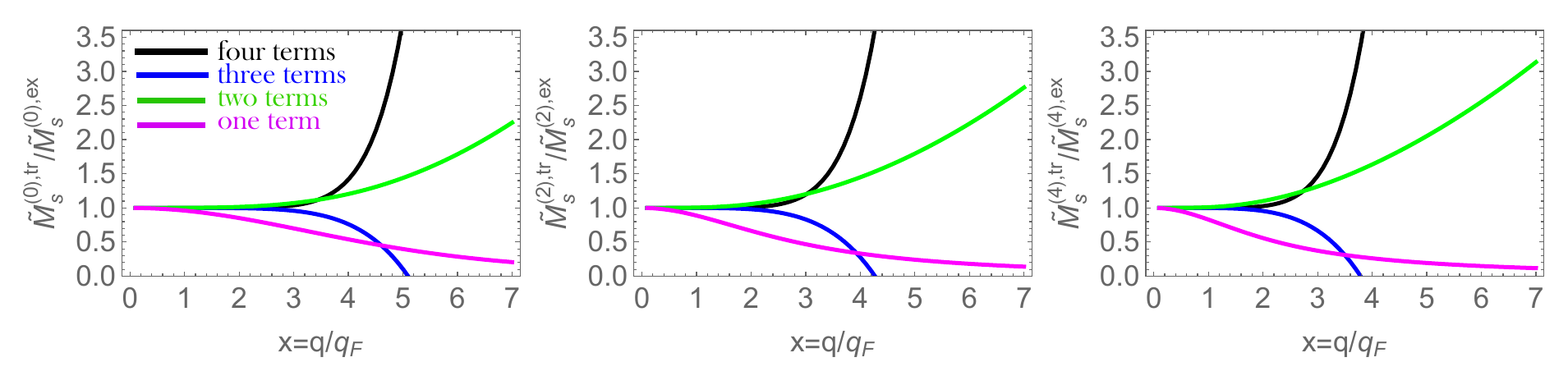}\\\vspace*{-0.25cm}
\includegraphics[width=1.0\textwidth]{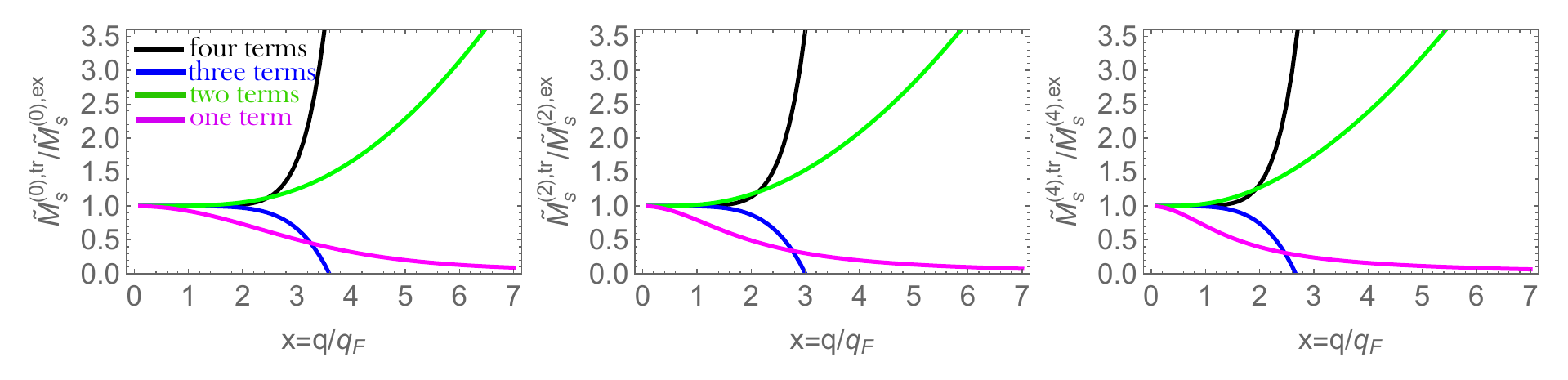}\caption{\label{fig:Fermiall} Results for the non-interacting Fermi gas ($r_{\mathrm{s}}\to0$) at $\Theta=32$ (top), $\Theta=16$ (second from top), $\Theta=8$ (third from top), $\Theta=4$ (fourth from top) and $\Theta=2$ (bottom). The ratio of the  truncated frequency moment over the exact frequency moment for $\alpha=0$ (left), $\alpha=2$ (center), $\alpha=4$ (right).}
\end{figure*}

\section{Results\label{sec:results}}

\subsection{Exact results for the non-interacting UEG\label{subsec:resultsFermi}}

We begin with the non-interacting UEG at finite temperatures. In Fig.\ref{fig:Fermiall}, we plot the ratio of truncated series approximations of the even frequency moments, see Eq.(\ref{momentseriesgennorm}), over the exact even frequency moments. The odd and even frequency moments of the non-interacting UEG are calculated via Eqs.(\ref{ITCFmomHFodd},\ref{ITCFmomHFeven}), respectively. The truncated series approximations feature one, two, three and four terms. The even frequency moments of interest are the zeroth, second and fourth. The degeneracy parameters of the examples are $\Theta=32,\,16,\,8,\,4,\,2$. It is evident that early truncations of the series representation will be very accurate for $\Theta\geq32$ and that the series representation fails for $\Theta\leq2$. The following observations are due:

(1) In general, the level of accuracy of the truncated series approximations strongly increases with the degeneracy parameter $\Theta$, decreases with the wavenumber $q$ and decreases with the order of the even frequency moment $2k$. On the other hand, higher orders of truncations $\ell$ do not always translate to more accurate approximations.

(2) Regardless of the degeneracy parameter and the even frequency order, the first term of the series representation is exact at the long wavelength limit $q/q_{\mathrm{F}}\to0$. To be more specific, in view of Eqs.(\ref{ITCFmomHFodd},\ref{ITCFmomHFeven}), it is straightforward to prove that
\begin{equation}
\widetilde{M}_{S_{\mathrm{HF}}}^{(2k)}(x\to0)=2\Theta\widetilde{M}_{S_{\mathrm{HF}}}^{(2k-1)}(x\to0)\,.
\end{equation}

(3) Regardless of the degeneracy parameter and the even frequency order, the first term of the series representation is accurate only near the long wavelength limit $q\lesssim{q}_{\mathrm{F}}$ and it progressively underestimates the exact frequency moment at high wavenumbers, see Fig.\ref{fig:Fermiall}. Therefore, in the non-interacting case, the first term of the series representation of Eq.(\ref{momentseriesgen}) serves as a lower bound to the even frequency moments. For the zero frequency moment, this holds in the general interacting case, for which a system-agnostic result of linear density response theory states that the classical limit of the static density response function $-n\beta{S}(\boldsymbol{q})$ serves as an upper bound (in absolute values) of the quantum static density response function $\chi(\boldsymbol{q})$\,\cite{quantum_theory,kugler_bounds,Dornheim_EPL_2024}.

(4) Irrespective of temperature, many electron systems always exhibit quantum mechanical behavior provided that the length scales of interest are sufficiently small. As aforementioned, the truncation of the exact series representation for the even frequency moments of the dynamic structure factor constitutes a semi-classical approximation. Therefore, regardless of the degeneracy parameter and even frequency order, any truncated summations are bound to fail at large enough wavenumbers. In fact, for the zero frequency moment, truncations featuring two or more terms begin to diverge from the exact result around $q\sim6q_{\mathrm{F}}$ at $\Theta=32$, $q\sim5q_{\mathrm{F}}$ at $\Theta=16$, $q\sim4q_{\mathrm{F}}$ at $\Theta=8$, $q\sim3q_{\mathrm{F}}$ at $\Theta=4$ and $q\sim2q_{\mathrm{F}}$ at $\Theta=2$ as discerned from the left panels of Fig.\ref{fig:Fermiall}. It is also evident that, as the even frequency order increases, the truncated summations fail at gradually smaller wavenumbers. For instance, at $\Theta=16$, the three term truncated sum begins to diverge from the exact results around $q\sim5q_{\mathrm{F}}$ for the zeroth moment, $q\sim4q_{\mathrm{F}}$ for the second moment and $q\sim3q_{\mathrm{F}}$ for the fourth moment, see the second row of Fig.\ref{fig:Fermiall}.

(5) Higher orders of truncation do not necessarily imply a better approximation of the exact result. This is not surprising given that the hyperbolic cotangent Laurent series only converges for $|\hbar\omega|\less2\pi{T}$. Therefore, the series representation does not constitute an ordinary perturbative series with respect to $1/\Theta$. For instance, at $\Theta=2,4,8,16$ and for any even frequency order, the two term truncated sum is more accurate than the three and four term truncated sums at the short wavelength limit, see the respective rows of Fig.\ref{fig:Fermiall}. As a more extreme example, at $\Theta=2,\,4$ and for any frequency order, the three term truncated sum acquires un-physical negative values at intermediate to large wavenumbers. Arithmetically, the origin of this anomalous behavior can be easily tracked down to the fact that the Bernoulli sequence also features negative rational numbers.

(6) Provided that the short wavelength limit is not considered, the two term truncated sums provide an accurate description of the zeroth, second and fourth frequency moments at $\Theta=8,\,16,\,32$. At $\Theta=2,4$, the first two terms of the series representation are only accurate within a restricted wavenumber range, especially for the fourth frequency moment.

\subsection{PIMC simulation results for the warm dense UEG\label{subsec:resultsPIMC}}

\begin{figure*}\centering
\includegraphics[width=0.44\textwidth]{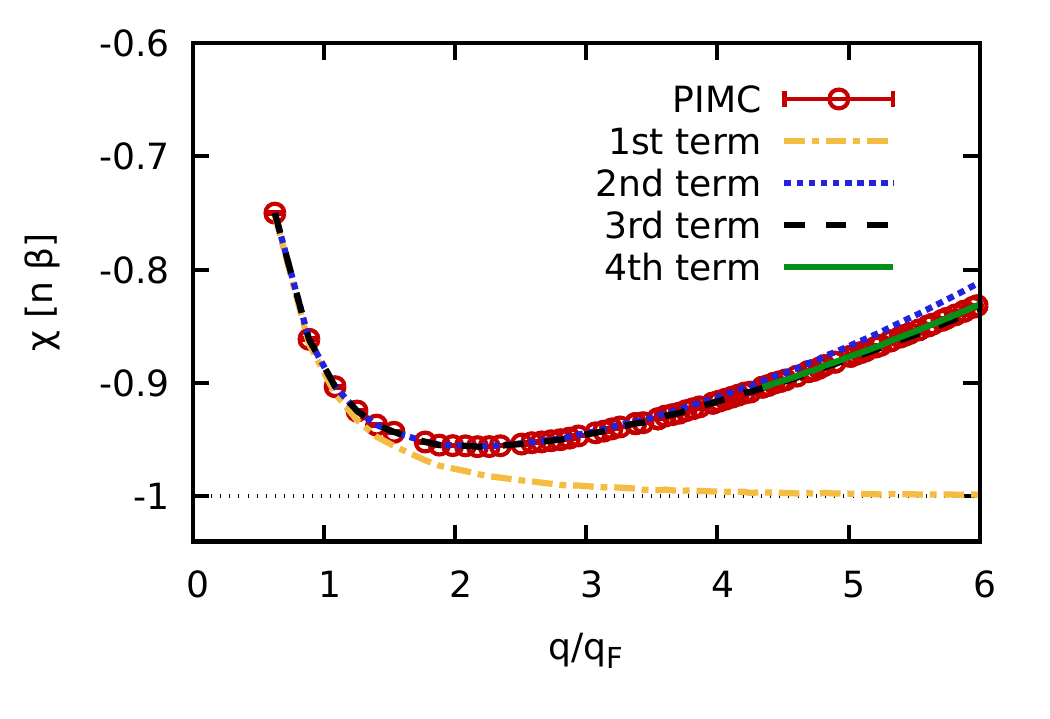}\includegraphics[width=0.44\textwidth]{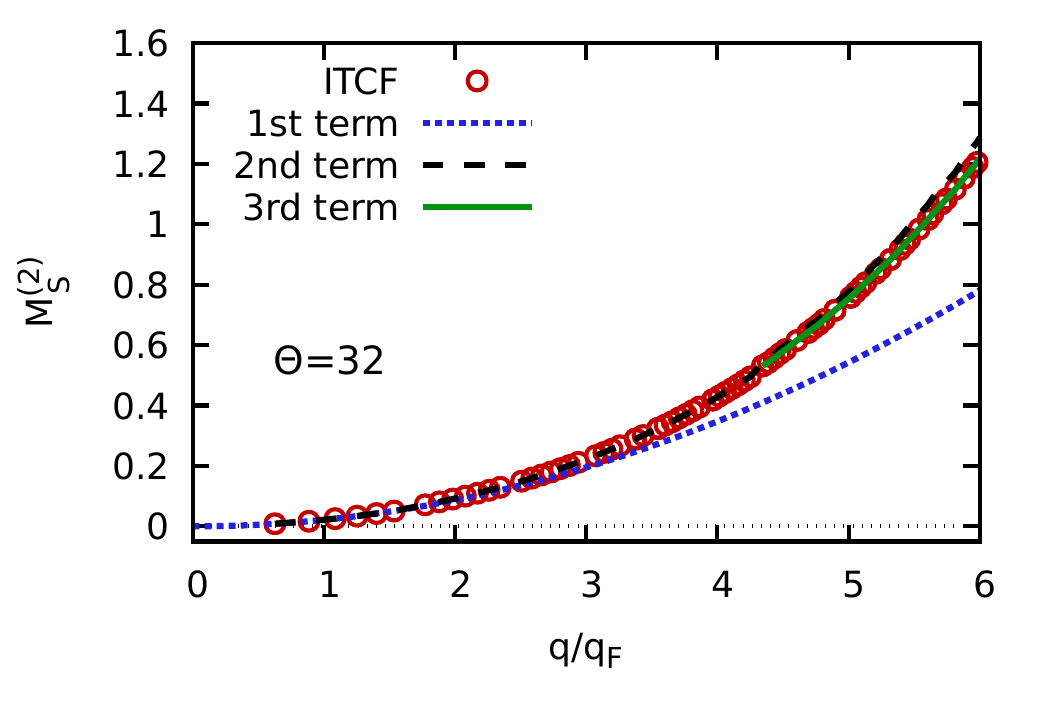}\\\vspace*{-1.25cm}
\includegraphics[width=0.44\textwidth]{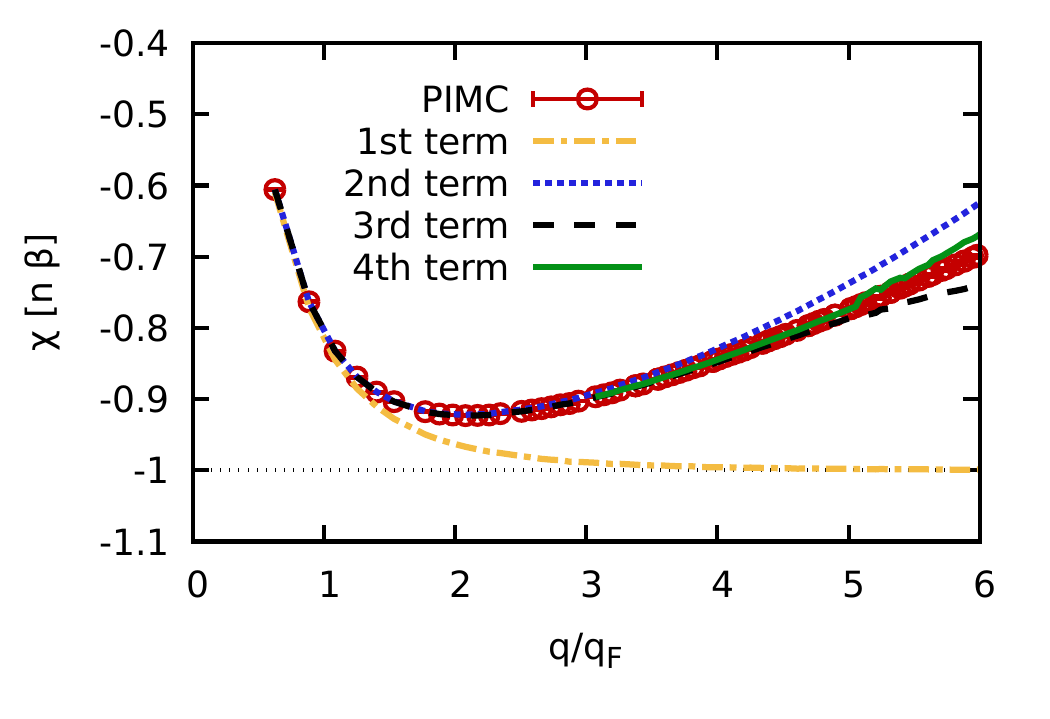}\includegraphics[width=0.44\textwidth]{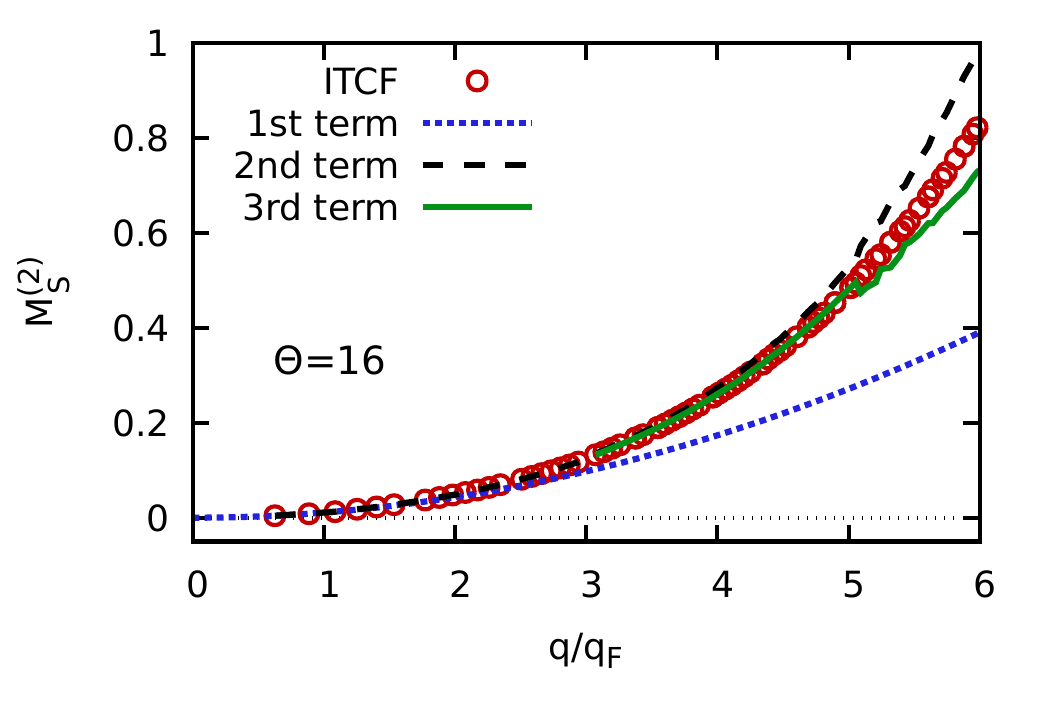}\\\vspace*{-1.25cm}
\includegraphics[width=0.44\textwidth]{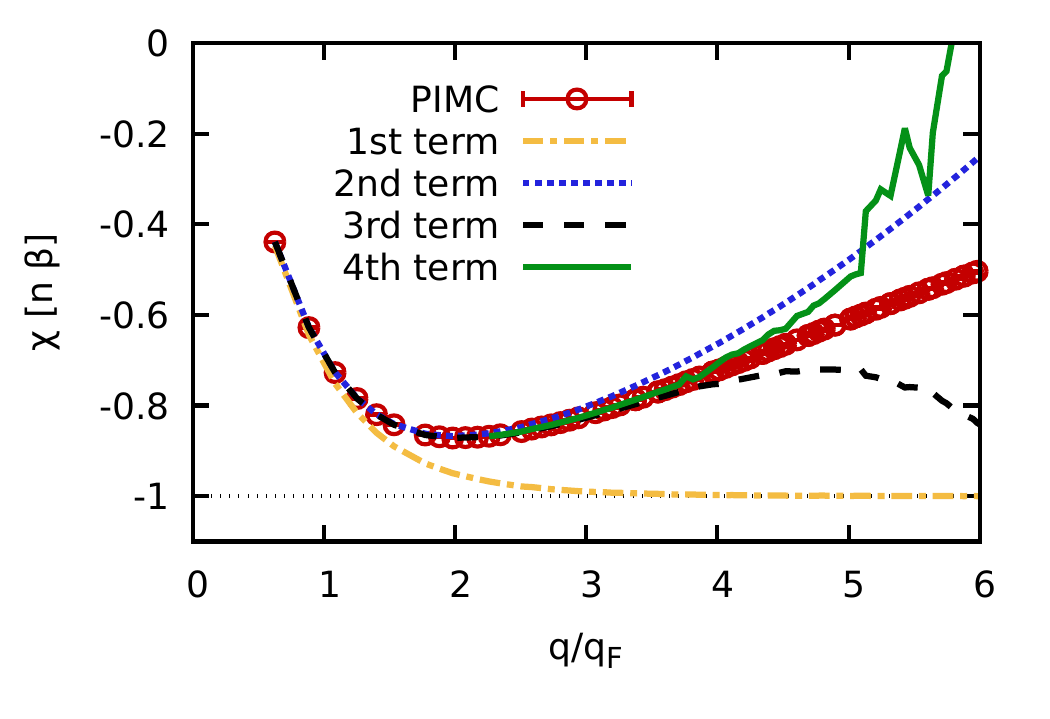}\includegraphics[width=0.44\textwidth]{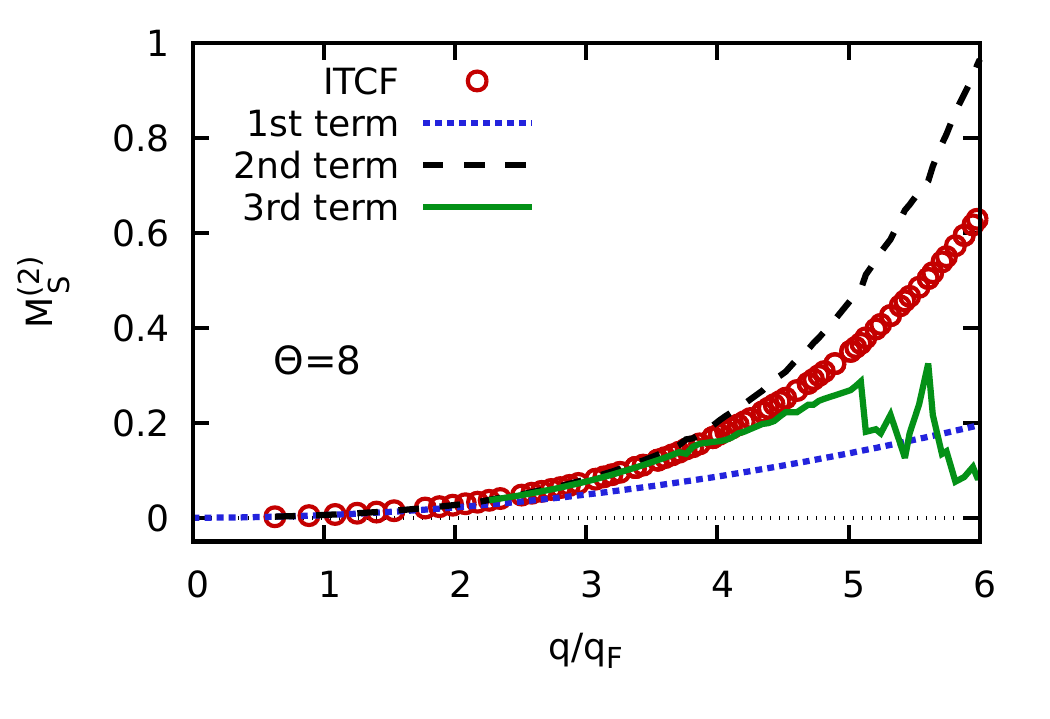}\\\vspace*{-1.25cm}
\includegraphics[width=0.44\textwidth]{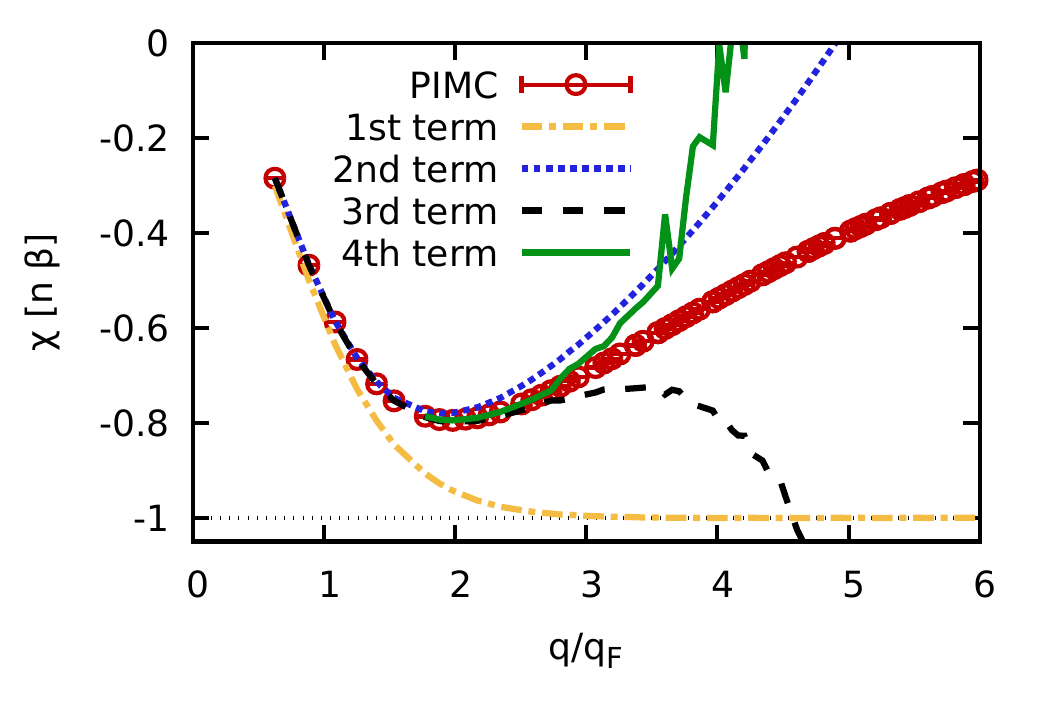}\includegraphics[width=0.44\textwidth]{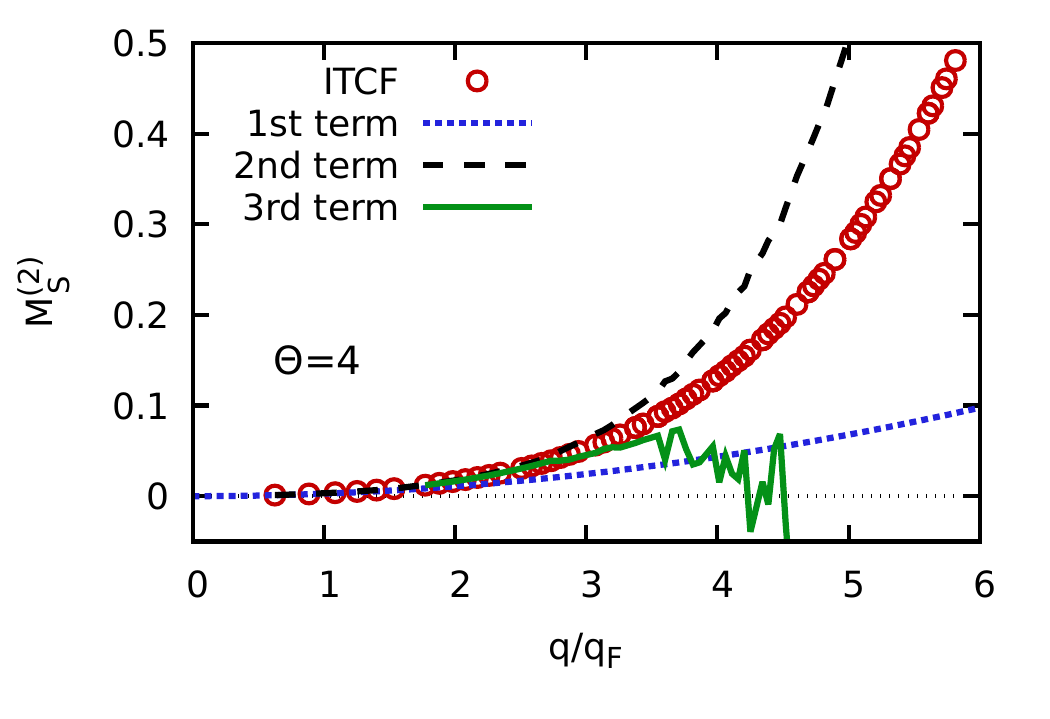}\\\vspace*{-1.25cm}
\includegraphics[width=0.44\textwidth]{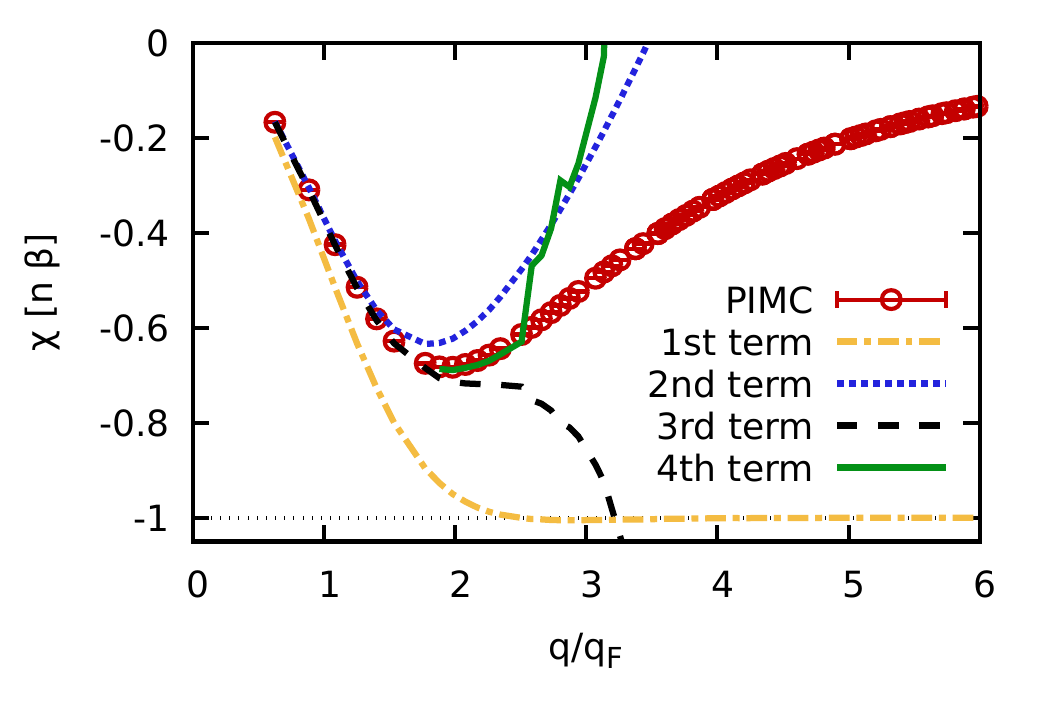}\includegraphics[width=0.44\textwidth]{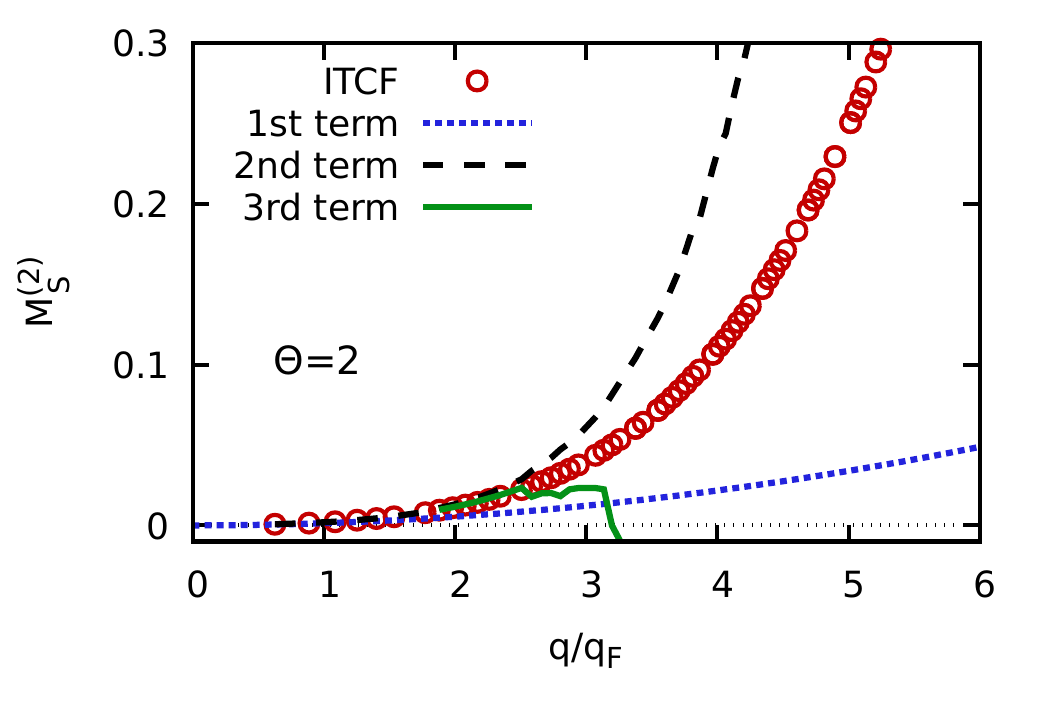}
\caption{\label{fig:UEG_rs10} \emph{Ab initio} PIMC results for the unpolarized UEG at $r_s=10$ with $N=34$ particles. From top to bottom, the rows correspond to $\Theta=32,16,8,4$ and $2$. The left column corresponds to the static density response function $\chi(\mathbf{q})$ that determines the inverse frequency moment $M^{(-1)}_S$ of the DSF [Eq.(\ref{DSFMOMinv})] with the red dots computed from the ITCF via Eq.(\ref{eq:static_chi}) and the solid lines corresponding to different truncations of the Eq.(\ref{eq:chi}) expansion. The right column corresponds to the second frequency moment, with the red dots computed from the ITCF via Eq.(\ref{ITCFmom4}) and the solid lines corresponding to different truncations of the Eq.(\ref{momentseries1}) expansion.}
\end{figure*} 

\begin{figure*}\centering
\includegraphics[width=0.44\textwidth]{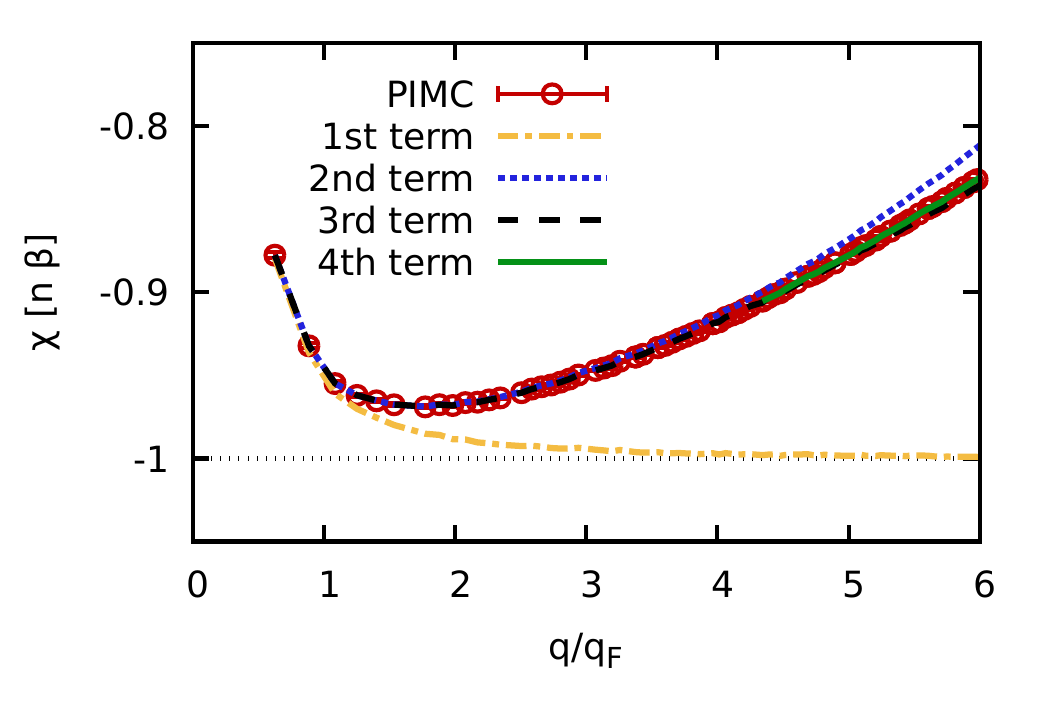}\includegraphics[width=0.44\textwidth]{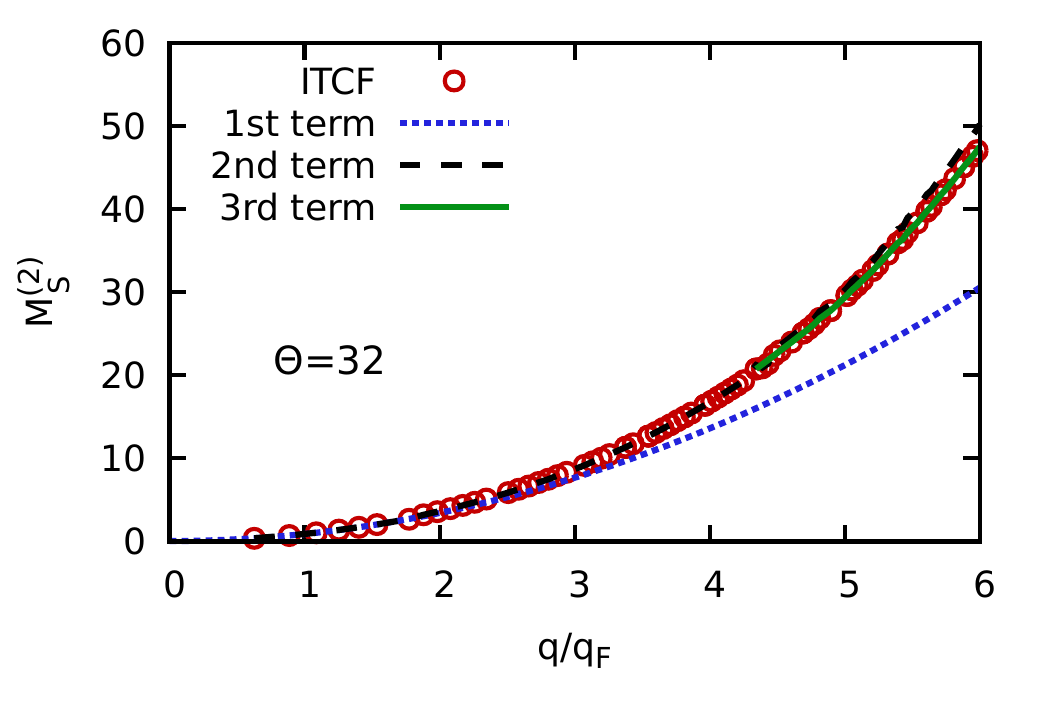}\\\vspace*{-1.25cm}
\includegraphics[width=0.44\textwidth]{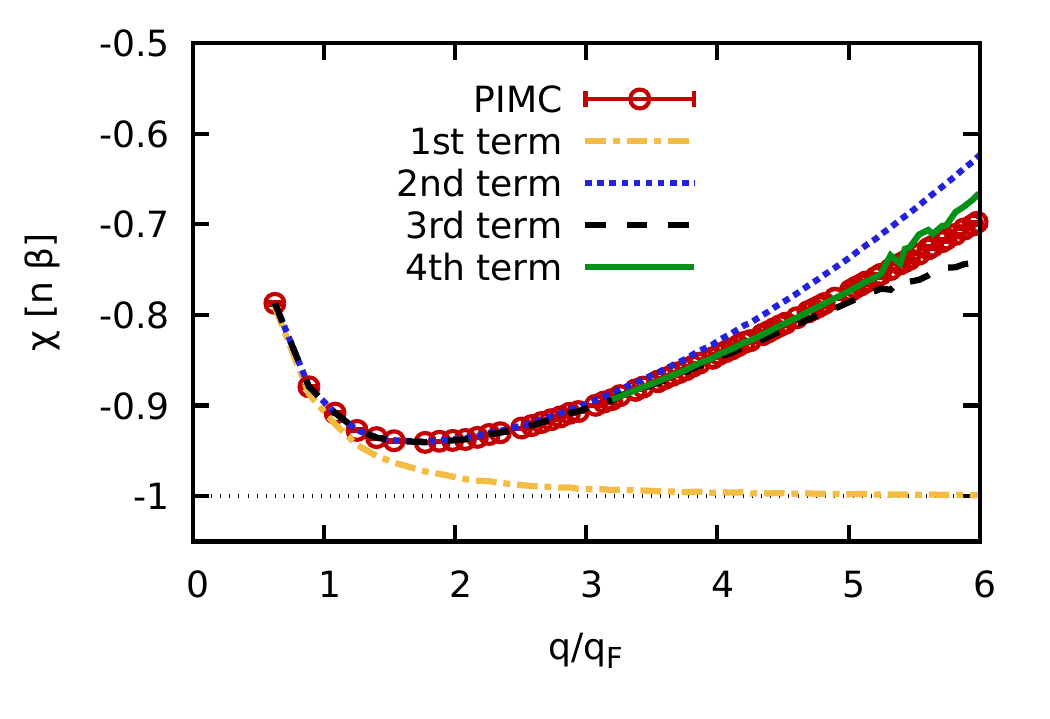}\includegraphics[width=0.44\textwidth]{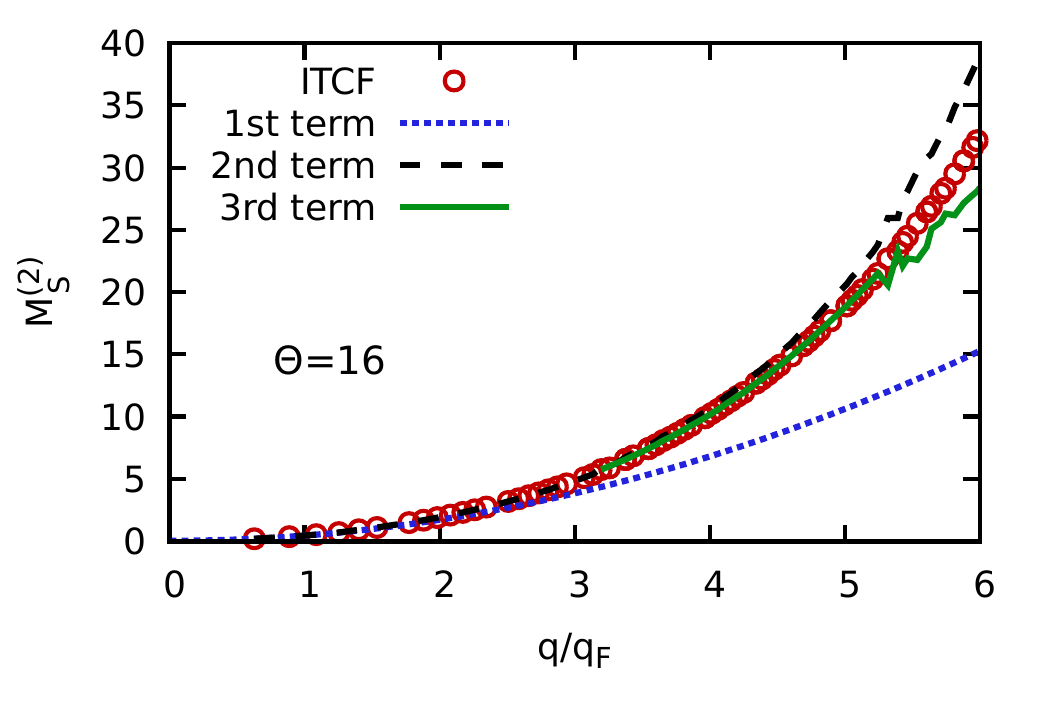}\\\vspace*{-1.25cm}
\includegraphics[width=0.44\textwidth]{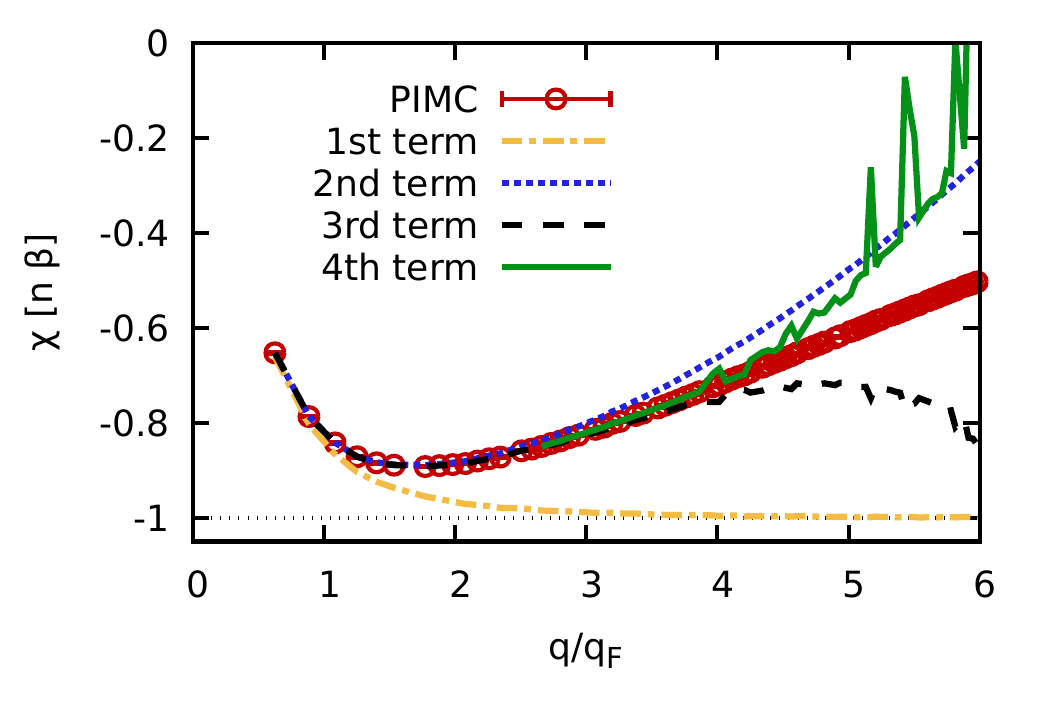}\includegraphics[width=0.44\textwidth]{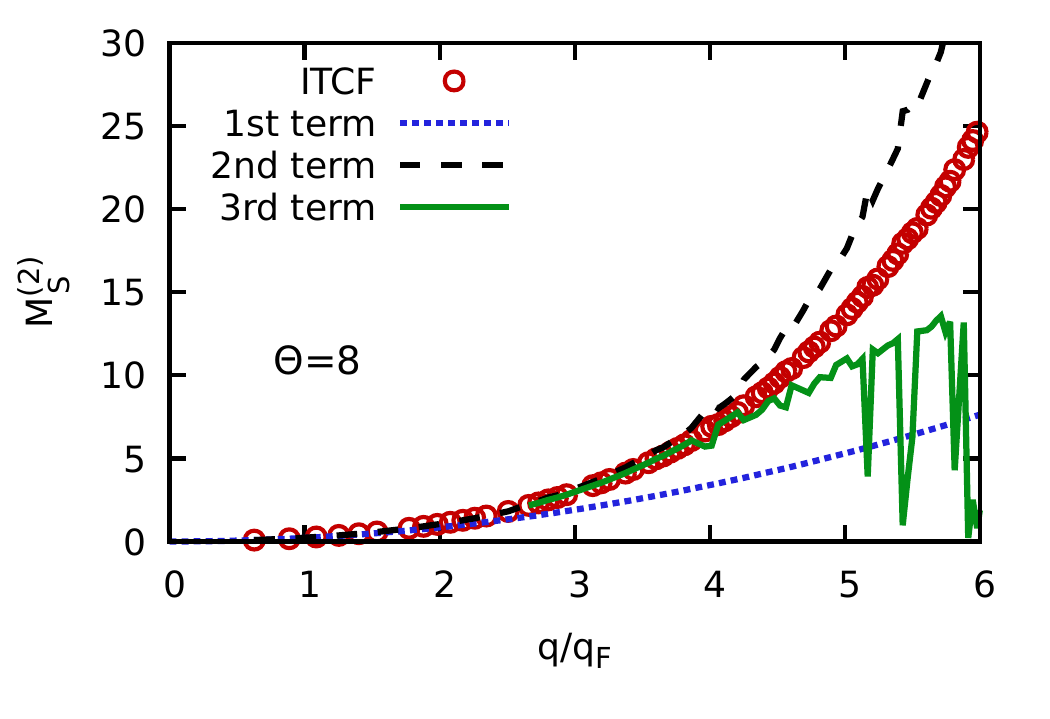}\\\vspace*{-1.25cm}
\includegraphics[width=0.44\textwidth]{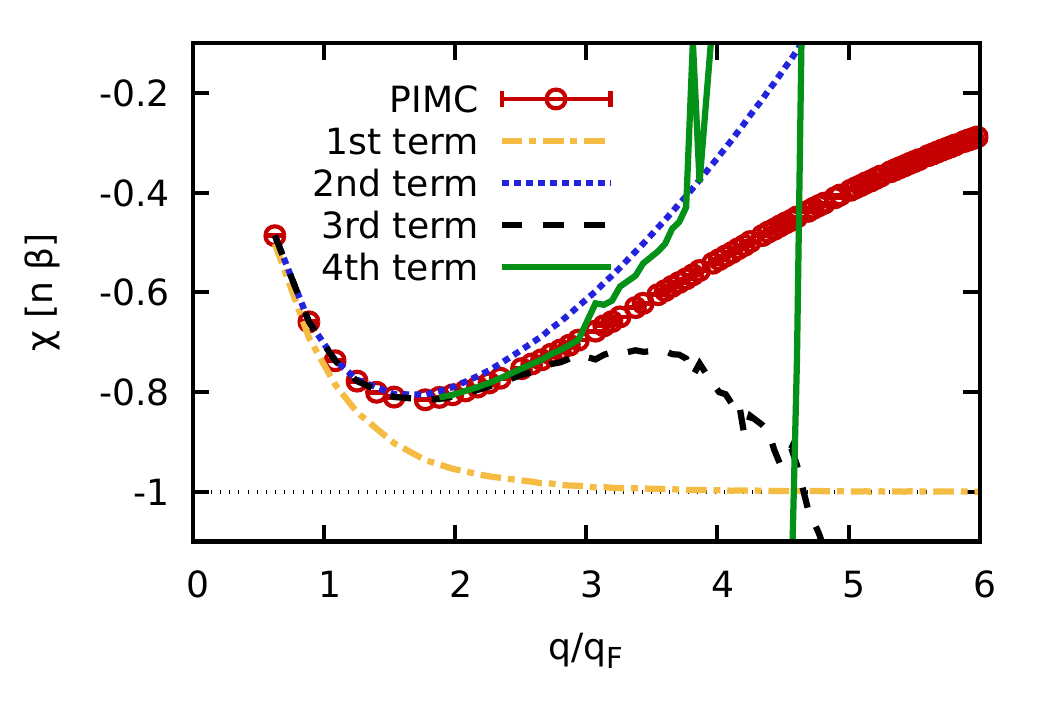}\includegraphics[width=0.44\textwidth]{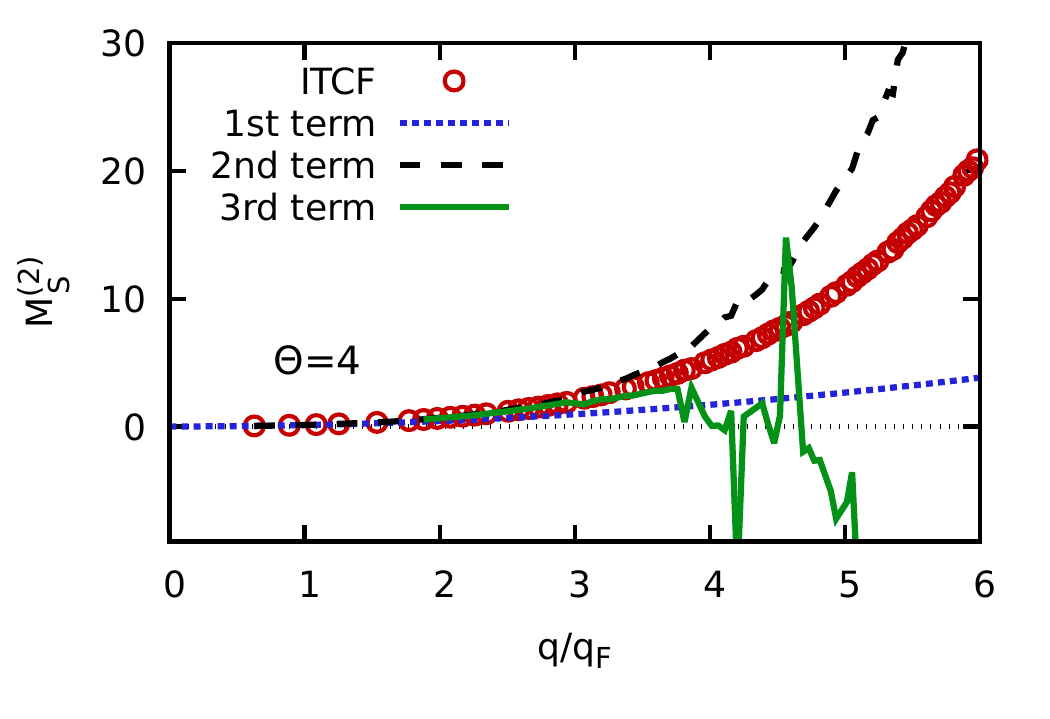}\\\vspace*{-1.25cm}
\includegraphics[width=0.44\textwidth]{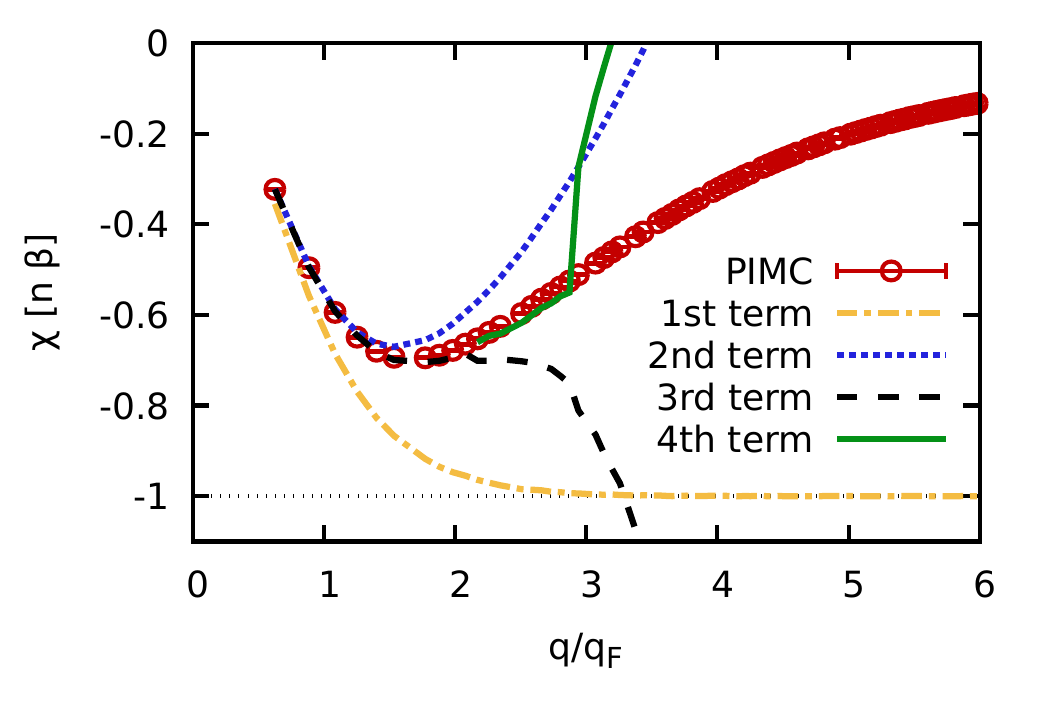}\includegraphics[width=0.44\textwidth]{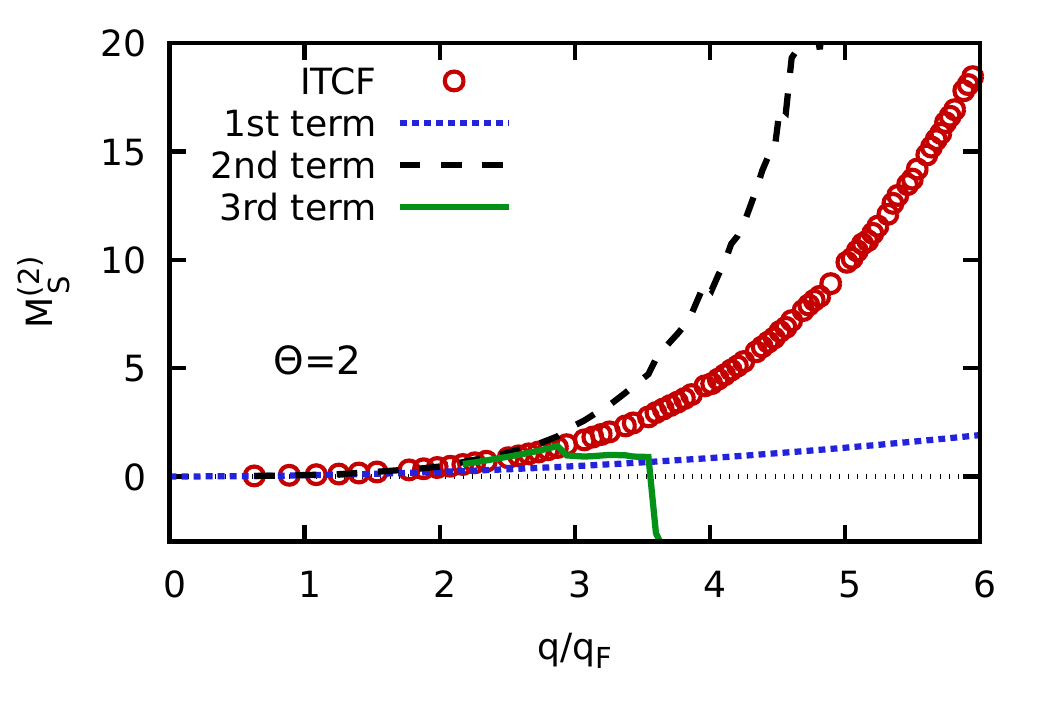}
\caption{\label{fig:UEG_rs4} \emph{Ab initio} PIMC results for the unpolarized UEG at $r_s=4$ with $N=34$ particles. From top to bottom, the rows correspond to $\Theta=32,16,8,4$ and $2$. The left column corresponds to the static density response function $\chi(\mathbf{q})$ that determines the inverse frequency moment $M^{(-1)}_S$ of the DSF [Eq.(\ref{DSFMOMinv})] with the red dots computed from the ITCF via Eq.(\ref{eq:static_chi}) and the solid lines corresponding to different truncations of the Eq.(\ref{eq:chi}) expansion. The right column corresponds to the second frequency moment, with the red dots computed from the ITCF via Eq.(\ref{ITCFmom4}) and the solid lines corresponding to different truncations of the Eq.(\ref{momentseries1}) expansion.}
\end{figure*} 

We next consider the interacting UEG based on quasi-exact, i.e, exact within the Monte Carlo error bars, results of PIMC simulations for the ITCF $F(\mathbf{q},\tau)$~\cite{Dornheim_MRE_2023,Dornheim_JCP_ITCF_2021}. All presented results have been obtained using the canonical extended ensemble adaption~\cite{Dornheim_PRB_nk_2021} of the worm algorithm by Boninsegni \emph{et al.}~\cite{boninsegni1,boninsegni2} as implemented into the open-source \texttt{ISHTAR} code~\cite{ISHTAR} and are freely available in an online repository~\cite{repo}. We note that the $N=34$ particles that are being considered throughout have negligible finite-size effects; this is a general feature of wavenumber resolved properties of the UEG at these parameters, see the extensive available literature~\cite{Chiesa_PRL_2006,Drummond_PRB_2008,dornheim_prl,review,Dornheim_JCP_2021,Holzmann_PRB_2016,Dornheim_PRE_2020}.

In Fig.\ref{fig:UEG_rs10}, we focus on $r_s=10$, which lies at the margin of the strongly correlated electron liquid regime when the degeneracy parameter is less than unity~\cite{dornheim_electron_liquid,dornheim_dynamic,Takada_PRB_2016}. The left column corresponds to the static linear density response function $\chi(\mathbf{q})$, which is directly related to the inverse frequency moment of the DSF, $M_S^{(-1)}$, cf.~Eq.(\ref{DSFMOMinv}). The red circles show our quasi-exact PIMC results, which have been obtained from the imaginary--time version of the fluctuation--dissipation theorem~\cite{bowen2,Dornheim_MRE_2023},
\begin{eqnarray}\label{eq:static_chi}
    \chi(\mathbf{q}) = -n\int_0^\beta F(\mathbf{q},\tau)\quad .
\end{eqnarray}
In addition, the continuous lines show the $\chi(\mathbf{q})$ expansion in terms of frequency moments,
\begin{eqnarray}\label{eq:chi}
    \chi(\mathbf{q}) &=& n\beta \left\{
-S(\mathbf{q}) + \frac{\beta\hbar^2{q}^2}{12} - \frac{\beta^3\hbar^3}{360}M^{(3)}_S(\mathbf{q})
    \right. \\  & &\nonumber \left. 
+ \frac{\beta^5\hbar^5}{15120}M_S^{(5)}(\mathbf{q}) + \dots
    \right\}\, ,
\end{eqnarray}
for different truncations. It is straightforward to discern that Eq.(\ref{eq:chi}) constitutes a simple re-ordering of the zero-moment expansion Eq.(\ref{momentseries0}) by solving for $\chi(\mathbf{q})$ instead of $S(\mathbf{q})$. Here, we choose to focus on Eq.(\ref{eq:chi}) based on the following practical considerations. The most convenient route to estimate $\chi(\mathbf{q})$ is based on the imaginary--time version of the fluctuation--dissipation theorem that requires full knowledge of $F(\mathbf{q},\tau)$. This is easily accessible for direct PIMC simulations, but tedious for ground-state QMC methods~\cite{anderson2007quantum} and PIMC simulations with restricted nodal surfaces~\cite{Ceperley1991}. Alternatively, one may electrostatically perturb the system and subsequently measure its density response~\cite{moroni,moroni2,dornheim_pre,groth_jcp,Dornheim_PRL_2020,Moldabekov_JCTC_2023,moldabekov2025density}. This is, in principle, straightforward for ground-state QMC and restricted PIMC implementations, but it requires a number of independent simulations for different perturbation amplitudes to determine $\chi(\mathbf{q})$ at a single wavenumber. In contrast, the static structure factor $S(\mathbf{q})$, which is estimated in the original Eq.(\ref{momentseries0}), is routinely estimated with either method over the entire $q$-range from a single simulation of the unperturbed system~\cite{Militzer_PRE_2001,Brown_PRL_2013,Drummond_PRB_2008,Chiesa_PRL_2006,Fraser_PRB_1996}. Thus, the evaluation of the first three terms on the RHS of Eq.(\ref{eq:chi}) might constitute a more convenient route to compute the static linear density response function when the direct access to the ITCF is prohibited.

As far as the PIMC based study of $\chi(\mathbf{q})$ -- shown in the left column of Fig.~\ref{fig:UEG_rs10} -- is concerned, we compare Eq.(\ref{eq:static_chi}) with the frequency moment based expansion Eq.(\ref{eq:chi}) using our PIMC results for $M_S^{(\alpha)}$ that have been extracted from $F(\mathbf{q},\tau)$ via Eq.(\ref{ITCFmom4}) utilizing a similar procedure as described in Ref.\cite{Dornheim_moments_2023}. We observe the following trends: (i) The long wavelength limit~\cite{kugler_bounds} 
\begin{eqnarray}
    \lim_{q\to0}\chi(\mathbf{q})=-\frac{q^2}{4\pi{e^2}}
\end{eqnarray}
that follows from perfect screening in the one-component plasma is satisfied both by the PIMC results [Eq.(\ref{eq:static_chi})] and by the expansion Eq.(\ref{eq:chi}) even using a single term. This is due to the fact that quantum delocalization and degeneracy effects do not play a role when the wavelength $\lambda=2\pi/q$ is much larger than the thermal de Broglie wavelength $\lambda_\beta=\sqrt{2\pi\beta\hbar^2/m_e}$. (ii) The truncation of Eq.(\ref{eq:chi}) after the first term corresponds to the classical relation $\chi_\textnormal{cl}(\mathbf{q}) = -n\beta S_\textnormal{cl}(\mathbf{q})$, which systematically overestimates the magnitude of the true density response for $q\gtrsim q_\textnormal{F}$ even at $\Theta=32$. This clearly demonstrates the importance of quantum delocalization effects for practical applications even at temperatures much larger than the Fermi temperature. (iii) The validity range of Eq.(\ref{eq:chi}) strongly depends on the considered value of $\Theta$. For $\Theta=32$, evaluating the first three terms is relatively accurate over the entire depicted $q$-range, whereas the same truncation fails to capture the minimum in $\chi(\mathbf{q})$ around $q=2q_\textnormal{F}$ for $\Theta=2$. Generally, evaluating the fourth term in Eq.(\ref{eq:chi}), which involves the fifth frequency moment of the DSF for which currently no sum rule exists in the quantum case, only leads to small improvements and a comparably small extension of the applicable $q$-range. (iv) Finally, we mention the well-known Matsubara representation of $S(\mathbf{q})$~\cite{stls,Tolias_JCP_2024}, which can be written as
\begin{eqnarray}\label{eq:chi_FM}
    \chi(\mathbf{q}) = - n \beta S(\mathbf{q}) - 2\sum_{\ell=1}^{\infty}\widetilde{\chi}(\mathbf{q},z_\ell)\ ,
\end{eqnarray}
where $\widetilde{\chi}(\mathbf{q},z_\ell)$ is the dynamic Matsubara density response function~\cite{Dornheim_PRB_2024,Dornheim_EPL_2024,moldabekov2025density} and $z_\ell=i2\pi\ell/\beta\hbar$
the bosonic imaginary Matsubara frequencies. Recently, quasi-exact results for $\widetilde{\chi}(\mathbf{q},z_\ell)$ of the UEG have been presented based on an exact Fourier-Matsubara expansion using PIMC results for the ITCF~\cite{Tolias_JCP_2024,Dornheim_PRB_2024,Dornheim_EPL_2024}, which has given some intuitive insights into the behavior of $\chi(\mathbf{q})$ at large $q$. In particular, it has been reported that the system reacts increasingly weakly to an external perturbation when the associated wavelength $\lambda=2\pi/q$ becomes much smaller than $\lambda_\beta$. This is a direct consequence of quantum delocalization effects, which are absent in the classical case where it holds $\lim_{q\to\infty}\chi(\mathbf{q})=-n\beta$. For the quantum case, the delocalization is captured by the sum over Matsubara frequencies in Eq.(\ref{eq:chi_FM}). Naturally, the sum over the odd frequency moments in Eq.(\ref{eq:chi}) is formally equivalent to the latter, but the explicit extraction of the frequency moments $M_S^{(\alpha)}$ is numerically less stable compared to the direct extraction of the $\widetilde{\chi}(\mathbf{q},z_l)$. Nevertheless, it can be concluded that Eq.(\ref{eq:chi}) allows for useful estimates of $\chi(\mathbf{q})$ based on available results for $S(\mathbf{q})$.

The right column of Fig.\ref{fig:UEG_rs10} shows a similar analysis for $M_S^{(2)}$ at the same conditions. Here, the red circles have been directly obtained from the PIMC results for $F(\mathbf{q},\tau)$ via Eq.(\ref{ITCFmom4}) and the other curves show Eq.(\ref{momentseries1}) considering the first three terms. Our findings can be summarized as follows: (i) In contrast to $\chi(\mathbf{q})$, $M^{(2)}_S$ monotonically increases with $q$~\cite{Dornheim_moments_2023}. (ii) The validity range is comparable to the expansion for $\chi(\mathbf{q})$, and to corresponding results for the ideal Fermi gas discussed in Sec.~\ref{subsec:resultsFermi} above. (iii) The high accuracy of Eq.(\ref{momentseries1}) in particular at small wavenumbers might be helpful for the interpretation and modeling of XRTS experiments in a forward scattering geometry~\cite{siegfried_review,Gawne_PRB_2024,Bellenbaum_APL_2025}.

\begin{figure*}\centering
\includegraphics[width=0.44\textwidth]{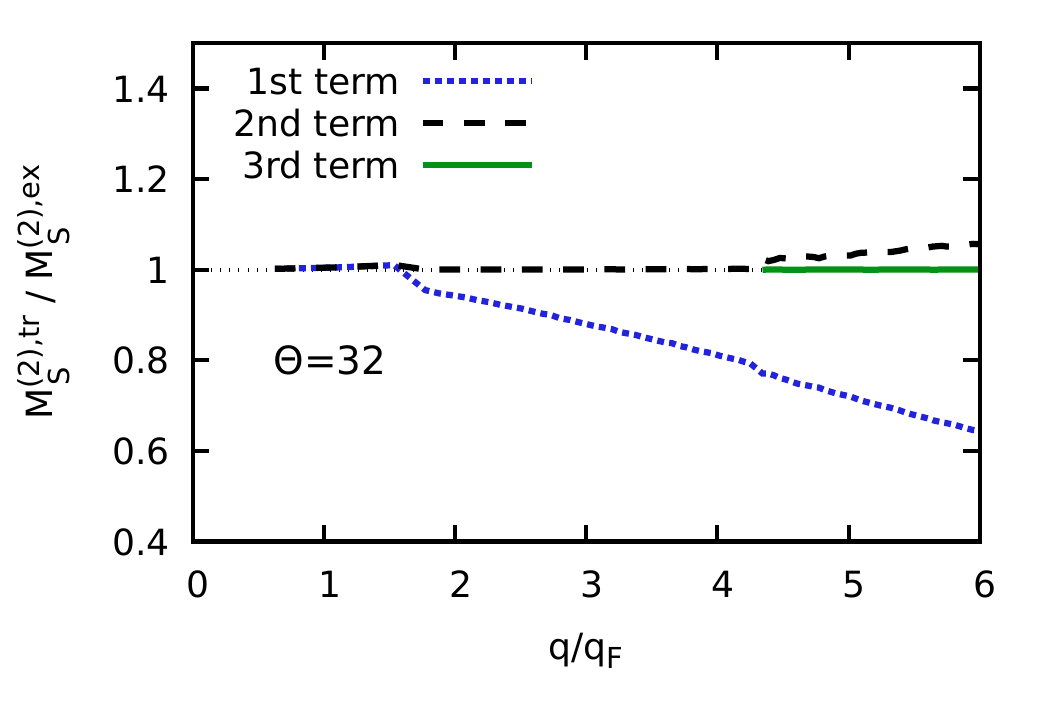}\includegraphics[width=0.44\textwidth]{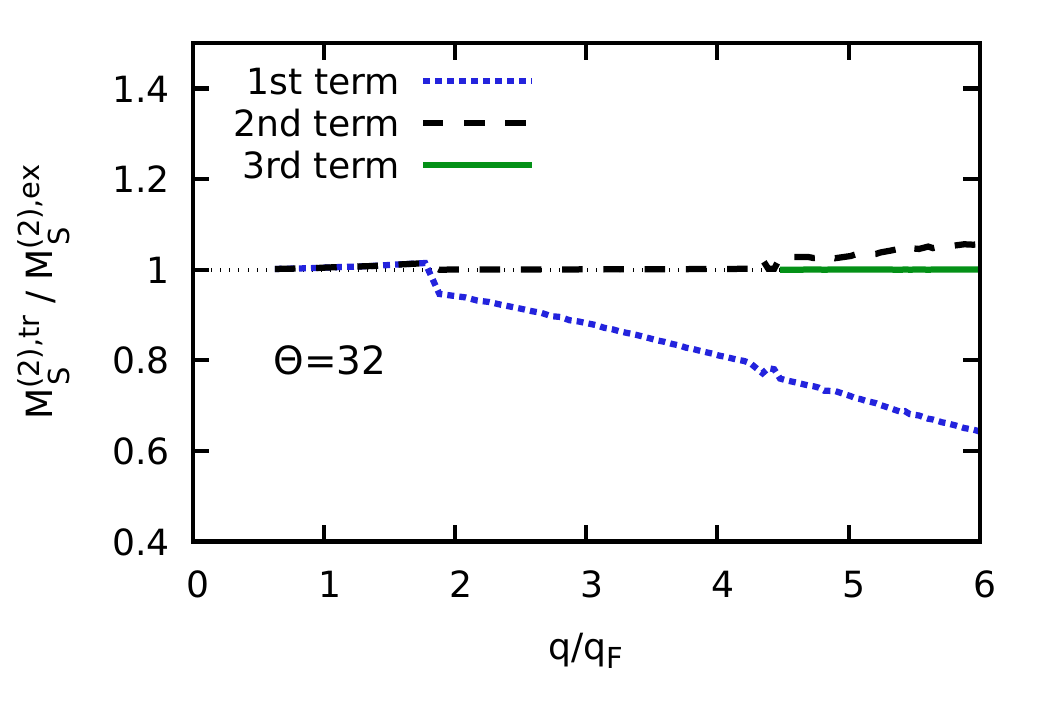}\\\vspace*{-1.25cm}
\includegraphics[width=0.44\textwidth]{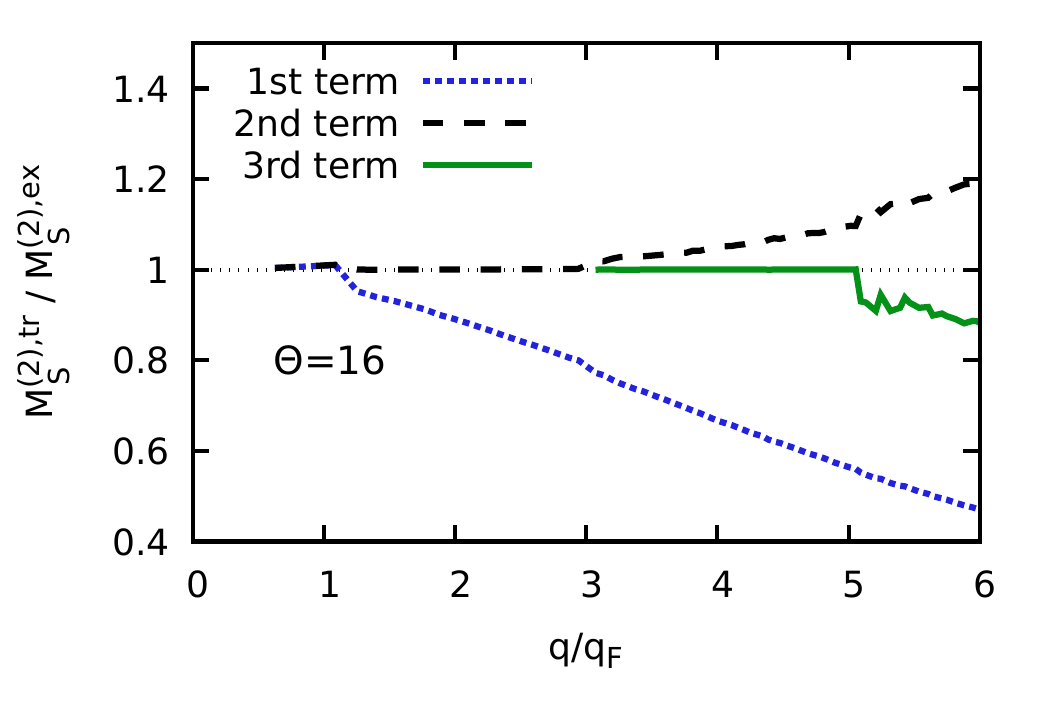}\includegraphics[width=0.44\textwidth]{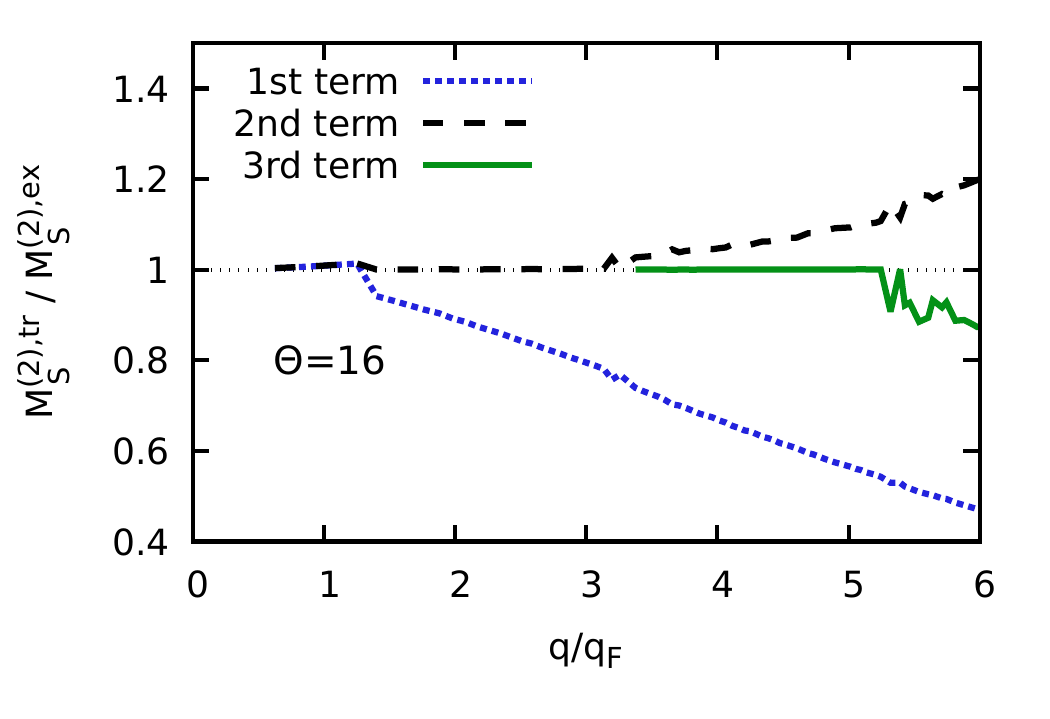}\\\vspace*{-1.25cm}
\includegraphics[width=0.44\textwidth]{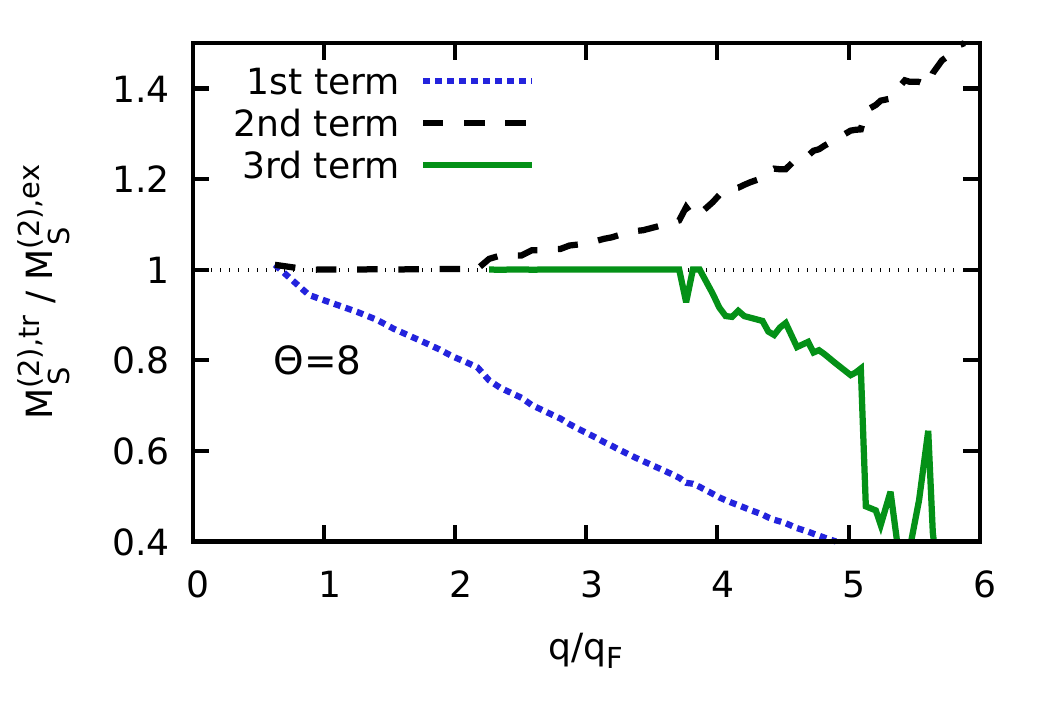}\includegraphics[width=0.44\textwidth]{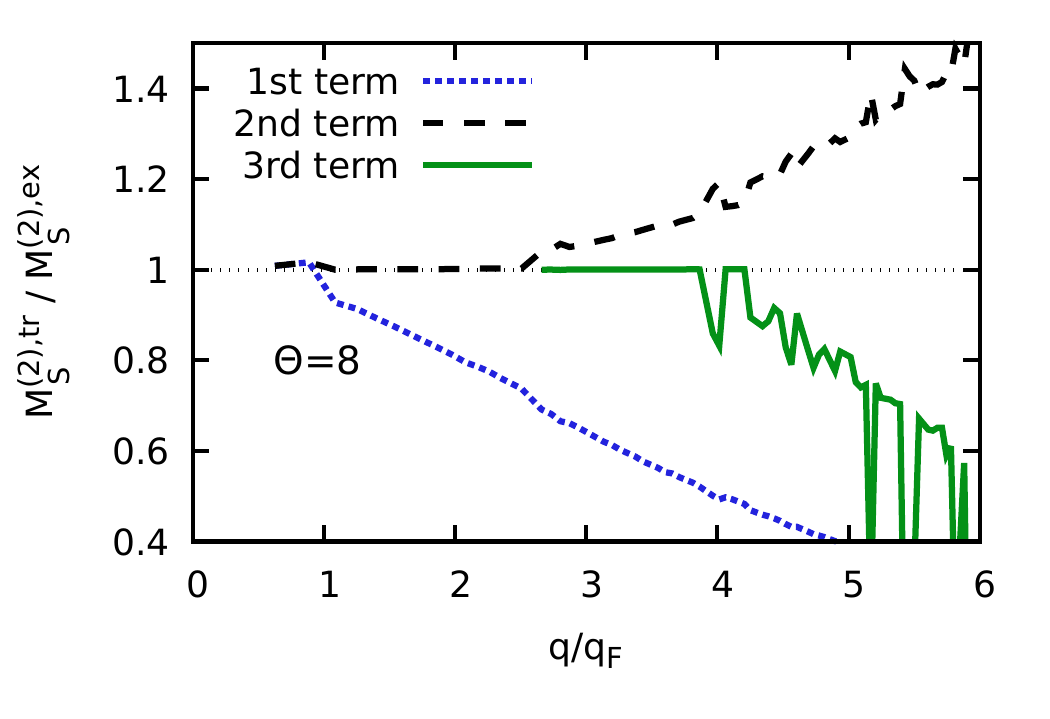}\\\vspace*{-1.25cm}
\includegraphics[width=0.44\textwidth]{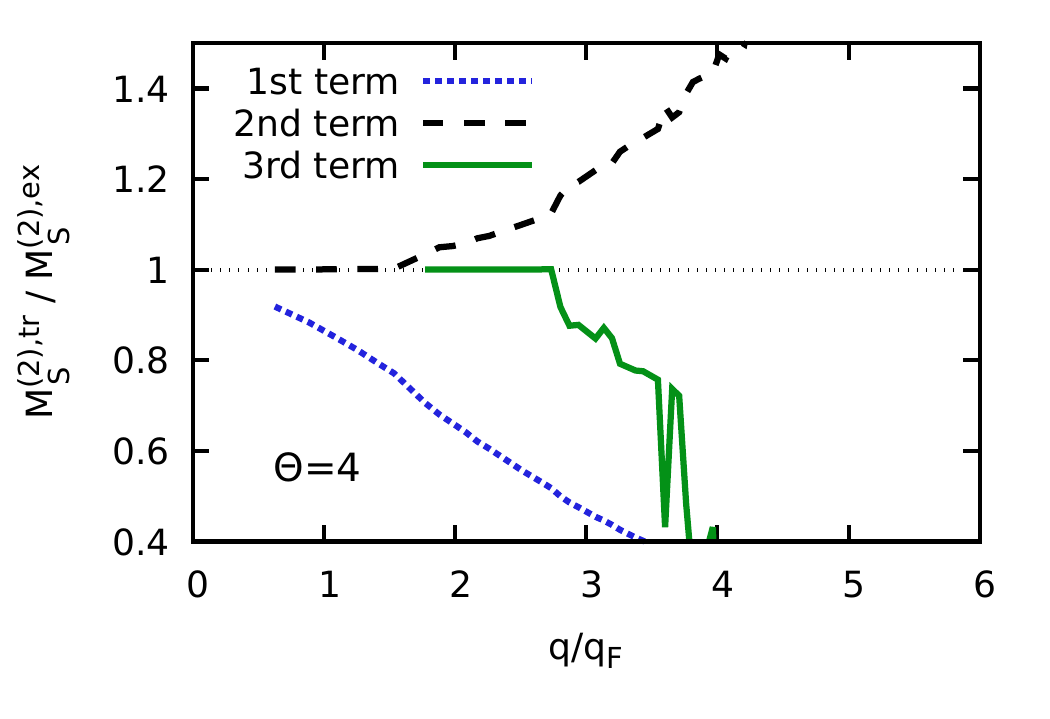}\includegraphics[width=0.44\textwidth]{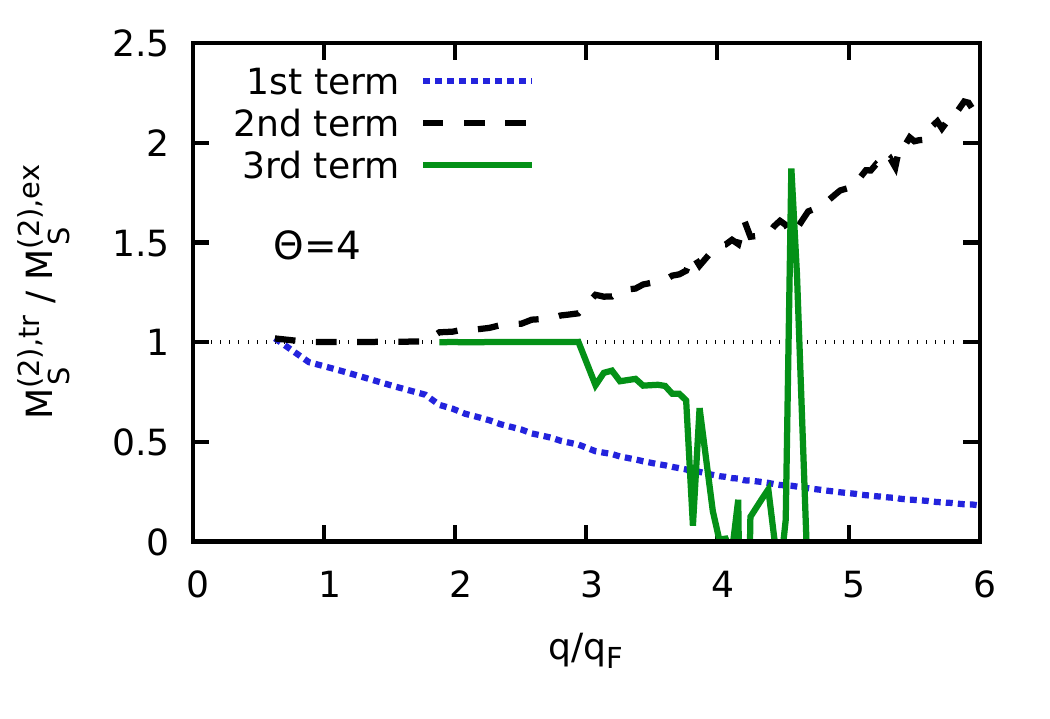}\\\vspace*{-1.25cm}
\includegraphics[width=0.44\textwidth]{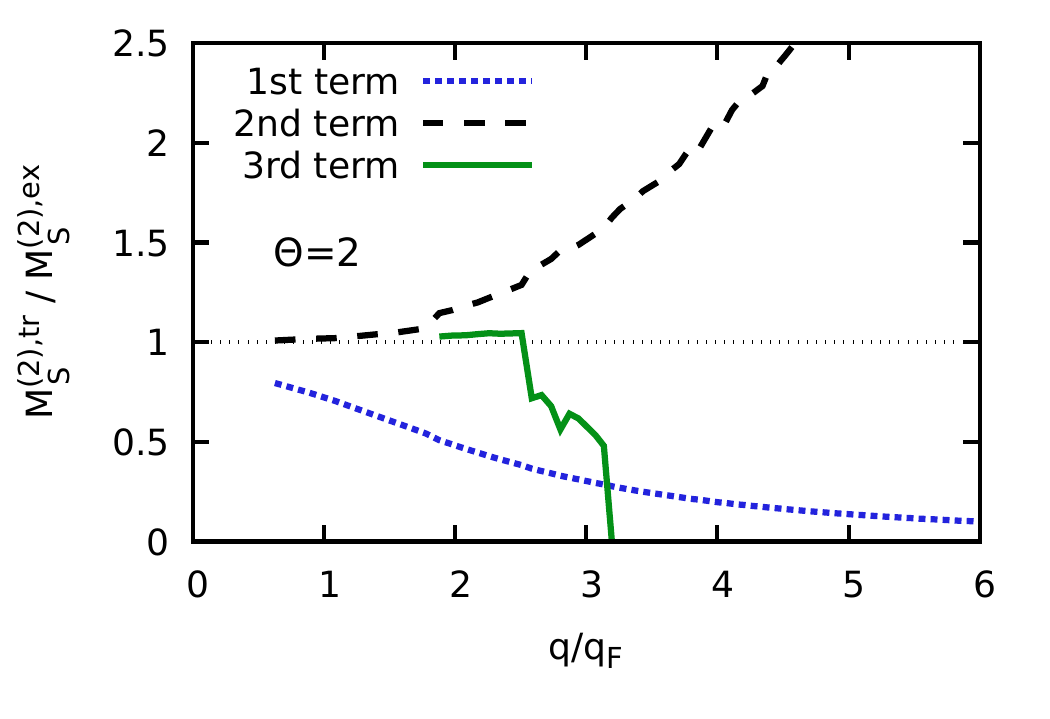}\includegraphics[width=0.44\textwidth]{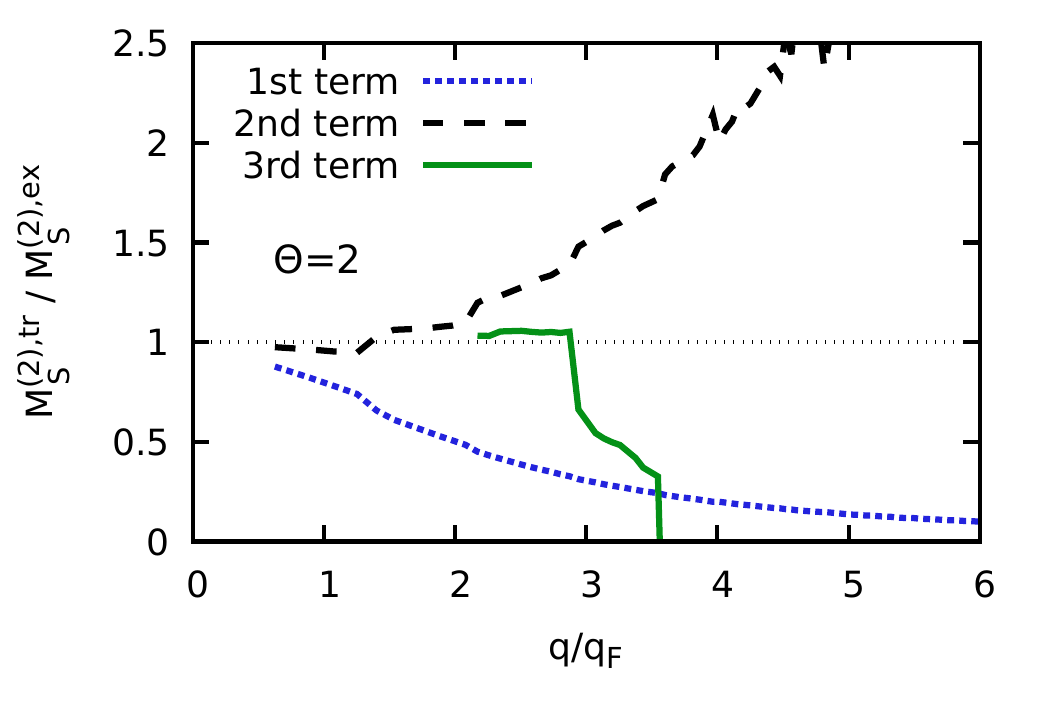}
\caption{\label{fig:UEG_relative} Results for the interacting UEG at $r_s=10$ (left panel), $r_s=4$ (right panel) and $\Theta=2,\,4,\,8,\,16,\,32$. The ratio of different truncated versions of the second frequency moment (one terms, two terms, three terms) over the quasi-exact PIMC extracted second frequency moment.}
\end{figure*} 

For completeness, we have repeated the above analysis for $r_s=4$, which is more typical for WDM and can be realized in experiments with sodium~\cite{Huotari}. Moreover, it is close to the electronic density in hydrogen jets~\cite{Fletcher_Frontiers_2022,Zastrau,Hamann_PRR_2023}, $r_s=3.23$. We find equivalent trends as for $r_s=10$ and $r_s\to0$.

Finally, similar to the middle panel of Fig.\ref{fig:Fermiall} that is dedicated to the second frequency moment for the non-interacting Fermi gas, in Fig.\ref{fig:UEG_relative}, we plot the ratio of different truncated series approximations of the second frequency moment over the PIMC extracted second frequency moment for two coupling parameters ($r_s=4,\,10$) and five degeneracy parameters ($\Theta=2,\,4,\,8,\,16,\,32$). Note that only one term, two term and three term truncations can be probed, since four term truncations based on PIMC results are excessively noisy. It is evident that all conclusions drawn for the non-interacting Fermi gas (see Sec.\ref{subsec:resultsFermi}) are also applicable to the interacting UEG.

\section{Summary and Outlook\label{sec:outlook}}

On the basis of linear density response theory for finite temperature quantum systems, a novel series representation has been derived that expresses the even frequency moments of the dynamic structure factor (for which no theoretical expression is known outside the ground state limit) through an infinite sum involving odd frequency moments of the dynamic structure factor (which are all implicitly known). It is emphasized that the exact series expansion is valid for arbitrary quantum systems at finite temperatures and that it can be straightforwardly generalized to arbitrary spectral functions beyond the dynamic structure factor.

Different truncations of this exact series representation have been analyzed that would allow for the evaluation of the explicitly unknown second, fourth and fifth frequency moments for the finite temperature UEG. Exploiting the availability of analytical results within the non-interacting limit as well as a recently proposed PIMC framework that allows the direct extraction of frequency moments of arbitrary order from the ITCF in the interacting case\,\cite{Dornheim_PRB_2023}, the applicability region of different truncations of the series has been established. In particular, it has been revealed that three term truncations constitute very accurate approximations of the even frequency moments of the dynamic structure factor away from the short-wavelength limit for degeneracy parameters $\Theta\gtrsim16$ regardless of the quantum coupling parameter $r_{\mathrm{s}}$.

The present results can drive the formulation of novel variants of the self-consistent method of moments\,\cite{Arkhipov_2017,Arkhipov_2020,Filinov_PRB_2023,Filinov_RSTA_2023} that investigate the dynamic properties of the weakly degenerate UEG. The present results can also constitute additional constraints that facilitate the accurate numerical inversion of PIMC data for the ITCF of the weakly degenerate UEG to obtain the dynamic structure factor\,\cite{dornheim_dynamic,dynamic_folgepaper,Hamann_PRB_2020}, see the notorious analytic continuation problem\,\cite{epstein2008badtruth,JARRELL1996133,chuna2025dualformulationmaximumentropy}. In addition, the truncated series for the known zero frequency moment constitutes a convenient route to compute the static linear density response function of the weakly degenerate UEG with simulation approaches that prohibit direct access to the ITCF, e.g. with the restricted PIMC method\,\cite{Ceperley1991}. Finally, as aforementioned, the high accuracy of the truncated series at small wavenumbers might prove helpful for the interpretation and modeling of XRTS experiments in a forward scattering geometry~\cite{siegfried_review,Gawne_PRB_2024,Bellenbaum_APL_2025}.

\begin{acknowledgements}

\noindent This work was partially supported by the Center for Advanced Systems Understanding (CASUS), financed by Germany’s Federal Ministry of Education and Research and the Saxon state government out of the State budget approved by the Saxon State Parliament. This work has received funding from the European Union's Just Transition Fund (JTF) within the project \emph{R\"ontgenlaser-Optimierung der Laserfusion} (ROLF), contract number 5086999001, co-financed by the Saxon state government out of the State budget approved by the Saxon State Parliament. 
This work has received funding from the European Research Council (ERC) under the European Union’s Horizon 2022 research and innovation programme (Grant agreement No. 101076233, "PREXTREME"). Views and opinions expressed are however those of the authors only and do not necessarily reflect those of the European Union or the European Research Council Executive Agency. Neither the European Union nor the granting authority can be held responsible for them. Computations were performed on a Bull Cluster at the Center for Information Services and High-Performance Computing (ZIH) at Technische Universit\"at Dresden and at the Norddeutscher Verbund f\"ur Hoch- und H\"ochstleistungsrechnen (HLRN) under grant mvp00024.
\end{acknowledgements}

\bibliography{bibliography}

\begin{thebibliography}{116}%
\makeatletter
\providecommand \@ifxundefined [1]{%
 \@ifx{#1\undefined}
}%
\providecommand \@ifnum [1]{%
 \ifnum #1\expandafter \@firstoftwo
 \else \expandafter \@secondoftwo
 \fi
}%
\providecommand \@ifx [1]{%
 \ifx #1\expandafter \@firstoftwo
 \else \expandafter \@secondoftwo
 \fi
}%
\providecommand \natexlab [1]{#1}%
\providecommand \enquote  [1]{``#1''}%
\providecommand \bibnamefont  [1]{#1}%
\providecommand \bibfnamefont [1]{#1}%
\providecommand \citenamefont [1]{#1}%
\providecommand \href@noop [0]{\@secondoftwo}%
\providecommand \href [0]{\begingroup \@sanitize@url \@href}%
\providecommand \@href[1]{\@@startlink{#1}\@@href}%
\providecommand \@@href[1]{\endgroup#1\@@endlink}%
\providecommand \@sanitize@url [0]{\catcode `\\12\catcode `\$12\catcode `\&12\catcode `\#12\catcode `\^12\catcode `\_12\catcode `\%12\relax}%
\providecommand \@@startlink[1]{}%
\providecommand \@@endlink[0]{}%
\providecommand \url  [0]{\begingroup\@sanitize@url \@url }%
\providecommand \@url [1]{\endgroup\@href {#1}{\urlprefix }}%
\providecommand \urlprefix  [0]{URL }%
\providecommand \Eprint [0]{\href }%
\providecommand \doibase [0]{http://dx.doi.org/}%
\providecommand \selectlanguage [0]{\@gobble}%
\providecommand \bibinfo  [0]{\@secondoftwo}%
\providecommand \bibfield  [0]{\@secondoftwo}%
\providecommand \translation [1]{[#1]}%
\providecommand \BibitemOpen [0]{}%
\providecommand \bibitemStop [0]{}%
\providecommand \bibitemNoStop [0]{.\EOS\space}%
\providecommand \EOS [0]{\spacefactor3000\relax}%
\providecommand \BibitemShut  [1]{\csname bibitem#1\endcsname}%
\let\auto@bib@innerbib\@empty
\bibitem [{\citenamefont {Stefanucci}\ and\ \citenamefont {van Leeuwen}(2013)}]{stefanucci2013nonequilibrium}%
  \BibitemOpen
  \bibfield  {author} {\bibinfo {author} {\bibfnamefont {G.}~\bibnamefont {Stefanucci}}\ and\ \bibinfo {author} {\bibfnamefont {R.}~\bibnamefont {van Leeuwen}},\ }\href {https://books.google.de/books?id=6GsrjPFXLDYC} {\emph {\bibinfo {title} {Nonequilibrium Many-Body Theory of Quantum Systems: A Modern Introduction}}}\ (\bibinfo  {publisher} {Cambridge University Press},\ \bibinfo {year} {2013})\BibitemShut {NoStop}%
\bibitem [{\citenamefont {Bonitz}(2016)}]{bonitz_book}%
  \BibitemOpen
  \bibfield  {author} {\bibinfo {author} {\bibfnamefont {M.}~\bibnamefont {Bonitz}},\ }\href@noop {} {\emph {\bibinfo {title} {Quantum kinetic theory}}}\ (\bibinfo  {publisher} {Springer},\ \bibinfo {address} {Heidelberg},\ \bibinfo {year} {2016})\BibitemShut {NoStop}%
\bibitem [{\citenamefont {A.~A.~Abrikosov}\ and\ \citenamefont {Dzyaloshinski}(1965)}]{abrikosov_book}%
  \BibitemOpen
  \bibfield  {author} {\bibinfo {author} {\bibfnamefont {L.~P.~Gorkov}\ \bibnamefont {A.~A.~Abrikosov}}\ and\ \bibinfo {author} {\bibfnamefont {I.~E.}\ \bibnamefont {Dzyaloshinski}},\ }\href@noop {} {\emph {\bibinfo {title} {Methods of quantum field theory in statistical physics}}}\ (\bibinfo  {publisher} {Pergamon Press, Oxford},\ \bibinfo {year} {1965})\BibitemShut {NoStop}%
\bibitem [{\citenamefont {Howard}\ \emph {et~al.}(2023)\citenamefont {Howard}, \citenamefont {Guillot}, \citenamefont {Bazot}, \citenamefont {Miguel}, \citenamefont {Stevenson}, \citenamefont {Galanti}, \citenamefont {Kaspi}, \citenamefont {Hubbard}, \citenamefont {Militzer}, \citenamefont {Helled} \emph {et~al.}}]{Howard2023}%
  \BibitemOpen
  \bibfield  {author} {\bibinfo {author} {\bibfnamefont {S.}~\bibnamefont {Howard}}, \bibinfo {author} {\bibfnamefont {T.}~\bibnamefont {Guillot}}, \bibinfo {author} {\bibfnamefont {M.}~\bibnamefont {Bazot}}, \bibinfo {author} {\bibfnamefont {Y.}~\bibnamefont {Miguel}}, \bibinfo {author} {\bibfnamefont {D.~J.}\ \bibnamefont {Stevenson}}, \bibinfo {author} {\bibfnamefont {E.}~\bibnamefont {Galanti}}, \bibinfo {author} {\bibfnamefont {Y.}~\bibnamefont {Kaspi}}, \bibinfo {author} {\bibfnamefont {W.~B.}\ \bibnamefont {Hubbard}}, \bibinfo {author} {\bibfnamefont {B.}~\bibnamefont {Militzer}}, \bibinfo {author} {\bibfnamefont {R.}~\bibnamefont {Helled}},  \emph {et~al.},\ }\bibfield  {title} {\enquote {\bibinfo {title} {Jupiter's interior from {Juno}: Equation-of-state uncertainties and dilute core extent},}\ }\href {\doibase 10.1051/0004-6361/202245625} {\bibfield  {journal} {\bibinfo  {journal} {A\&A}\ }\textbf {\bibinfo {volume} {672}},\ \bibinfo {pages} {A33} (\bibinfo {year} {2023})}\BibitemShut {NoStop}%
\bibitem [{\citenamefont {Hernandez}\ \emph {et~al.}(2023)\citenamefont {Hernandez}, \citenamefont {Bethkenhagen}, \citenamefont {Ninet}, \citenamefont {French}, \citenamefont {Benuzzi-Mounaix}, \citenamefont {Datchi}, \citenamefont {Guarguaglini}, \citenamefont {Lefevre}, \citenamefont {Occelli}, \citenamefont {Redmer}, \citenamefont {Vinci},\ and\ \citenamefont {Ravasio}}]{Hernandez2023}%
  \BibitemOpen
  \bibfield  {author} {\bibinfo {author} {\bibfnamefont {J.~A.}\ \bibnamefont {Hernandez}}, \bibinfo {author} {\bibfnamefont {M.}~\bibnamefont {Bethkenhagen}}, \bibinfo {author} {\bibfnamefont {S.}~\bibnamefont {Ninet}}, \bibinfo {author} {\bibfnamefont {M.}~\bibnamefont {French}}, \bibinfo {author} {\bibfnamefont {A.}~\bibnamefont {Benuzzi-Mounaix}}, \bibinfo {author} {\bibfnamefont {F.}~\bibnamefont {Datchi}}, \bibinfo {author} {\bibfnamefont {M.}~\bibnamefont {Guarguaglini}}, \bibinfo {author} {\bibfnamefont {F.}~\bibnamefont {Lefevre}}, \bibinfo {author} {\bibfnamefont {F.}~\bibnamefont {Occelli}}, \bibinfo {author} {\bibfnamefont {R.}~\bibnamefont {Redmer}}, \bibinfo {author} {\bibfnamefont {T.}~\bibnamefont {Vinci}}, \ and\ \bibinfo {author} {\bibfnamefont {A.}~\bibnamefont {Ravasio}},\ }\bibfield  {title} {\enquote {\bibinfo {title} {Melting curve of superionic ammonia at planetary interior conditions},}\ }\href {\doibase 10.1038/s41567-023-02074-8} {\bibfield  {journal} {\bibinfo  {journal} {Nature
  Physics}\ }\textbf {\bibinfo {volume} {19}},\ \bibinfo {pages} {1280--1285} (\bibinfo {year} {2023})}\BibitemShut {NoStop}%
\bibitem [{\citenamefont {Breuer}\ and\ \citenamefont {Spohn}(2023)}]{Breuer2023}%
  \BibitemOpen
  \bibfield  {author} {\bibinfo {author} {\bibfnamefont {Doris}\ \bibnamefont {Breuer}}\ and\ \bibinfo {author} {\bibfnamefont {Tilman}\ \bibnamefont {Spohn}},\ }\href {\doibase 10.1093/acrefore/9780190647926.013.28} {\enquote {\bibinfo {title} {Terrestrial planets: Interior structure, dynamics, and evolution},}\ } (\bibinfo {year} {2023})\BibitemShut {NoStop}%
\bibitem [{\citenamefont {Kritcher}\ \emph {et~al.}(2020)\citenamefont {Kritcher}, \citenamefont {Swift}, \citenamefont {Döppner}, \citenamefont {Bachmann}, \citenamefont {Benedict}, \citenamefont {Collins}, \citenamefont {DuBois}, \citenamefont {Elsner}, \citenamefont {Fontaine}, \citenamefont {Gaffney}, \citenamefont {Hamel}, \citenamefont {Lazicki} \emph {et~al.}}]{Kritcher2020}%
  \BibitemOpen
  \bibfield  {author} {\bibinfo {author} {\bibfnamefont {Andrea~L.}\ \bibnamefont {Kritcher}}, \bibinfo {author} {\bibfnamefont {Damian~C.}\ \bibnamefont {Swift}}, \bibinfo {author} {\bibfnamefont {Tilo}\ \bibnamefont {Döppner}}, \bibinfo {author} {\bibfnamefont {Benjamin}\ \bibnamefont {Bachmann}}, \bibinfo {author} {\bibfnamefont {Lorin~X.}\ \bibnamefont {Benedict}}, \bibinfo {author} {\bibfnamefont {Gilbert~W.}\ \bibnamefont {Collins}}, \bibinfo {author} {\bibfnamefont {Jonathan~L.}\ \bibnamefont {DuBois}}, \bibinfo {author} {\bibfnamefont {Fred}\ \bibnamefont {Elsner}}, \bibinfo {author} {\bibfnamefont {Gilles}\ \bibnamefont {Fontaine}}, \bibinfo {author} {\bibfnamefont {Jim~A.}\ \bibnamefont {Gaffney}}, \bibinfo {author} {\bibfnamefont {Sebastien}\ \bibnamefont {Hamel}}, \bibinfo {author} {\bibfnamefont {Amy}\ \bibnamefont {Lazicki}},  \emph {et~al.},\ }\bibfield  {title} {\enquote {\bibinfo {title} {A measurement of the equation of state of carbon envelopes of white dwarfs},}\ }\href {\doibase
  10.1038/s41586-020-2535-y} {\bibfield  {journal} {\bibinfo  {journal} {Nature}\ }\textbf {\bibinfo {volume} {584}},\ \bibinfo {pages} {51--54} (\bibinfo {year} {2020})}\BibitemShut {NoStop}%
\bibitem [{\citenamefont {Saumon}\ \emph {et~al.}(2022)\citenamefont {Saumon}, \citenamefont {Blouin},\ and\ \citenamefont {Tremblay}}]{SAUMON20221}%
  \BibitemOpen
  \bibfield  {author} {\bibinfo {author} {\bibfnamefont {Didier}\ \bibnamefont {Saumon}}, \bibinfo {author} {\bibfnamefont {Simon}\ \bibnamefont {Blouin}}, \ and\ \bibinfo {author} {\bibfnamefont {Pier-Emmanuel}\ \bibnamefont {Tremblay}},\ }\bibfield  {title} {\enquote {\bibinfo {title} {Current challenges in the physics of white dwarf stars},}\ }\href {\doibase https://doi.org/10.1016/j.physrep.2022.09.001} {\bibfield  {journal} {\bibinfo  {journal} {Phys. Rep.}\ }\textbf {\bibinfo {volume} {988}},\ \bibinfo {pages} {1--63} (\bibinfo {year} {2022})}\BibitemShut {NoStop}%
\bibitem [{\citenamefont {{Chamel}}\ and\ \citenamefont {{Haensel}}(2008)}]{chamel_lrr_2008}%
  \BibitemOpen
  \bibfield  {author} {\bibinfo {author} {\bibfnamefont {Nicolas}\ \bibnamefont {{Chamel}}}\ and\ \bibinfo {author} {\bibfnamefont {Pawel}\ \bibnamefont {{Haensel}}},\ }\bibfield  {title} {\enquote {\bibinfo {title} {{Physics of Neutron Star Crusts}},}\ }\href {\doibase 10.12942/lrr-2008-10} {\bibfield  {journal} {\bibinfo  {journal} {Living Reviews in Relativity}\ }\textbf {\bibinfo {volume} {11}},\ \bibinfo {eid} {10} (\bibinfo {year} {2008})},\ \Eprint {http://arxiv.org/abs/0812.3955} {arXiv:0812.3955 [astro-ph]} \BibitemShut {NoStop}%
\bibitem [{\citenamefont {Beznogov}\ \emph {et~al.}(2021)\citenamefont {Beznogov}, \citenamefont {Potekhin},\ and\ \citenamefont {Yakovlev}}]{neutronstar_2021}%
  \BibitemOpen
  \bibfield  {author} {\bibinfo {author} {\bibfnamefont {M.~V.}\ \bibnamefont {Beznogov}}, \bibinfo {author} {\bibfnamefont {A.~Y.}\ \bibnamefont {Potekhin}}, \ and\ \bibinfo {author} {\bibfnamefont {D.~G.}\ \bibnamefont {Yakovlev}},\ }\bibfield  {title} {\enquote {\bibinfo {title} {Heat blanketing envelopes of neutron stars},}\ }\href {\doibase 10.1016/j.physrep.2021.03.004} {\bibfield  {journal} {\bibinfo  {journal} {Phys. Rep.}\ }\textbf {\bibinfo {volume} {919}},\ \bibinfo {pages} {1--68} (\bibinfo {year} {2021})}\BibitemShut {NoStop}%
\bibitem [{\citenamefont {Hurricane}\ \emph {et~al.}(2023)\citenamefont {Hurricane}, \citenamefont {Patel}, \citenamefont {Betti}, \citenamefont {Froula}, \citenamefont {Regan}, \citenamefont {Slutz}, \citenamefont {Gomez},\ and\ \citenamefont {Sweeney}}]{Hurricane_RevModPhys_2023}%
  \BibitemOpen
  \bibfield  {author} {\bibinfo {author} {\bibfnamefont {O.~A.}\ \bibnamefont {Hurricane}}, \bibinfo {author} {\bibfnamefont {P.~K.}\ \bibnamefont {Patel}}, \bibinfo {author} {\bibfnamefont {R.}~\bibnamefont {Betti}}, \bibinfo {author} {\bibfnamefont {D.~H.}\ \bibnamefont {Froula}}, \bibinfo {author} {\bibfnamefont {S.~P.}\ \bibnamefont {Regan}}, \bibinfo {author} {\bibfnamefont {S.~A.}\ \bibnamefont {Slutz}}, \bibinfo {author} {\bibfnamefont {M.~R.}\ \bibnamefont {Gomez}}, \ and\ \bibinfo {author} {\bibfnamefont {M.~A.}\ \bibnamefont {Sweeney}},\ }\bibfield  {title} {\enquote {\bibinfo {title} {Physics principles of inertial confinement fusion and {U.S.} program overview},}\ }\href {\doibase 10.1103/RevModPhys.95.025005} {\bibfield  {journal} {\bibinfo  {journal} {Rev. Mod. Phys.}\ }\textbf {\bibinfo {volume} {95}},\ \bibinfo {pages} {025005} (\bibinfo {year} {2023})}\BibitemShut {NoStop}%
\bibitem [{\citenamefont {Hu}\ \emph {et~al.}(2011)\citenamefont {Hu}, \citenamefont {Militzer}, \citenamefont {Goncharov},\ and\ \citenamefont {Skupsky}}]{hu_ICF}%
  \BibitemOpen
  \bibfield  {author} {\bibinfo {author} {\bibfnamefont {S.~X.}\ \bibnamefont {Hu}}, \bibinfo {author} {\bibfnamefont {B.}~\bibnamefont {Militzer}}, \bibinfo {author} {\bibfnamefont {V.~N.}\ \bibnamefont {Goncharov}}, \ and\ \bibinfo {author} {\bibfnamefont {S.}~\bibnamefont {Skupsky}},\ }\bibfield  {title} {\enquote {\bibinfo {title} {First-principles equation-of-state table of deuterium for inertial confinement fusion applications},}\ }\href {https://journals.aps.org/prb/abstract/10.1103/PhysRevB.84.224109} {\bibfield  {journal} {\bibinfo  {journal} {Phys. Rev. B}\ }\textbf {\bibinfo {volume} {84}},\ \bibinfo {pages} {224109} (\bibinfo {year} {2011})}\BibitemShut {NoStop}%
\bibitem [{\citenamefont {Remington}(2005)}]{Remington_2005}%
  \BibitemOpen
  \bibfield  {author} {\bibinfo {author} {\bibfnamefont {Bruce~A}\ \bibnamefont {Remington}},\ }\bibfield  {title} {\enquote {\bibinfo {title} {High energy density laboratory astrophysics},}\ }\href {\doibase 10.1088/0741-3335/47/5A/014} {\bibfield  {journal} {\bibinfo  {journal} {Plasma Phys. Control. Fusion}\ }\textbf {\bibinfo {volume} {47}},\ \bibinfo {pages} {A191} (\bibinfo {year} {2005})}\BibitemShut {NoStop}%
\bibitem [{\citenamefont {Graziani}\ \emph {et~al.}(2014)\citenamefont {Graziani}, \citenamefont {Desjarlais}, \citenamefont {Redmer},\ and\ \citenamefont {Trickey}}]{wdm_book}%
  \BibitemOpen
  \bibinfo {editor} {\bibfnamefont {F.}~\bibnamefont {Graziani}}, \bibinfo {editor} {\bibfnamefont {M.~P.}\ \bibnamefont {Desjarlais}}, \bibinfo {editor} {\bibfnamefont {R.}~\bibnamefont {Redmer}}, \ and\ \bibinfo {editor} {\bibfnamefont {S.~B.}\ \bibnamefont {Trickey}},\ eds.,\ \href@noop {} {\emph {\bibinfo {title} {Frontiers and Challenges in Warm Dense Matter}}}\ (\bibinfo  {publisher} {Springer},\ \bibinfo {address} {International Publishing},\ \bibinfo {year} {2014})\BibitemShut {NoStop}%
\bibitem [{\citenamefont {Riley}(2021)}]{Riley_2021}%
  \BibitemOpen
  \bibfield  {author} {\bibinfo {author} {\bibfnamefont {David}\ \bibnamefont {Riley}},\ }\href {\doibase 10.1088/978-0-7503-2348-2} {\emph {\bibinfo {title} {Warm Dense Matter}}},\ 2053-2563\ (\bibinfo  {publisher} {IOP Publishing},\ \bibinfo {year} {2021})\BibitemShut {NoStop}%
\bibitem [{\citenamefont {Vorberger}\ \emph {et~al.}(2025)\citenamefont {Vorberger}, \citenamefont {Graziani}, \citenamefont {Riley}, \citenamefont {Baczewski}, \citenamefont {Baraffe}, \citenamefont {Bethkenhagen}, \citenamefont {Blouin}, \citenamefont {Böhme}, \citenamefont {Bonitz}, \citenamefont {Bussmann}, \citenamefont {Casner}, \citenamefont {Cayzac}, \citenamefont {Celliers}, \citenamefont {Chabrier}, \citenamefont {Chamel}, \citenamefont {Chapman}, \citenamefont {Chen}, \citenamefont {Clérouin}, \citenamefont {Collins}, \citenamefont {Coppari}, \citenamefont {Döppner}, \citenamefont {Dornheim}, \citenamefont {Fletcher}, \citenamefont {Gericke}, \citenamefont {Glenzer}, \citenamefont {Goncharov}, \citenamefont {Gregori}, \citenamefont {Hamel}, \citenamefont {Hansen}, \citenamefont {Hartley}, \citenamefont {Hu}, \citenamefont {Hurricane}, \citenamefont {Karasiev}, \citenamefont {Kas}, \citenamefont {Kettle}, \citenamefont {Kluge}, \citenamefont {Knudson}, \citenamefont {Kononov}, \citenamefont
  {Konôpková}, \citenamefont {Kraus}, \citenamefont {Kritcher}, \citenamefont {Malko}, \citenamefont {Massacrier}, \citenamefont {Militzer}, \citenamefont {Moldabekov}, \citenamefont {Murillo}, \citenamefont {Nagler}, \citenamefont {Nettelmann}, \citenamefont {Neumayer}, \citenamefont {Ofori-Okai}, \citenamefont {Oleynik}, \citenamefont {Preising}, \citenamefont {Pribram-Jones}, \citenamefont {Ramazanov}, \citenamefont {Ravasio}, \citenamefont {Redmer}, \citenamefont {Rethfeld}, \citenamefont {Robinson}, \citenamefont {Röpke}, \citenamefont {Soubiran}, \citenamefont {Starrett}, \citenamefont {Steinle-Neumann}, \citenamefont {Sterne}, \citenamefont {Tanaka}, \citenamefont {Thompson}, \citenamefont {Trickey}, \citenamefont {Vinci}, \citenamefont {Vinko}, \citenamefont {Wang}, \citenamefont {White}, \citenamefont {White}, \citenamefont {Zastrau}, \citenamefont {Zurek},\ and\ \citenamefont {Tolias}}]{vorberger2025roadmapwarmdensematter}%
  \BibitemOpen
  \bibfield  {author} {\bibinfo {author} {\bibfnamefont {Jan}\ \bibnamefont {Vorberger}}, \bibinfo {author} {\bibfnamefont {Frank}\ \bibnamefont {Graziani}}, \bibinfo {author} {\bibfnamefont {David}\ \bibnamefont {Riley}}, \bibinfo {author} {\bibfnamefont {Andrew~D.}\ \bibnamefont {Baczewski}}, \bibinfo {author} {\bibfnamefont {Isabelle}\ \bibnamefont {Baraffe}}, \bibinfo {author} {\bibfnamefont {Mandy}\ \bibnamefont {Bethkenhagen}}, \bibinfo {author} {\bibfnamefont {Simon}\ \bibnamefont {Blouin}}, \bibinfo {author} {\bibfnamefont {Maximilian~P.}\ \bibnamefont {Böhme}}, \bibinfo {author} {\bibfnamefont {Michael}\ \bibnamefont {Bonitz}}, \bibinfo {author} {\bibfnamefont {Michael}\ \bibnamefont {Bussmann}}, \bibinfo {author} {\bibfnamefont {Alexis}\ \bibnamefont {Casner}}, \bibinfo {author} {\bibfnamefont {Witold}\ \bibnamefont {Cayzac}}, \bibinfo {author} {\bibfnamefont {Peter}\ \bibnamefont {Celliers}}, \bibinfo {author} {\bibfnamefont {Gilles}\ \bibnamefont {Chabrier}}, \bibinfo {author} {\bibfnamefont
  {Nicolas}\ \bibnamefont {Chamel}}, \bibinfo {author} {\bibfnamefont {Dave}\ \bibnamefont {Chapman}}, \bibinfo {author} {\bibfnamefont {Mohan}\ \bibnamefont {Chen}}, \bibinfo {author} {\bibfnamefont {Jean}\ \bibnamefont {Clérouin}}, \bibinfo {author} {\bibfnamefont {Gilbert}\ \bibnamefont {Collins}}, \bibinfo {author} {\bibfnamefont {Federica}\ \bibnamefont {Coppari}}, \bibinfo {author} {\bibfnamefont {Tilo}\ \bibnamefont {Döppner}}, \bibinfo {author} {\bibfnamefont {Tobias}\ \bibnamefont {Dornheim}}, \bibinfo {author} {\bibfnamefont {Luke~B.}\ \bibnamefont {Fletcher}}, \bibinfo {author} {\bibfnamefont {Dirk~O.}\ \bibnamefont {Gericke}}, \bibinfo {author} {\bibfnamefont {Siegfried}\ \bibnamefont {Glenzer}}, \bibinfo {author} {\bibfnamefont {Alexander~F.}\ \bibnamefont {Goncharov}}, \bibinfo {author} {\bibfnamefont {Gianluca}\ \bibnamefont {Gregori}}, \bibinfo {author} {\bibfnamefont {Sebastien}\ \bibnamefont {Hamel}}, \bibinfo {author} {\bibfnamefont {Stephanie~B.}\ \bibnamefont {Hansen}}, \bibinfo
  {author} {\bibfnamefont {Nicholas~J.}\ \bibnamefont {Hartley}}, \bibinfo {author} {\bibfnamefont {Suxing}\ \bibnamefont {Hu}}, \bibinfo {author} {\bibfnamefont {Omar~A.}\ \bibnamefont {Hurricane}}, \bibinfo {author} {\bibfnamefont {Valentin~V.}\ \bibnamefont {Karasiev}}, \bibinfo {author} {\bibfnamefont {Joshua~J.}\ \bibnamefont {Kas}}, \bibinfo {author} {\bibfnamefont {Brendan}\ \bibnamefont {Kettle}}, \bibinfo {author} {\bibfnamefont {Thomas}\ \bibnamefont {Kluge}}, \bibinfo {author} {\bibfnamefont {Marcus~D.}\ \bibnamefont {Knudson}}, \bibinfo {author} {\bibfnamefont {Alina}\ \bibnamefont {Kononov}}, \bibinfo {author} {\bibfnamefont {Zuzana}\ \bibnamefont {Konôpková}}, \bibinfo {author} {\bibfnamefont {Dominik}\ \bibnamefont {Kraus}}, \bibinfo {author} {\bibfnamefont {Andrea}\ \bibnamefont {Kritcher}}, \bibinfo {author} {\bibfnamefont {Sophia}\ \bibnamefont {Malko}}, \bibinfo {author} {\bibfnamefont {Gérard}\ \bibnamefont {Massacrier}}, \bibinfo {author} {\bibfnamefont {Burkhard}\ \bibnamefont
  {Militzer}}, \bibinfo {author} {\bibfnamefont {Zhandos~A.}\ \bibnamefont {Moldabekov}}, \bibinfo {author} {\bibfnamefont {Michael~S.}\ \bibnamefont {Murillo}}, \bibinfo {author} {\bibfnamefont {Bob}\ \bibnamefont {Nagler}}, \bibinfo {author} {\bibfnamefont {Nadine}\ \bibnamefont {Nettelmann}}, \bibinfo {author} {\bibfnamefont {Paul}\ \bibnamefont {Neumayer}}, \bibinfo {author} {\bibfnamefont {Benjamin~K.}\ \bibnamefont {Ofori-Okai}}, \bibinfo {author} {\bibfnamefont {Ivan~I.}\ \bibnamefont {Oleynik}}, \bibinfo {author} {\bibfnamefont {Martin}\ \bibnamefont {Preising}}, \bibinfo {author} {\bibfnamefont {Aurora}\ \bibnamefont {Pribram-Jones}}, \bibinfo {author} {\bibfnamefont {Tlekkabul}\ \bibnamefont {Ramazanov}}, \bibinfo {author} {\bibfnamefont {Alessandra}\ \bibnamefont {Ravasio}}, \bibinfo {author} {\bibfnamefont {Ronald}\ \bibnamefont {Redmer}}, \bibinfo {author} {\bibfnamefont {Baerbel}\ \bibnamefont {Rethfeld}}, \bibinfo {author} {\bibfnamefont {Alex P.~L.}\ \bibnamefont {Robinson}}, \bibinfo {author}
  {\bibfnamefont {Gerd}\ \bibnamefont {Röpke}}, \bibinfo {author} {\bibfnamefont {François}\ \bibnamefont {Soubiran}}, \bibinfo {author} {\bibfnamefont {Charles~E.}\ \bibnamefont {Starrett}}, \bibinfo {author} {\bibfnamefont {Gerd}\ \bibnamefont {Steinle-Neumann}}, \bibinfo {author} {\bibfnamefont {Phillip~A.}\ \bibnamefont {Sterne}}, \bibinfo {author} {\bibfnamefont {Shigenori}\ \bibnamefont {Tanaka}}, \bibinfo {author} {\bibfnamefont {Aidan~P.}\ \bibnamefont {Thompson}}, \bibinfo {author} {\bibfnamefont {Samuel~B.}\ \bibnamefont {Trickey}}, \bibinfo {author} {\bibfnamefont {Tommaso}\ \bibnamefont {Vinci}}, \bibinfo {author} {\bibfnamefont {Sam~M.}\ \bibnamefont {Vinko}}, \bibinfo {author} {\bibfnamefont {Lei}\ \bibnamefont {Wang}}, \bibinfo {author} {\bibfnamefont {Alexander~J.}\ \bibnamefont {White}}, \bibinfo {author} {\bibfnamefont {Thomas~G.}\ \bibnamefont {White}}, \bibinfo {author} {\bibfnamefont {Ulf}\ \bibnamefont {Zastrau}}, \bibinfo {author} {\bibfnamefont {Eva}\ \bibnamefont {Zurek}}, \ and\
  \bibinfo {author} {\bibfnamefont {Panagiotis}\ \bibnamefont {Tolias}},\ }\href {https://arxiv.org/abs/2505.02494} {\enquote {\bibinfo {title} {Roadmap for warm dense matter physics},}\ } (\bibinfo {year} {2025}),\ \Eprint {http://arxiv.org/abs/2505.02494} {arXiv:2505.02494 [physics.plasm-ph]} \BibitemShut {NoStop}%
\bibitem [{\citenamefont {Baus}\ and\ \citenamefont {Hansen}(1980)}]{Baus_Hansen_OCP}%
  \BibitemOpen
  \bibfield  {author} {\bibinfo {author} {\bibfnamefont {Marc}\ \bibnamefont {Baus}}\ and\ \bibinfo {author} {\bibfnamefont {Jean-Pierre}\ \bibnamefont {Hansen}},\ }\bibfield  {title} {\enquote {\bibinfo {title} {Statistical mechanics of simple {Coulomb} systems},}\ }\href {https://www.sciencedirect.com/science/article/pii/0370157380900228} {\bibfield  {journal} {\bibinfo  {journal} {Phys. Rep.}\ }\textbf {\bibinfo {volume} {59}},\ \bibinfo {pages} {1--94} (\bibinfo {year} {1980})}\BibitemShut {NoStop}%
\bibitem [{\citenamefont {Ichimaru}(1982)}]{plasma2}%
  \BibitemOpen
  \bibfield  {author} {\bibinfo {author} {\bibfnamefont {S.}~\bibnamefont {Ichimaru}},\ }\bibfield  {title} {\enquote {\bibinfo {title} {Strongly coupled plasmas: high-density classical plasmas and degenerate electron liquids},}\ }\href {https://journals.aps.org/rmp/abstract/10.1103/RevModPhys.54.1017} {\bibfield  {journal} {\bibinfo  {journal} {Rev. Mod. Phys}\ }\textbf {\bibinfo {volume} {54}},\ \bibinfo {pages} {1017} (\bibinfo {year} {1982})}\BibitemShut {NoStop}%
\bibitem [{\citenamefont {Ichimaru}\ \emph {et~al.}(1987)\citenamefont {Ichimaru}, \citenamefont {Iyetomi},\ and\ \citenamefont {Tanaka}}]{IIT}%
  \BibitemOpen
  \bibfield  {author} {\bibinfo {author} {\bibfnamefont {Setsuo}\ \bibnamefont {Ichimaru}}, \bibinfo {author} {\bibfnamefont {Hiroshi}\ \bibnamefont {Iyetomi}}, \ and\ \bibinfo {author} {\bibfnamefont {Shigenori}\ \bibnamefont {Tanaka}},\ }\bibfield  {title} {\enquote {\bibinfo {title} {Statistical physics of dense plasmas: Thermodynamics, transport coefficients and dynamic correlations},}\ }\href {\doibase https://doi.org/10.1016/0370-1573(87)90125-6} {\bibfield  {journal} {\bibinfo  {journal} {Phys. Rep.}\ }\textbf {\bibinfo {volume} {149}},\ \bibinfo {pages} {91--205} (\bibinfo {year} {1987})}\BibitemShut {NoStop}%
\bibitem [{\citenamefont {Dornheim}\ \emph {et~al.}(2018{\natexlab{a}})\citenamefont {Dornheim}, \citenamefont {Groth},\ and\ \citenamefont {Bonitz}}]{review}%
  \BibitemOpen
  \bibfield  {author} {\bibinfo {author} {\bibfnamefont {T.}~\bibnamefont {Dornheim}}, \bibinfo {author} {\bibfnamefont {S.}~\bibnamefont {Groth}}, \ and\ \bibinfo {author} {\bibfnamefont {M.}~\bibnamefont {Bonitz}},\ }\bibfield  {title} {\enquote {\bibinfo {title} {The uniform electron gas at warm dense matter conditions},}\ }\href {https://www.sciencedirect.com/science/article/abs/pii/S0370157318300516} {\bibfield  {journal} {\bibinfo  {journal} {Phys. Rep.}\ }\textbf {\bibinfo {volume} {744}},\ \bibinfo {pages} {1--86} (\bibinfo {year} {2018}{\natexlab{a}})}\BibitemShut {NoStop}%
\bibitem [{\citenamefont {Dornheim}\ \emph {et~al.}(2023{\natexlab{a}})\citenamefont {Dornheim}, \citenamefont {Moldabekov}, \citenamefont {Ramakrishna}, \citenamefont {Tolias}, \citenamefont {Baczewski}, \citenamefont {Kraus}, \citenamefont {Preston}, \citenamefont {Chapman}, \citenamefont {Böhme}, \citenamefont {Döppner}, \citenamefont {Graziani}, \citenamefont {Bonitz}, \citenamefont {Cangi},\ and\ \citenamefont {Vorberger}}]{Dornheim_review}%
  \BibitemOpen
  \bibfield  {author} {\bibinfo {author} {\bibfnamefont {Tobias}\ \bibnamefont {Dornheim}}, \bibinfo {author} {\bibfnamefont {Zhandos~A.}\ \bibnamefont {Moldabekov}}, \bibinfo {author} {\bibfnamefont {Kushal}\ \bibnamefont {Ramakrishna}}, \bibinfo {author} {\bibfnamefont {Panagiotis}\ \bibnamefont {Tolias}}, \bibinfo {author} {\bibfnamefont {Andrew~D.}\ \bibnamefont {Baczewski}}, \bibinfo {author} {\bibfnamefont {Dominik}\ \bibnamefont {Kraus}}, \bibinfo {author} {\bibfnamefont {Thomas~R.}\ \bibnamefont {Preston}}, \bibinfo {author} {\bibfnamefont {David~A.}\ \bibnamefont {Chapman}}, \bibinfo {author} {\bibfnamefont {Maximilian~P.}\ \bibnamefont {Böhme}}, \bibinfo {author} {\bibfnamefont {Tilo}\ \bibnamefont {Döppner}}, \bibinfo {author} {\bibfnamefont {Frank}\ \bibnamefont {Graziani}}, \bibinfo {author} {\bibfnamefont {Michael}\ \bibnamefont {Bonitz}}, \bibinfo {author} {\bibfnamefont {Attila}\ \bibnamefont {Cangi}}, \ and\ \bibinfo {author} {\bibfnamefont {Jan}\ \bibnamefont {Vorberger}},\ }\bibfield
  {title} {\enquote {\bibinfo {title} {Electronic density response of warm dense matter},}\ }\href {\doibase 10.1063/5.0138955} {\bibfield  {journal} {\bibinfo  {journal} {Phys. Plasmas}\ }\textbf {\bibinfo {volume} {30}},\ \bibinfo {pages} {032705} (\bibinfo {year} {2023}{\natexlab{a}})}\BibitemShut {NoStop}%
\bibitem [{\citenamefont {Dornheim}\ \emph {et~al.}(2017{\natexlab{a}})\citenamefont {Dornheim}, \citenamefont {Groth}, \citenamefont {Malone}, \citenamefont {Schoof}, \citenamefont {Sjostrom}, \citenamefont {Foulkes},\ and\ \citenamefont {Bonitz}}]{Dornheim_POP_2017}%
  \BibitemOpen
  \bibfield  {author} {\bibinfo {author} {\bibfnamefont {Tobias}\ \bibnamefont {Dornheim}}, \bibinfo {author} {\bibfnamefont {Simon}\ \bibnamefont {Groth}}, \bibinfo {author} {\bibfnamefont {Fionn~D.}\ \bibnamefont {Malone}}, \bibinfo {author} {\bibfnamefont {Tim}\ \bibnamefont {Schoof}}, \bibinfo {author} {\bibfnamefont {Travis}\ \bibnamefont {Sjostrom}}, \bibinfo {author} {\bibfnamefont {W.~M.~C.}\ \bibnamefont {Foulkes}}, \ and\ \bibinfo {author} {\bibfnamefont {Michael}\ \bibnamefont {Bonitz}},\ }\bibfield  {title} {\enquote {\bibinfo {title} {Ab initio quantum {Monte Carlo} simulation of the warm dense electron gas},}\ }\href {\doibase 10.1063/1.4977920} {\bibfield  {journal} {\bibinfo  {journal} {Phys. Plasmas}\ }\textbf {\bibinfo {volume} {24}},\ \bibinfo {pages} {056303} (\bibinfo {year} {2017}{\natexlab{a}})}\BibitemShut {NoStop}%
\bibitem [{\citenamefont {Bonitz}\ \emph {et~al.}(2020)\citenamefont {Bonitz}, \citenamefont {Dornheim}, \citenamefont {Moldabekov}, \citenamefont {Zhang}, \citenamefont {Hamann}, \citenamefont {Kählert}, \citenamefont {Filinov}, \citenamefont {Ramakrishna},\ and\ \citenamefont {Vorberger}}]{new_POP}%
  \BibitemOpen
  \bibfield  {author} {\bibinfo {author} {\bibfnamefont {M.}~\bibnamefont {Bonitz}}, \bibinfo {author} {\bibfnamefont {T.}~\bibnamefont {Dornheim}}, \bibinfo {author} {\bibfnamefont {Zh.~A.}\ \bibnamefont {Moldabekov}}, \bibinfo {author} {\bibfnamefont {S.}~\bibnamefont {Zhang}}, \bibinfo {author} {\bibfnamefont {P.}~\bibnamefont {Hamann}}, \bibinfo {author} {\bibfnamefont {H.}~\bibnamefont {Kählert}}, \bibinfo {author} {\bibfnamefont {A.}~\bibnamefont {Filinov}}, \bibinfo {author} {\bibfnamefont {K.}~\bibnamefont {Ramakrishna}}, \ and\ \bibinfo {author} {\bibfnamefont {J.}~\bibnamefont {Vorberger}},\ }\bibfield  {title} {\enquote {\bibinfo {title} {Ab initio simulation of warm dense matter},}\ }\href {\doibase 10.1063/1.5143225} {\bibfield  {journal} {\bibinfo  {journal} {Phys. Plasmas}\ }\textbf {\bibinfo {volume} {27}},\ \bibinfo {pages} {042710} (\bibinfo {year} {2020})}\BibitemShut {NoStop}%
\bibitem [{\citenamefont {Dornheim}(2019)}]{dornheim_sign_problem}%
  \BibitemOpen
  \bibfield  {author} {\bibinfo {author} {\bibfnamefont {T.}~\bibnamefont {Dornheim}},\ }\bibfield  {title} {\enquote {\bibinfo {title} {Fermion sign problem in path integral {M}onte {C}arlo simulations: Quantum dots, ultracold atoms, and warm dense matter},}\ }\href {https://journals.aps.org/pre/abstract/10.1103/PhysRevE.100.023307} {\bibfield  {journal} {\bibinfo  {journal} {Phys. Rev. E}\ }\textbf {\bibinfo {volume} {100}},\ \bibinfo {pages} {023307} (\bibinfo {year} {2019})}\BibitemShut {NoStop}%
\bibitem [{\citenamefont {Dornheim}(2021)}]{Dornheim_2021}%
  \BibitemOpen
  \bibfield  {author} {\bibinfo {author} {\bibfnamefont {Tobias}\ \bibnamefont {Dornheim}},\ }\bibfield  {title} {\enquote {\bibinfo {title} {Fermion sign problem in path integral monte carlo simulations: grand-canonical ensemble},}\ }\href {\doibase 10.1088/1751-8121/ac1481} {\bibfield  {journal} {\bibinfo  {journal} {J. Phys. A: Math. Theor.}\ }\textbf {\bibinfo {volume} {54}},\ \bibinfo {pages} {335001} (\bibinfo {year} {2021})}\BibitemShut {NoStop}%
\bibitem [{\citenamefont {Dornheim}\ \emph {et~al.}(2016)\citenamefont {Dornheim}, \citenamefont {Groth}, \citenamefont {Sjostrom}, \citenamefont {Malone}, \citenamefont {Foulkes},\ and\ \citenamefont {Bonitz}}]{dornheim_prl}%
  \BibitemOpen
  \bibfield  {author} {\bibinfo {author} {\bibfnamefont {T.}~\bibnamefont {Dornheim}}, \bibinfo {author} {\bibfnamefont {S.}~\bibnamefont {Groth}}, \bibinfo {author} {\bibfnamefont {T.}~\bibnamefont {Sjostrom}}, \bibinfo {author} {\bibfnamefont {F.~D.}\ \bibnamefont {Malone}}, \bibinfo {author} {\bibfnamefont {W.~M.~C.}\ \bibnamefont {Foulkes}}, \ and\ \bibinfo {author} {\bibfnamefont {M.}~\bibnamefont {Bonitz}},\ }\bibfield  {title} {\enquote {\bibinfo {title} {Ab initio quantum {M}onte {C}arlo simulation of the warm dense electron gas in the thermodynamic limit},}\ }\href {http://link.aps.org/doi/10.1103/PhysRevLett.117.156403} {\bibfield  {journal} {\bibinfo  {journal} {Phys. Rev. Lett.}\ }\textbf {\bibinfo {volume} {117}},\ \bibinfo {pages} {156403} (\bibinfo {year} {2016})}\BibitemShut {NoStop}%
\bibitem [{\citenamefont {Dornheim}\ \emph {et~al.}(2025{\natexlab{a}})\citenamefont {Dornheim}, \citenamefont {Moldabekov}, \citenamefont {Schwalbe},\ and\ \citenamefont {Vorberger}}]{dornheim2024directfreeenergycalculation}%
  \BibitemOpen
  \bibfield  {author} {\bibinfo {author} {\bibfnamefont {Tobias}\ \bibnamefont {Dornheim}}, \bibinfo {author} {\bibfnamefont {Zhandos~A.}\ \bibnamefont {Moldabekov}}, \bibinfo {author} {\bibfnamefont {Sebastian}\ \bibnamefont {Schwalbe}}, \ and\ \bibinfo {author} {\bibfnamefont {Jan}\ \bibnamefont {Vorberger}},\ }\bibfield  {title} {\enquote {\bibinfo {title} {Direct free energy calculation from ab initio path integral {Monte Carlo} simulations of warm dense matter},}\ }\href {\doibase 10.1103/PhysRevB.111.L041114} {\bibfield  {journal} {\bibinfo  {journal} {Phys. Rev. B}\ }\textbf {\bibinfo {volume} {111}},\ \bibinfo {pages} {L041114} (\bibinfo {year} {2025}{\natexlab{a}})}\BibitemShut {NoStop}%
\bibitem [{\citenamefont {Dornheim}\ \emph {et~al.}(2025{\natexlab{b}})\citenamefont {Dornheim}, \citenamefont {Bonitz}, \citenamefont {Moldabekov}, \citenamefont {Schwalbe}, \citenamefont {Tolias},\ and\ \citenamefont {Vorberger}}]{dornheim2024chemicalpotentialwarmdense}%
  \BibitemOpen
  \bibfield  {author} {\bibinfo {author} {\bibfnamefont {Tobias}\ \bibnamefont {Dornheim}}, \bibinfo {author} {\bibfnamefont {Michael}\ \bibnamefont {Bonitz}}, \bibinfo {author} {\bibfnamefont {Zhandos~A.}\ \bibnamefont {Moldabekov}}, \bibinfo {author} {\bibfnamefont {Sebastian}\ \bibnamefont {Schwalbe}}, \bibinfo {author} {\bibfnamefont {Panagiotis}\ \bibnamefont {Tolias}}, \ and\ \bibinfo {author} {\bibfnamefont {Jan}\ \bibnamefont {Vorberger}},\ }\bibfield  {title} {\enquote {\bibinfo {title} {Chemical potential of the warm dense electron gas from ab initio path integral {Monte Carlo} simulations},}\ }\href {\doibase 10.1103/PhysRevB.111.115149} {\bibfield  {journal} {\bibinfo  {journal} {Phys. Rev. B}\ }\textbf {\bibinfo {volume} {111}},\ \bibinfo {pages} {115149} (\bibinfo {year} {2025}{\natexlab{b}})}\BibitemShut {NoStop}%
\bibitem [{\citenamefont {Dornheim}\ \emph {et~al.}(2020{\natexlab{a}})\citenamefont {Dornheim}, \citenamefont {Cangi}, \citenamefont {Ramakrishna}, \citenamefont {B\"ohme}, \citenamefont {Tanaka},\ and\ \citenamefont {Vorberger}}]{Dornheim_PRL_2020_ESA}%
  \BibitemOpen
  \bibfield  {author} {\bibinfo {author} {\bibfnamefont {Tobias}\ \bibnamefont {Dornheim}}, \bibinfo {author} {\bibfnamefont {Attila}\ \bibnamefont {Cangi}}, \bibinfo {author} {\bibfnamefont {Kushal}\ \bibnamefont {Ramakrishna}}, \bibinfo {author} {\bibfnamefont {Maximilian}\ \bibnamefont {B\"ohme}}, \bibinfo {author} {\bibfnamefont {Shigenori}\ \bibnamefont {Tanaka}}, \ and\ \bibinfo {author} {\bibfnamefont {Jan}\ \bibnamefont {Vorberger}},\ }\bibfield  {title} {\enquote {\bibinfo {title} {Effective static approximation: A fast and reliable tool for warm-dense matter theory},}\ }\href {\doibase 10.1103/PhysRevLett.125.235001} {\bibfield  {journal} {\bibinfo  {journal} {Phys. Rev. Lett.}\ }\textbf {\bibinfo {volume} {125}},\ \bibinfo {pages} {235001} (\bibinfo {year} {2020}{\natexlab{a}})}\BibitemShut {NoStop}%
\bibitem [{\citenamefont {Dornheim}\ \emph {et~al.}(2019)\citenamefont {Dornheim}, \citenamefont {Vorberger}, \citenamefont {Groth}, \citenamefont {Hoffmann}, \citenamefont {Moldabekov},\ and\ \citenamefont {Bonitz}}]{dornheim_ML}%
  \BibitemOpen
  \bibfield  {author} {\bibinfo {author} {\bibfnamefont {T.}~\bibnamefont {Dornheim}}, \bibinfo {author} {\bibfnamefont {J.}~\bibnamefont {Vorberger}}, \bibinfo {author} {\bibfnamefont {S.}~\bibnamefont {Groth}}, \bibinfo {author} {\bibfnamefont {N.}~\bibnamefont {Hoffmann}}, \bibinfo {author} {\bibfnamefont {Zh.A.}\ \bibnamefont {Moldabekov}}, \ and\ \bibinfo {author} {\bibfnamefont {M.}~\bibnamefont {Bonitz}},\ }\bibfield  {title} {\enquote {\bibinfo {title} {The static local field correction of the warm dense electron gas: An ab initio path integral {M}onte {C}arlo study and machine learning representation},}\ }\href {https://aip.scitation.org/doi/full/10.1063/1.5123013} {\bibfield  {journal} {\bibinfo  {journal} {J. Chem. Phys}\ }\textbf {\bibinfo {volume} {151}},\ \bibinfo {pages} {194104} (\bibinfo {year} {2019})}\BibitemShut {NoStop}%
\bibitem [{\citenamefont {Dornheim}\ \emph {et~al.}(2021{\natexlab{a}})\citenamefont {Dornheim}, \citenamefont {Moldabekov},\ and\ \citenamefont {Tolias}}]{Dornheim_PRB_ESA_2021}%
  \BibitemOpen
  \bibfield  {author} {\bibinfo {author} {\bibfnamefont {Tobias}\ \bibnamefont {Dornheim}}, \bibinfo {author} {\bibfnamefont {Zhandos~A.}\ \bibnamefont {Moldabekov}}, \ and\ \bibinfo {author} {\bibfnamefont {Panagiotis}\ \bibnamefont {Tolias}},\ }\bibfield  {title} {\enquote {\bibinfo {title} {Analytical representation of the local field correction of the uniform electron gas within the effective static approximation},}\ }\href {\doibase 10.1103/PhysRevB.103.165102} {\bibfield  {journal} {\bibinfo  {journal} {Phys. Rev. B}\ }\textbf {\bibinfo {volume} {103}},\ \bibinfo {pages} {165102} (\bibinfo {year} {2021}{\natexlab{a}})}\BibitemShut {NoStop}%
\bibitem [{\citenamefont {Dornheim}\ \emph {et~al.}(2022{\natexlab{a}})\citenamefont {Dornheim}, \citenamefont {Vorberger}, \citenamefont {Moldabekov}, \citenamefont {Röpke},\ and\ \citenamefont {Kraeft}}]{Dornheim_HEDP_2022}%
  \BibitemOpen
  \bibfield  {author} {\bibinfo {author} {\bibfnamefont {Tobias}\ \bibnamefont {Dornheim}}, \bibinfo {author} {\bibfnamefont {Jan}\ \bibnamefont {Vorberger}}, \bibinfo {author} {\bibfnamefont {Zhandos}\ \bibnamefont {Moldabekov}}, \bibinfo {author} {\bibfnamefont {Gerd}\ \bibnamefont {Röpke}}, \ and\ \bibinfo {author} {\bibfnamefont {Wolf-Dietrich}\ \bibnamefont {Kraeft}},\ }\bibfield  {title} {\enquote {\bibinfo {title} {The uniform electron gas at high temperatures: ab initio path integral {Monte Carlo} simulations and analytical theory},}\ }\href {\doibase https://doi.org/10.1016/j.hedp.2022.101015} {\bibfield  {journal} {\bibinfo  {journal} {High Energy Density Phys.}\ }\textbf {\bibinfo {volume} {45}},\ \bibinfo {pages} {101015} (\bibinfo {year} {2022}{\natexlab{a}})}\BibitemShut {NoStop}%
\bibitem [{\citenamefont {Dornheim}\ \emph {et~al.}(2024{\natexlab{a}})\citenamefont {Dornheim}, \citenamefont {Schwalbe}, \citenamefont {Moldabekov}, \citenamefont {Vorberger},\ and\ \citenamefont {Tolias}}]{Dornheim_JPCL_2024}%
  \BibitemOpen
  \bibfield  {author} {\bibinfo {author} {\bibfnamefont {Tobias}\ \bibnamefont {Dornheim}}, \bibinfo {author} {\bibfnamefont {Sebastian}\ \bibnamefont {Schwalbe}}, \bibinfo {author} {\bibfnamefont {Zhandos~A.}\ \bibnamefont {Moldabekov}}, \bibinfo {author} {\bibfnamefont {Jan}\ \bibnamefont {Vorberger}}, \ and\ \bibinfo {author} {\bibfnamefont {Panagiotis}\ \bibnamefont {Tolias}},\ }\bibfield  {title} {\enquote {\bibinfo {title} {Ab initio path integral {Monte Carlo} simulations of the uniform electron gas on large length scales},}\ }\href {\doibase 10.1021/acs.jpclett.3c03193} {\bibfield  {journal} {\bibinfo  {journal} {J. Phys. Chem. Lett.}\ }\textbf {\bibinfo {volume} {15}},\ \bibinfo {pages} {1305--1313} (\bibinfo {year} {2024}{\natexlab{a}})}\BibitemShut {NoStop}%
\bibitem [{\citenamefont {Dornheim}\ \emph {et~al.}(2018{\natexlab{b}})\citenamefont {Dornheim}, \citenamefont {Groth}, \citenamefont {Vorberger},\ and\ \citenamefont {Bonitz}}]{dornheim_dynamic}%
  \BibitemOpen
  \bibfield  {author} {\bibinfo {author} {\bibfnamefont {T.}~\bibnamefont {Dornheim}}, \bibinfo {author} {\bibfnamefont {S.}~\bibnamefont {Groth}}, \bibinfo {author} {\bibfnamefont {J.}~\bibnamefont {Vorberger}}, \ and\ \bibinfo {author} {\bibfnamefont {M.}~\bibnamefont {Bonitz}},\ }\bibfield  {title} {\enquote {\bibinfo {title} {Ab initio path integral {M}onte {C}arlo results for the dynamic structure factor of correlated electrons: From the electron liquid to warm dense matter},}\ }\href {https://journals.aps.org/prl/abstract/10.1103/PhysRevLett.121.255001} {\bibfield  {journal} {\bibinfo  {journal} {Phys. Rev. Lett.}\ }\textbf {\bibinfo {volume} {121}},\ \bibinfo {pages} {255001} (\bibinfo {year} {2018}{\natexlab{b}})}\BibitemShut {NoStop}%
\bibitem [{\citenamefont {Groth}\ \emph {et~al.}(2019)\citenamefont {Groth}, \citenamefont {Dornheim},\ and\ \citenamefont {Vorberger}}]{dynamic_folgepaper}%
  \BibitemOpen
  \bibfield  {author} {\bibinfo {author} {\bibfnamefont {S.}~\bibnamefont {Groth}}, \bibinfo {author} {\bibfnamefont {T.}~\bibnamefont {Dornheim}}, \ and\ \bibinfo {author} {\bibfnamefont {J.}~\bibnamefont {Vorberger}},\ }\bibfield  {title} {\enquote {\bibinfo {title} {Ab initio path integral {M}onte {C}arlo approach to the static and dynamic density response of the uniform electron gas},}\ }\href {https://link.aps.org/doi/10.1103/PhysRevB.99.235122} {\bibfield  {journal} {\bibinfo  {journal} {Phys. Rev. B}\ }\textbf {\bibinfo {volume} {99}},\ \bibinfo {pages} {235122} (\bibinfo {year} {2019})}\BibitemShut {NoStop}%
\bibitem [{\citenamefont {Hamann}\ \emph {et~al.}(2020)\citenamefont {Hamann}, \citenamefont {Dornheim}, \citenamefont {Vorberger}, \citenamefont {Moldabekov},\ and\ \citenamefont {Bonitz}}]{Hamann_PRB_2020}%
  \BibitemOpen
  \bibfield  {author} {\bibinfo {author} {\bibfnamefont {Paul}\ \bibnamefont {Hamann}}, \bibinfo {author} {\bibfnamefont {Tobias}\ \bibnamefont {Dornheim}}, \bibinfo {author} {\bibfnamefont {Jan}\ \bibnamefont {Vorberger}}, \bibinfo {author} {\bibfnamefont {Zhandos~A.}\ \bibnamefont {Moldabekov}}, \ and\ \bibinfo {author} {\bibfnamefont {Michael}\ \bibnamefont {Bonitz}},\ }\bibfield  {title} {\enquote {\bibinfo {title} {Dynamic properties of the warm dense electron gas based on ab initio path integral {Monte Carlo} simulations},}\ }\href {\doibase 10.1103/PhysRevB.102.125150} {\bibfield  {journal} {\bibinfo  {journal} {Phys. Rev. B}\ }\textbf {\bibinfo {volume} {102}},\ \bibinfo {pages} {125150} (\bibinfo {year} {2020})}\BibitemShut {NoStop}%
\bibitem [{\citenamefont {Dornheim}\ \emph {et~al.}(2023{\natexlab{b}})\citenamefont {Dornheim}, \citenamefont {Tolias}, \citenamefont {Groth}, \citenamefont {Moldabekov}, \citenamefont {Vorberger},\ and\ \citenamefont {Hirshberg}}]{Dornheim_JCP_xi_2023}%
  \BibitemOpen
  \bibfield  {author} {\bibinfo {author} {\bibfnamefont {Tobias}\ \bibnamefont {Dornheim}}, \bibinfo {author} {\bibfnamefont {Panagiotis}\ \bibnamefont {Tolias}}, \bibinfo {author} {\bibfnamefont {Simon}\ \bibnamefont {Groth}}, \bibinfo {author} {\bibfnamefont {Zhandos~A.}\ \bibnamefont {Moldabekov}}, \bibinfo {author} {\bibfnamefont {Jan}\ \bibnamefont {Vorberger}}, \ and\ \bibinfo {author} {\bibfnamefont {Barak}\ \bibnamefont {Hirshberg}},\ }\bibfield  {title} {\enquote {\bibinfo {title} {{Fermionic physics from ab initio path integral Monte Carlo simulations of fictitious identical particles}},}\ }\href {\doibase 10.1063/5.0171930} {\bibfield  {journal} {\bibinfo  {journal} {J. Chem. Phys.}\ }\textbf {\bibinfo {volume} {159}},\ \bibinfo {pages} {164113} (\bibinfo {year} {2023}{\natexlab{b}})}\BibitemShut {NoStop}%
\bibitem [{\citenamefont {Dornheim}\ \emph {et~al.}(2024{\natexlab{b}})\citenamefont {Dornheim}, \citenamefont {Tolias}, \citenamefont {Kalkavouras}, \citenamefont {Moldabekov},\ and\ \citenamefont {Vorberger}}]{Dornheim_PRB_2024}%
  \BibitemOpen
  \bibfield  {author} {\bibinfo {author} {\bibfnamefont {Tobias}\ \bibnamefont {Dornheim}}, \bibinfo {author} {\bibfnamefont {Panagiotis}\ \bibnamefont {Tolias}}, \bibinfo {author} {\bibfnamefont {Fotios}\ \bibnamefont {Kalkavouras}}, \bibinfo {author} {\bibfnamefont {Zhandos~A.}\ \bibnamefont {Moldabekov}}, \ and\ \bibinfo {author} {\bibfnamefont {Jan}\ \bibnamefont {Vorberger}},\ }\bibfield  {title} {\enquote {\bibinfo {title} {Dynamic exchange correlation effects in the strongly coupled electron liquid},}\ }\href {\doibase 10.1103/PhysRevB.110.075137} {\bibfield  {journal} {\bibinfo  {journal} {Phys. Rev. B}\ }\textbf {\bibinfo {volume} {110}},\ \bibinfo {pages} {075137} (\bibinfo {year} {2024}{\natexlab{b}})}\BibitemShut {NoStop}%
\bibitem [{\citenamefont {Dornheim}\ \emph {et~al.}(2024{\natexlab{c}})\citenamefont {Dornheim}, \citenamefont {Tolias}, \citenamefont {Vorberger},\ and\ \citenamefont {Moldabekov}}]{Dornheim_EPL_2024}%
  \BibitemOpen
  \bibfield  {author} {\bibinfo {author} {\bibfnamefont {Tobias}\ \bibnamefont {Dornheim}}, \bibinfo {author} {\bibfnamefont {Panagiotis}\ \bibnamefont {Tolias}}, \bibinfo {author} {\bibfnamefont {Jan}\ \bibnamefont {Vorberger}}, \ and\ \bibinfo {author} {\bibfnamefont {Zhandos~A.}\ \bibnamefont {Moldabekov}},\ }\bibfield  {title} {\enquote {\bibinfo {title} {Quantum delocalization, structural order, and density response of the strongly coupled electron liquid},}\ }\href {\doibase 10.1209/0295-5075/ad5d88} {\bibfield  {journal} {\bibinfo  {journal} {EPL (Europhysics Letters)}\ }\textbf {\bibinfo {volume} {147}},\ \bibinfo {pages} {36001} (\bibinfo {year} {2024}{\natexlab{c}})}\BibitemShut {NoStop}%
\bibitem [{\citenamefont {Dornheim}\ \emph {et~al.}(2020{\natexlab{b}})\citenamefont {Dornheim}, \citenamefont {Vorberger},\ and\ \citenamefont {Bonitz}}]{Dornheim_PRL_2020}%
  \BibitemOpen
  \bibfield  {author} {\bibinfo {author} {\bibfnamefont {Tobias}\ \bibnamefont {Dornheim}}, \bibinfo {author} {\bibfnamefont {Jan}\ \bibnamefont {Vorberger}}, \ and\ \bibinfo {author} {\bibfnamefont {Michael}\ \bibnamefont {Bonitz}},\ }\bibfield  {title} {\enquote {\bibinfo {title} {Nonlinear electronic density response in warm dense matter},}\ }\href {\doibase 10.1103/PhysRevLett.125.085001} {\bibfield  {journal} {\bibinfo  {journal} {Phys. Rev. Lett.}\ }\textbf {\bibinfo {volume} {125}},\ \bibinfo {pages} {085001} (\bibinfo {year} {2020}{\natexlab{b}})}\BibitemShut {NoStop}%
\bibitem [{\citenamefont {Dornheim}\ \emph {et~al.}(2021{\natexlab{b}})\citenamefont {Dornheim}, \citenamefont {Moldabekov},\ and\ \citenamefont {Vorberger}}]{Dornheim_JCP_ITCF_2021}%
  \BibitemOpen
  \bibfield  {author} {\bibinfo {author} {\bibfnamefont {Tobias}\ \bibnamefont {Dornheim}}, \bibinfo {author} {\bibfnamefont {Zhandos~A.}\ \bibnamefont {Moldabekov}}, \ and\ \bibinfo {author} {\bibfnamefont {Jan}\ \bibnamefont {Vorberger}},\ }\bibfield  {title} {\enquote {\bibinfo {title} {Nonlinear density response from imaginary-time correlation functions: Ab initio path integral {Monte Carlo} simulations of the warm dense electron gas},}\ }\href {\doibase 10.1063/5.0058988} {\bibfield  {journal} {\bibinfo  {journal} {J. Chem. Phys.}\ }\textbf {\bibinfo {volume} {155}},\ \bibinfo {pages} {054110} (\bibinfo {year} {2021}{\natexlab{b}})}\BibitemShut {NoStop}%
\bibitem [{\citenamefont {Tolias}\ \emph {et~al.}(2023)\citenamefont {Tolias}, \citenamefont {Dornheim}, \citenamefont {Moldabekov},\ and\ \citenamefont {Vorberger}}]{Tolias_2023}%
  \BibitemOpen
  \bibfield  {author} {\bibinfo {author} {\bibfnamefont {Panagiotis}\ \bibnamefont {Tolias}}, \bibinfo {author} {\bibfnamefont {Tobias}\ \bibnamefont {Dornheim}}, \bibinfo {author} {\bibfnamefont {Zhandos~A.}\ \bibnamefont {Moldabekov}}, \ and\ \bibinfo {author} {\bibfnamefont {Jan}\ \bibnamefont {Vorberger}},\ }\bibfield  {title} {\enquote {\bibinfo {title} {Unravelling the nonlinear ideal density response of many-body systems},}\ }\href {\doibase 10.1209/0295-5075/acd3a6} {\bibfield  {journal} {\bibinfo  {journal} {EPL (Europhysics Letters)}\ }\textbf {\bibinfo {volume} {142}},\ \bibinfo {pages} {44001} (\bibinfo {year} {2023})}\BibitemShut {NoStop}%
\bibitem [{\citenamefont {Tkachenko}\ \emph {et~al.}(2012)\citenamefont {Tkachenko}, \citenamefont {Arkhipov},\ and\ \citenamefont {Askaruly}}]{tkachenko_book}%
  \BibitemOpen
  \bibfield  {author} {\bibinfo {author} {\bibfnamefont {Igor~M.}\ \bibnamefont {Tkachenko}}, \bibinfo {author} {\bibfnamefont {Yuriy~V.}\ \bibnamefont {Arkhipov}}, \ and\ \bibinfo {author} {\bibfnamefont {Adil}\ \bibnamefont {Askaruly}},\ }\href@noop {} {\emph {\bibinfo {title} {The Method of Moments and its Applications in Plasma Physics}}}\ (\bibinfo  {publisher} {Akademikerverlag, Saarbr\"ucken, Germany},\ \bibinfo {year} {2012})\BibitemShut {NoStop}%
\bibitem [{\citenamefont {Giuliani}\ and\ \citenamefont {Vignale}(2008)}]{quantum_theory}%
  \BibitemOpen
  \bibfield  {author} {\bibinfo {author} {\bibfnamefont {G.}~\bibnamefont {Giuliani}}\ and\ \bibinfo {author} {\bibfnamefont {G.}~\bibnamefont {Vignale}},\ }\href@noop {} {\emph {\bibinfo {title} {Quantum Theory of the Electron Liquid}}}\ (\bibinfo  {publisher} {Cambridge University Press},\ \bibinfo {address} {Cambridge},\ \bibinfo {year} {2008})\BibitemShut {NoStop}%
\bibitem [{\citenamefont {Ichimaru}(2018)}]{Ichimaru_BookII}%
  \BibitemOpen
  \bibfield  {author} {\bibinfo {author} {\bibfnamefont {S.}~\bibnamefont {Ichimaru}},\ }\href@noop {} {\emph {\bibinfo {title} {Statistical Plasma Physics Vol. II: Condensed Plasmas}}}\ (\bibinfo  {publisher} {CRC Press},\ \bibinfo {address} {Boca Raton, Florida},\ \bibinfo {year} {2018})\BibitemShut {NoStop}%
\bibitem [{\citenamefont {Singwi}\ and\ \citenamefont {Tosi}(1981)}]{SingwiTosi_Review}%
  \BibitemOpen
  \bibfield  {author} {\bibinfo {author} {\bibfnamefont {K.~S.}\ \bibnamefont {Singwi}}\ and\ \bibinfo {author} {\bibfnamefont {M.~P.}\ \bibnamefont {Tosi}},\ }\bibfield  {title} {\enquote {\bibinfo {title} {Correlations in electron liquids},}\ }\href {\doibase 10.1016/S0081-1947(08)60116-2} {\bibfield  {journal} {\bibinfo  {journal} {Solid State Physics}\ }\textbf {\bibinfo {volume} {36}},\ \bibinfo {pages} {177--266} (\bibinfo {year} {1981})}\BibitemShut {NoStop}%
\bibitem [{\citenamefont {Placzek}(1952)}]{Placzek_1952}%
  \BibitemOpen
  \bibfield  {author} {\bibinfo {author} {\bibfnamefont {G.}~\bibnamefont {Placzek}},\ }\bibfield  {title} {\enquote {\bibinfo {title} {The scattering of neutrons by systems of heavy nuclei},}\ }\href {\doibase 10.1103/PhysRev.86.377} {\bibfield  {journal} {\bibinfo  {journal} {Phys. Rev.}\ }\textbf {\bibinfo {volume} {86}},\ \bibinfo {pages} {377} (\bibinfo {year} {1952})}\BibitemShut {NoStop}%
\bibitem [{\citenamefont {Puff}(1965)}]{Puff_1965}%
  \BibitemOpen
  \bibfield  {author} {\bibinfo {author} {\bibfnamefont {R.~D.}\ \bibnamefont {Puff}},\ }\bibfield  {title} {\enquote {\bibinfo {title} {Application of sum rules to the low-temperature interacting boson system},}\ }\href {\doibase 10.1103/PhysRev.137.A406} {\bibfield  {journal} {\bibinfo  {journal} {Phys. Rev.}\ }\textbf {\bibinfo {volume} {137}},\ \bibinfo {pages} {A406} (\bibinfo {year} {1965})}\BibitemShut {NoStop}%
\bibitem [{\citenamefont {Dornheim}\ \emph {et~al.}(2023{\natexlab{c}})\citenamefont {Dornheim}, \citenamefont {Wicaksono}, \citenamefont {Suarez-Cardona}, \citenamefont {Tolias}, \citenamefont {B\"ohme}, \citenamefont {Moldabekov}, \citenamefont {Hecht},\ and\ \citenamefont {Vorberger}}]{Dornheim_PRB_2023}%
  \BibitemOpen
  \bibfield  {author} {\bibinfo {author} {\bibfnamefont {Tobias}\ \bibnamefont {Dornheim}}, \bibinfo {author} {\bibfnamefont {Damar~C.}\ \bibnamefont {Wicaksono}}, \bibinfo {author} {\bibfnamefont {Juan~E.}\ \bibnamefont {Suarez-Cardona}}, \bibinfo {author} {\bibfnamefont {Panagiotis}\ \bibnamefont {Tolias}}, \bibinfo {author} {\bibfnamefont {Maximilian~P.}\ \bibnamefont {B\"ohme}}, \bibinfo {author} {\bibfnamefont {Zhandos~A.}\ \bibnamefont {Moldabekov}}, \bibinfo {author} {\bibfnamefont {Michael}\ \bibnamefont {Hecht}}, \ and\ \bibinfo {author} {\bibfnamefont {Jan}\ \bibnamefont {Vorberger}},\ }\bibfield  {title} {\enquote {\bibinfo {title} {Extraction of the frequency moments of spectral densities from imaginary-time correlation function data},}\ }\href {\doibase 10.1103/PhysRevB.107.155148} {\bibfield  {journal} {\bibinfo  {journal} {Phys. Rev. B}\ }\textbf {\bibinfo {volume} {107}},\ \bibinfo {pages} {155148} (\bibinfo {year} {2023}{\natexlab{c}})}\BibitemShut {NoStop}%
\bibitem [{\citenamefont {Kugler}(1973)}]{kugler_classical}%
  \BibitemOpen
  \bibfield  {author} {\bibinfo {author} {\bibfnamefont {A.~A.}\ \bibnamefont {Kugler}},\ }\bibfield  {title} {\enquote {\bibinfo {title} {Collective modes, damping, and the scattering function in classical liquids},}\ }\href {https://link.springer.com/article/10.1007/BF01008535} {\bibfield  {journal} {\bibinfo  {journal} {J. Stat. Phys}\ }\textbf {\bibinfo {volume} {8}},\ \bibinfo {pages} {107} (\bibinfo {year} {1973})}\BibitemShut {NoStop}%
\bibitem [{\citenamefont {Kugler}(1970)}]{kugler_bounds}%
  \BibitemOpen
  \bibfield  {author} {\bibinfo {author} {\bibfnamefont {A.~A.}\ \bibnamefont {Kugler}},\ }\bibfield  {title} {\enquote {\bibinfo {title} {Bounds for some equilibrium properties of an electron gas},}\ }\href {https://journals.aps.org/pra/abstract/10.1103/PhysRevA.1.1688} {\bibfield  {journal} {\bibinfo  {journal} {Phys. Rev. A}\ }\textbf {\bibinfo {volume} {1}},\ \bibinfo {pages} {1688} (\bibinfo {year} {1970})}\BibitemShut {NoStop}%
\bibitem [{\citenamefont {Arkhipov}\ \emph {et~al.}(2017)\citenamefont {Arkhipov}, \citenamefont {Askaruly}, \citenamefont {Davletov}, \citenamefont {Dubovtsev}, \citenamefont {Donk\'o}, \citenamefont {Hartmann}, \citenamefont {Korolov}, \citenamefont {Conde},\ and\ \citenamefont {Tkachenko}}]{Arkhipov_2017}%
  \BibitemOpen
  \bibfield  {author} {\bibinfo {author} {\bibfnamefont {Yu.~V.}\ \bibnamefont {Arkhipov}}, \bibinfo {author} {\bibfnamefont {A.}~\bibnamefont {Askaruly}}, \bibinfo {author} {\bibfnamefont {A.~E.}\ \bibnamefont {Davletov}}, \bibinfo {author} {\bibfnamefont {D.~Yu.}\ \bibnamefont {Dubovtsev}}, \bibinfo {author} {\bibfnamefont {Z.}~\bibnamefont {Donk\'o}}, \bibinfo {author} {\bibfnamefont {P.}~\bibnamefont {Hartmann}}, \bibinfo {author} {\bibfnamefont {I.}~\bibnamefont {Korolov}}, \bibinfo {author} {\bibfnamefont {L.}~\bibnamefont {Conde}}, \ and\ \bibinfo {author} {\bibfnamefont {I.~M.}\ \bibnamefont {Tkachenko}},\ }\bibfield  {title} {\enquote {\bibinfo {title} {Direct determination of dynamic properties of {Coulomb and Yukawa} classical one-component plasmas},}\ }\href {\doibase 10.1103/PhysRevLett.119.045001} {\bibfield  {journal} {\bibinfo  {journal} {Phys. Rev. Lett.}\ }\textbf {\bibinfo {volume} {119}},\ \bibinfo {pages} {045001} (\bibinfo {year} {2017})}\BibitemShut {NoStop}%
\bibitem [{\citenamefont {Forster}\ \emph {et~al.}(1968)\citenamefont {Forster}, \citenamefont {Martin},\ and\ \citenamefont {Yip}}]{Forster_1968}%
  \BibitemOpen
  \bibfield  {author} {\bibinfo {author} {\bibfnamefont {Dieter}\ \bibnamefont {Forster}}, \bibinfo {author} {\bibfnamefont {Paul~C.}\ \bibnamefont {Martin}}, \ and\ \bibinfo {author} {\bibfnamefont {Sidney}\ \bibnamefont {Yip}},\ }\bibfield  {title} {\enquote {\bibinfo {title} {Moments of the momentum density correlation functions in simple liquids},}\ }\href {\doibase 10.1103/PhysRev.170.155} {\bibfield  {journal} {\bibinfo  {journal} {Phys. Rev.}\ }\textbf {\bibinfo {volume} {170}},\ \bibinfo {pages} {155--159} (\bibinfo {year} {1968})}\BibitemShut {NoStop}%
\bibitem [{\citenamefont {Bansal}\ and\ \citenamefont {Pathak}(1974)}]{Bansal_1974}%
  \BibitemOpen
  \bibfield  {author} {\bibinfo {author} {\bibfnamefont {Ravinder}\ \bibnamefont {Bansal}}\ and\ \bibinfo {author} {\bibfnamefont {K.~N.}\ \bibnamefont {Pathak}},\ }\bibfield  {title} {\enquote {\bibinfo {title} {Sum rules and atomic correlations in classical liquids},}\ }\href {\doibase 10.1103/PhysRevA.9.2773} {\bibfield  {journal} {\bibinfo  {journal} {Phys. Rev. A}\ }\textbf {\bibinfo {volume} {9}},\ \bibinfo {pages} {2773--2782} (\bibinfo {year} {1974})}\BibitemShut {NoStop}%
\bibitem [{\citenamefont {Ichimaru}\ \emph {et~al.}(1975)\citenamefont {Ichimaru}, \citenamefont {Totsuji}, \citenamefont {Tange},\ and\ \citenamefont {Pines}}]{Ichimaru_1975}%
  \BibitemOpen
  \bibfield  {author} {\bibinfo {author} {\bibfnamefont {Setsuo}\ \bibnamefont {Ichimaru}}, \bibinfo {author} {\bibfnamefont {Hiroo}\ \bibnamefont {Totsuji}}, \bibinfo {author} {\bibfnamefont {Toshio}\ \bibnamefont {Tange}}, \ and\ \bibinfo {author} {\bibfnamefont {David}\ \bibnamefont {Pines}},\ }\bibfield  {title} {\enquote {\bibinfo {title} {{Sum-Rule Analysis of Long-Wavelength Excitations in Electron Liquids}},}\ }\href {\doibase 10.1143/PTP.54.1077} {\bibfield  {journal} {\bibinfo  {journal} {Prog. Theor. Phys.}\ }\textbf {\bibinfo {volume} {54}},\ \bibinfo {pages} {1077--1092} (\bibinfo {year} {1975})}\BibitemShut {NoStop}%
\bibitem [{\citenamefont {Ailawadi}(1980)}]{Ailawadi_1980}%
  \BibitemOpen
  \bibfield  {author} {\bibinfo {author} {\bibfnamefont {N.K.}\ \bibnamefont {Ailawadi}},\ }\bibfield  {title} {\enquote {\bibinfo {title} {Equilibrium theories of simple liquids},}\ }\href {\doibase https://doi.org/10.1016/0370-1573(80)90063-0} {\bibfield  {journal} {\bibinfo  {journal} {Phys. Rep.}\ }\textbf {\bibinfo {volume} {57}},\ \bibinfo {pages} {241--306} (\bibinfo {year} {1980})}\BibitemShut {NoStop}%
\bibitem [{\citenamefont {Hansen}\ and\ \citenamefont {McDonald}(2013)}]{hansen2013theory}%
  \BibitemOpen
  \bibfield  {author} {\bibinfo {author} {\bibfnamefont {J.P.}\ \bibnamefont {Hansen}}\ and\ \bibinfo {author} {\bibfnamefont {I.R.}\ \bibnamefont {McDonald}},\ }\href {https://books.google.de/books?id=agbEswEACAAJ} {\emph {\bibinfo {title} {Theory of simple liquids : with applications to soft matter}}}\ (\bibinfo  {publisher} {Academic Press},\ \bibinfo {year} {2013})\BibitemShut {NoStop}%
\bibitem [{\citenamefont {Feenberg}(1969)}]{Feenberg_Book}%
  \BibitemOpen
  \bibfield  {author} {\bibinfo {author} {\bibfnamefont {E.}~\bibnamefont {Feenberg}},\ }\href@noop {} {\emph {\bibinfo {title} {Theory of Quantum Fluids}}}\ (\bibinfo  {publisher} {Academic Press},\ \bibinfo {address} {New York},\ \bibinfo {year} {1969})\BibitemShut {NoStop}%
\bibitem [{\citenamefont {Huang}\ and\ \citenamefont {Klein}(1964)}]{Huang_1964}%
  \BibitemOpen
  \bibfield  {author} {\bibinfo {author} {\bibfnamefont {K.}~\bibnamefont {Huang}}\ and\ \bibinfo {author} {\bibfnamefont {A.}~\bibnamefont {Klein}},\ }\bibfield  {title} {\enquote {\bibinfo {title} {Phonons in liquid helium},}\ }\href {\doibase 10.1016/0003-4916(64)90116-2} {\bibfield  {journal} {\bibinfo  {journal} {Ann. Phys.}\ }\textbf {\bibinfo {volume} {30}},\ \bibinfo {pages} {203} (\bibinfo {year} {1964})}\BibitemShut {NoStop}%
\bibitem [{\citenamefont {Hall}\ and\ \citenamefont {Feenberg}(1971)}]{Hall_1971}%
  \BibitemOpen
  \bibfield  {author} {\bibinfo {author} {\bibfnamefont {D.}~\bibnamefont {Hall}}\ and\ \bibinfo {author} {\bibfnamefont {E.}~\bibnamefont {Feenberg}},\ }\bibfield  {title} {\enquote {\bibinfo {title} {Sum rules, dynamic form factor, and elementary excitations in liquid {4He}},}\ }\href {\doibase 10.1016/B978-0-12-508201-3.50021-1} {\bibfield  {journal} {\bibinfo  {journal} {Ann. Phys.}\ }\textbf {\bibinfo {volume} {63}},\ \bibinfo {pages} {335} (\bibinfo {year} {1971})}\BibitemShut {NoStop}%
\bibitem [{\citenamefont {Dalfovo}\ and\ \citenamefont {Stringari}(1992)}]{Dalfovo_1992}%
  \BibitemOpen
  \bibfield  {author} {\bibinfo {author} {\bibfnamefont {F.}~\bibnamefont {Dalfovo}}\ and\ \bibinfo {author} {\bibfnamefont {S.}~\bibnamefont {Stringari}},\ }\bibfield  {title} {\enquote {\bibinfo {title} {Static response function for longitudinal and transverse excitations in superfiuid helium},}\ }\href {\doibase 10.1103/PhysRevB.46.13991} {\bibfield  {journal} {\bibinfo  {journal} {Phys. Rev. B}\ }\textbf {\bibinfo {volume} {46}},\ \bibinfo {pages} {13991} (\bibinfo {year} {1992})}\BibitemShut {NoStop}%
\bibitem [{\citenamefont {Stringari}(1992)}]{Stringari_1992}%
  \BibitemOpen
  \bibfield  {author} {\bibinfo {author} {\bibfnamefont {S.}~\bibnamefont {Stringari}},\ }\bibfield  {title} {\enquote {\bibinfo {title} {Sum rules for density and particle excitations in bose superfluids},}\ }\href {\doibase 10.1103/PhysRevB.46.2974} {\bibfield  {journal} {\bibinfo  {journal} {Phys. Rev. B}\ }\textbf {\bibinfo {volume} {46}},\ \bibinfo {pages} {2974} (\bibinfo {year} {1992})}\BibitemShut {NoStop}%
\bibitem [{\citenamefont {Boronat}\ \emph {et~al.}(1995)\citenamefont {Boronat}, \citenamefont {Casulleras}, \citenamefont {Dalfovo}, \citenamefont {Stringari},\ and\ \citenamefont {Moroni}}]{Boronat_1995}%
  \BibitemOpen
  \bibfield  {author} {\bibinfo {author} {\bibfnamefont {J.}~\bibnamefont {Boronat}}, \bibinfo {author} {\bibfnamefont {J.}~\bibnamefont {Casulleras}}, \bibinfo {author} {\bibfnamefont {F.}~\bibnamefont {Dalfovo}}, \bibinfo {author} {\bibfnamefont {S.}~\bibnamefont {Stringari}}, \ and\ \bibinfo {author} {\bibfnamefont {S.}~\bibnamefont {Moroni}},\ }\bibfield  {title} {\enquote {\bibinfo {title} {Bounds for the phonon-roton dispersion in superfluid {He}},}\ }\href {\doibase 10.1103/PhysRevB.52.1236} {\bibfield  {journal} {\bibinfo  {journal} {Phys. Rev. B}\ }\textbf {\bibinfo {volume} {52}},\ \bibinfo {pages} {1236} (\bibinfo {year} {1995})}\BibitemShut {NoStop}%
\bibitem [{\citenamefont {Feynman}\ and\ \citenamefont {Cohen}(1956)}]{Feynman_ansatz}%
  \BibitemOpen
  \bibfield  {author} {\bibinfo {author} {\bibfnamefont {R.~P.}\ \bibnamefont {Feynman}}\ and\ \bibinfo {author} {\bibfnamefont {Michael}\ \bibnamefont {Cohen}},\ }\bibfield  {title} {\enquote {\bibinfo {title} {Energy spectrum of the excitations in liquid helium},}\ }\href {\doibase 10.1103/PhysRev.102.1189} {\bibfield  {journal} {\bibinfo  {journal} {Phys. Rev.}\ }\textbf {\bibinfo {volume} {102}},\ \bibinfo {pages} {1189--1204} (\bibinfo {year} {1956})}\BibitemShut {NoStop}%
\bibitem [{\citenamefont {Silvestri}\ \emph {et~al.}(2019)\citenamefont {Silvestri}, \citenamefont {Kalman}, \citenamefont {Donk\'o}, \citenamefont {Hartmann}, \citenamefont {Rosenberg}, \citenamefont {Golden},\ and\ \citenamefont {Kyrkos}}]{Silvestri_2019}%
  \BibitemOpen
  \bibfield  {author} {\bibinfo {author} {\bibfnamefont {Luciano~G.}\ \bibnamefont {Silvestri}}, \bibinfo {author} {\bibfnamefont {Gabor~J.}\ \bibnamefont {Kalman}}, \bibinfo {author} {\bibfnamefont {Zolt\'an}\ \bibnamefont {Donk\'o}}, \bibinfo {author} {\bibfnamefont {Peter}\ \bibnamefont {Hartmann}}, \bibinfo {author} {\bibfnamefont {Marlene}\ \bibnamefont {Rosenberg}}, \bibinfo {author} {\bibfnamefont {Kenneth~I.}\ \bibnamefont {Golden}}, \ and\ \bibinfo {author} {\bibfnamefont {Stamatios}\ \bibnamefont {Kyrkos}},\ }\bibfield  {title} {\enquote {\bibinfo {title} {Sound speed in {Yukawa} one-component plasmas across coupling regimes},}\ }\href {\doibase 10.1103/PhysRevE.100.063206} {\bibfield  {journal} {\bibinfo  {journal} {Phys. Rev. E}\ }\textbf {\bibinfo {volume} {100}},\ \bibinfo {pages} {063206} (\bibinfo {year} {2019})}\BibitemShut {NoStop}%
\bibitem [{\citenamefont {Kalman}\ \emph {et~al.}(2010)\citenamefont {Kalman}, \citenamefont {Hartmann}, \citenamefont {Golden}, \citenamefont {Filinov},\ and\ \citenamefont {Donk{\'{o}}}}]{Kalman_2010}%
  \BibitemOpen
  \bibfield  {author} {\bibinfo {author} {\bibfnamefont {G.~J.}\ \bibnamefont {Kalman}}, \bibinfo {author} {\bibfnamefont {P.}~\bibnamefont {Hartmann}}, \bibinfo {author} {\bibfnamefont {K.~I.}\ \bibnamefont {Golden}}, \bibinfo {author} {\bibfnamefont {A.}~\bibnamefont {Filinov}}, \ and\ \bibinfo {author} {\bibfnamefont {Z.}~\bibnamefont {Donk{\'{o}}}},\ }\bibfield  {title} {\enquote {\bibinfo {title} {Correlational origin of the roton minimum},}\ }\href {\doibase 10.1209/0295-5075/90/55002} {\bibfield  {journal} {\bibinfo  {journal} {{EPL} (Europhysics Letters)}\ }\textbf {\bibinfo {volume} {90}},\ \bibinfo {pages} {55002} (\bibinfo {year} {2010})}\BibitemShut {NoStop}%
\bibitem [{\citenamefont {Arkhipov}\ \emph {et~al.}(2020)\citenamefont {Arkhipov}, \citenamefont {Ashikbayeva}, \citenamefont {Askaruly}, \citenamefont {Davletov}, \citenamefont {Dubovtsev}, \citenamefont {Santybayev}, \citenamefont {Syzganbayeva}, \citenamefont {Conde},\ and\ \citenamefont {Tkachenko}}]{Arkhipov_2020}%
  \BibitemOpen
  \bibfield  {author} {\bibinfo {author} {\bibfnamefont {Yu.~V.}\ \bibnamefont {Arkhipov}}, \bibinfo {author} {\bibfnamefont {A.}~\bibnamefont {Ashikbayeva}}, \bibinfo {author} {\bibfnamefont {A.}~\bibnamefont {Askaruly}}, \bibinfo {author} {\bibfnamefont {A.~E.}\ \bibnamefont {Davletov}}, \bibinfo {author} {\bibfnamefont {D.~Yu.}\ \bibnamefont {Dubovtsev}}, \bibinfo {author} {\bibfnamefont {Kh.~S.}\ \bibnamefont {Santybayev}}, \bibinfo {author} {\bibfnamefont {S.~A.}\ \bibnamefont {Syzganbayeva}}, \bibinfo {author} {\bibfnamefont {L.}~\bibnamefont {Conde}}, \ and\ \bibinfo {author} {\bibfnamefont {I.~M.}\ \bibnamefont {Tkachenko}},\ }\bibfield  {title} {\enquote {\bibinfo {title} {Dynamic characteristics of three-dimensional strongly coupled plasmas},}\ }\href {\doibase 10.1103/PhysRevE.102.053215} {\bibfield  {journal} {\bibinfo  {journal} {Phys. Rev. E}\ }\textbf {\bibinfo {volume} {102}},\ \bibinfo {pages} {053215} (\bibinfo {year} {2020})}\BibitemShut {NoStop}%
\bibitem [{\citenamefont {Filinov}\ \emph {et~al.}(2023{\natexlab{a}})\citenamefont {Filinov}, \citenamefont {Ara},\ and\ \citenamefont {Tkachenko}}]{Filinov_PRB_2023}%
  \BibitemOpen
  \bibfield  {author} {\bibinfo {author} {\bibfnamefont {A.~V.}\ \bibnamefont {Filinov}}, \bibinfo {author} {\bibfnamefont {J.}~\bibnamefont {Ara}}, \ and\ \bibinfo {author} {\bibfnamefont {I.~M.}\ \bibnamefont {Tkachenko}},\ }\bibfield  {title} {\enquote {\bibinfo {title} {Dynamical response in strongly coupled uniform electron liquids: Observation of plasmon-roton coexistence using nine sum rules, {Shannon} information entropy, and path-integral {Monte Carlo} simulations},}\ }\href {\doibase 10.1103/PhysRevB.107.195143} {\bibfield  {journal} {\bibinfo  {journal} {Phys. Rev. B}\ }\textbf {\bibinfo {volume} {107}},\ \bibinfo {pages} {195143} (\bibinfo {year} {2023}{\natexlab{a}})}\BibitemShut {NoStop}%
\bibitem [{\citenamefont {Filinov}\ \emph {et~al.}(2023{\natexlab{b}})\citenamefont {Filinov}, \citenamefont {Ara},\ and\ \citenamefont {Tkachenko}}]{Filinov_RSTA_2023}%
  \BibitemOpen
  \bibfield  {author} {\bibinfo {author} {\bibfnamefont {A.~V.}\ \bibnamefont {Filinov}}, \bibinfo {author} {\bibfnamefont {J.}~\bibnamefont {Ara}}, \ and\ \bibinfo {author} {\bibfnamefont {I.~M.}\ \bibnamefont {Tkachenko}},\ }\bibfield  {title} {\enquote {\bibinfo {title} {Dynamic properties and the roton mode attenuation in liquid {3He}: ab initio study within the self-consistent method of moments},}\ }\href {\doibase 10.1098/rsta.2022.0324} {\bibfield  {journal} {\bibinfo  {journal} {Phil. Trans. R. Soc. A}\ }\textbf {\bibinfo {volume} {381}},\ \bibinfo {pages} {20220324} (\bibinfo {year} {2023}{\natexlab{b}})}\BibitemShut {NoStop}%
\bibitem [{\citenamefont {Dornheim}\ \emph {et~al.}(2023{\natexlab{d}})\citenamefont {Dornheim}, \citenamefont {Moldabekov}, \citenamefont {Tolias}, \citenamefont {Böhme},\ and\ \citenamefont {Vorberger}}]{Dornheim_MRE_2023}%
  \BibitemOpen
  \bibfield  {author} {\bibinfo {author} {\bibfnamefont {Tobias}\ \bibnamefont {Dornheim}}, \bibinfo {author} {\bibfnamefont {Zhandos}\ \bibnamefont {Moldabekov}}, \bibinfo {author} {\bibfnamefont {Panagiotis}\ \bibnamefont {Tolias}}, \bibinfo {author} {\bibfnamefont {Maximilian}\ \bibnamefont {Böhme}}, \ and\ \bibinfo {author} {\bibfnamefont {Jan}\ \bibnamefont {Vorberger}},\ }\bibfield  {title} {\enquote {\bibinfo {title} {Physical insights from imaginary-time density--density correlation functions},}\ }\href {\doibase 10.1063/5.0149638} {\bibfield  {journal} {\bibinfo  {journal} {Matter Radiat. Extrem.}\ }\textbf {\bibinfo {volume} {8}},\ \bibinfo {pages} {056601} (\bibinfo {year} {2023}{\natexlab{d}})}\BibitemShut {NoStop}%
\bibitem [{\citenamefont {Thirumalai}\ and\ \citenamefont {Berne}(1983)}]{Berne_JCP_1983}%
  \BibitemOpen
  \bibfield  {author} {\bibinfo {author} {\bibfnamefont {Devarajan}\ \bibnamefont {Thirumalai}}\ and\ \bibinfo {author} {\bibfnamefont {Bruce~J.}\ \bibnamefont {Berne}},\ }\bibfield  {title} {\enquote {\bibinfo {title} {On the calculation of time correlation functions in quantum systems: Path integral techniques},}\ }\href {\doibase 10.1063/1.445597} {\bibfield  {journal} {\bibinfo  {journal} {J. Chem. Phys.}\ }\textbf {\bibinfo {volume} {79}},\ \bibinfo {pages} {5029--5033} (\bibinfo {year} {1983})}\BibitemShut {NoStop}%
\bibitem [{\citenamefont {Ceperley}(1995)}]{cep}%
  \BibitemOpen
  \bibfield  {author} {\bibinfo {author} {\bibfnamefont {D.~M.}\ \bibnamefont {Ceperley}},\ }\bibfield  {title} {\enquote {\bibinfo {title} {Path integrals in the theory of condensed helium},}\ }\href {https://journals.aps.org/rmp/abstract/10.1103/RevModPhys.67.279} {\bibfield  {journal} {\bibinfo  {journal} {Rev. Mod. Phys}\ }\textbf {\bibinfo {volume} {67}},\ \bibinfo {pages} {279} (\bibinfo {year} {1995})}\BibitemShut {NoStop}%
\bibitem [{\citenamefont {Dornheim}\ \emph {et~al.}(2022{\natexlab{b}})\citenamefont {Dornheim}, \citenamefont {B{\"o}hme}, \citenamefont {Kraus}, \citenamefont {D{\"o}ppner}, \citenamefont {Preston}, \citenamefont {Moldabekov},\ and\ \citenamefont {Vorberger}}]{Dornheim_T_2022}%
  \BibitemOpen
  \bibfield  {author} {\bibinfo {author} {\bibfnamefont {Tobias}\ \bibnamefont {Dornheim}}, \bibinfo {author} {\bibfnamefont {Maximilian}\ \bibnamefont {B{\"o}hme}}, \bibinfo {author} {\bibfnamefont {Dominik}\ \bibnamefont {Kraus}}, \bibinfo {author} {\bibfnamefont {Tilo}\ \bibnamefont {D{\"o}ppner}}, \bibinfo {author} {\bibfnamefont {Thomas~R.}\ \bibnamefont {Preston}}, \bibinfo {author} {\bibfnamefont {Zhandos~A.}\ \bibnamefont {Moldabekov}}, \ and\ \bibinfo {author} {\bibfnamefont {Jan}\ \bibnamefont {Vorberger}},\ }\bibfield  {title} {\enquote {\bibinfo {title} {Accurate temperature diagnostics for matter under extreme conditions},}\ }\href {\doibase 10.1038/s41467-022-35578-7} {\bibfield  {journal} {\bibinfo  {journal} {Nature Communications}\ }\textbf {\bibinfo {volume} {13}},\ \bibinfo {pages} {7911} (\bibinfo {year} {2022}{\natexlab{b}})}\BibitemShut {NoStop}%
\bibitem [{\citenamefont {Dornheim}\ \emph {et~al.}(2025{\natexlab{c}})\citenamefont {Dornheim}, \citenamefont {Döppner}, \citenamefont {Tolias}, \citenamefont {Böhme}, \citenamefont {Fletcher}, \citenamefont {Gawne}, \citenamefont {Graziani}, \citenamefont {Kraus}, \citenamefont {MacDonald}, \citenamefont {Moldabekov}, \citenamefont {Schwalbe}, \citenamefont {Gericke},\ and\ \citenamefont {Vorberger}}]{Dornheim_Nature_2025}%
  \BibitemOpen
  \bibfield  {author} {\bibinfo {author} {\bibfnamefont {Tobias}\ \bibnamefont {Dornheim}}, \bibinfo {author} {\bibfnamefont {Tilo}\ \bibnamefont {Döppner}}, \bibinfo {author} {\bibfnamefont {Panagiotis}\ \bibnamefont {Tolias}}, \bibinfo {author} {\bibfnamefont {Maximilian}\ \bibnamefont {Böhme}}, \bibinfo {author} {\bibfnamefont {Luke}\ \bibnamefont {Fletcher}}, \bibinfo {author} {\bibfnamefont {Thomas}\ \bibnamefont {Gawne}}, \bibinfo {author} {\bibfnamefont {Frank}\ \bibnamefont {Graziani}}, \bibinfo {author} {\bibfnamefont {Dominik}\ \bibnamefont {Kraus}}, \bibinfo {author} {\bibfnamefont {Michael}\ \bibnamefont {MacDonald}}, \bibinfo {author} {\bibfnamefont {Zhandos}\ \bibnamefont {Moldabekov}}, \bibinfo {author} {\bibfnamefont {Sebastian}\ \bibnamefont {Schwalbe}}, \bibinfo {author} {\bibfnamefont {Dirk}\ \bibnamefont {Gericke}}, \ and\ \bibinfo {author} {\bibfnamefont {Jan}\ \bibnamefont {Vorberger}},\ }\bibfield  {title} {\enquote {\bibinfo {title} {Unraveling electronic correlations in warm dense
  quantum plasmas},}\ }\href {\doibase 10.1038/s41467-025-60278-3} {\bibfield  {journal} {\bibinfo  {journal} {Nat. Commun.}\ }\textbf {\bibinfo {volume} {16}},\ \bibinfo {pages} {5103} (\bibinfo {year} {2025}{\natexlab{c}})}\BibitemShut {NoStop}%
\bibitem [{\citenamefont {Dornheim}\ \emph {et~al.}(2025{\natexlab{d}})\citenamefont {Dornheim}, \citenamefont {Bellenbaum}, \citenamefont {Bethkenhagen}, \citenamefont {Hansen}, \citenamefont {Böhme}, \citenamefont {Döppner}, \citenamefont {Fletcher}, \citenamefont {Gawne}, \citenamefont {Gericke}, \citenamefont {Hamel}, \citenamefont {Kraus}, \citenamefont {MacDonald}, \citenamefont {Moldabekov}, \citenamefont {Preston}, \citenamefont {Redmer}, \citenamefont {Schörner}, \citenamefont {Schwalbe}, \citenamefont {Tolias},\ and\ \citenamefont {Vorberger}}]{dornheim_POP_2025}%
  \BibitemOpen
  \bibfield  {author} {\bibinfo {author} {\bibfnamefont {Tobias}\ \bibnamefont {Dornheim}}, \bibinfo {author} {\bibfnamefont {Hannah~M.}\ \bibnamefont {Bellenbaum}}, \bibinfo {author} {\bibfnamefont {Mandy}\ \bibnamefont {Bethkenhagen}}, \bibinfo {author} {\bibfnamefont {Stephanie~B.}\ \bibnamefont {Hansen}}, \bibinfo {author} {\bibfnamefont {Maximilian~P.}\ \bibnamefont {Böhme}}, \bibinfo {author} {\bibfnamefont {Tilo}\ \bibnamefont {Döppner}}, \bibinfo {author} {\bibfnamefont {Luke~B.}\ \bibnamefont {Fletcher}}, \bibinfo {author} {\bibfnamefont {Thomas}\ \bibnamefont {Gawne}}, \bibinfo {author} {\bibfnamefont {Dirk~O.}\ \bibnamefont {Gericke}}, \bibinfo {author} {\bibfnamefont {Sebastien}\ \bibnamefont {Hamel}}, \bibinfo {author} {\bibfnamefont {Dominik}\ \bibnamefont {Kraus}}, \bibinfo {author} {\bibfnamefont {Michael~J.}\ \bibnamefont {MacDonald}}, \bibinfo {author} {\bibfnamefont {Zhandos~A.}\ \bibnamefont {Moldabekov}}, \bibinfo {author} {\bibfnamefont {Thomas~R.}\ \bibnamefont {Preston}}, \bibinfo
  {author} {\bibfnamefont {Ronald}\ \bibnamefont {Redmer}}, \bibinfo {author} {\bibfnamefont {Maximilian}\ \bibnamefont {Schörner}}, \bibinfo {author} {\bibfnamefont {Sebastian}\ \bibnamefont {Schwalbe}}, \bibinfo {author} {\bibfnamefont {Panagiotis}\ \bibnamefont {Tolias}}, \ and\ \bibinfo {author} {\bibfnamefont {Jan}\ \bibnamefont {Vorberger}},\ }\bibfield  {title} {\enquote {\bibinfo {title} {Model-free {Rayleigh weight from x-ray Thomson scattering measurements}},}\ }\href {\doibase 10.1063/5.0238630} {\bibfield  {journal} {\bibinfo  {journal} {Phys. Plasmas}\ }\textbf {\bibinfo {volume} {32}},\ \bibinfo {pages} {052712} (\bibinfo {year} {2025}{\natexlab{d}})}\BibitemShut {NoStop}%
\bibitem [{\citenamefont {Schwalbe}\ \emph {et~al.}(2025)\citenamefont {Schwalbe}, \citenamefont {Bellenbaum}, \citenamefont {Döppner}, \citenamefont {Böhme}, \citenamefont {Gawne}, \citenamefont {Kraus}, \citenamefont {MacDonald}, \citenamefont {Moldabekov}, \citenamefont {Tolias}, \citenamefont {Vorberger},\ and\ \citenamefont {Dornheim}}]{schwalbe2025staticlineardensityresponse}%
  \BibitemOpen
  \bibfield  {author} {\bibinfo {author} {\bibfnamefont {Sebastian}\ \bibnamefont {Schwalbe}}, \bibinfo {author} {\bibfnamefont {Hannah}\ \bibnamefont {Bellenbaum}}, \bibinfo {author} {\bibfnamefont {Tilo}\ \bibnamefont {Döppner}}, \bibinfo {author} {\bibfnamefont {Maximilian}\ \bibnamefont {Böhme}}, \bibinfo {author} {\bibfnamefont {Thomas}\ \bibnamefont {Gawne}}, \bibinfo {author} {\bibfnamefont {Dominik}\ \bibnamefont {Kraus}}, \bibinfo {author} {\bibfnamefont {Michael~J.}\ \bibnamefont {MacDonald}}, \bibinfo {author} {\bibfnamefont {Zhandos}\ \bibnamefont {Moldabekov}}, \bibinfo {author} {\bibfnamefont {Panagiotis}\ \bibnamefont {Tolias}}, \bibinfo {author} {\bibfnamefont {Jan}\ \bibnamefont {Vorberger}}, \ and\ \bibinfo {author} {\bibfnamefont {Tobias}\ \bibnamefont {Dornheim}},\ }\href {https://arxiv.org/abs/2504.13611} {\enquote {\bibinfo {title} {Static linear density response from {X-ray Thomson} scattering measurements: a case study of warm dense beryllium},}\ } (\bibinfo {year} {2025}),\ \Eprint
  {http://arxiv.org/abs/2504.13611} {arXiv:2504.13611 [physics.plasm-ph]} \BibitemShut {NoStop}%
\bibitem [{\citenamefont {Epstein}\ and\ \citenamefont {Schotland}(2008)}]{epstein2008badtruth}%
  \BibitemOpen
  \bibfield  {author} {\bibinfo {author} {\bibfnamefont {Charles~L}\ \bibnamefont {Epstein}}\ and\ \bibinfo {author} {\bibfnamefont {John}\ \bibnamefont {Schotland}},\ }\bibfield  {title} {\enquote {\bibinfo {title} {The bad truth about {Laplace's transform}},}\ }\href {\doibase 10.1137/060657273} {\bibfield  {journal} {\bibinfo  {journal} {SIAM Rev.}\ }\textbf {\bibinfo {volume} {50}},\ \bibinfo {pages} {504--520} (\bibinfo {year} {2008})}\BibitemShut {NoStop}%
\bibitem [{\citenamefont {Jarrell}\ and\ \citenamefont {Gubernatis}(1996)}]{JARRELL1996133}%
  \BibitemOpen
  \bibfield  {author} {\bibinfo {author} {\bibfnamefont {Mark}\ \bibnamefont {Jarrell}}\ and\ \bibinfo {author} {\bibfnamefont {J.E.}\ \bibnamefont {Gubernatis}},\ }\bibfield  {title} {\enquote {\bibinfo {title} {Bayesian inference and the analytic continuation of imaginary-time quantum {Monte Carlo} data},}\ }\href {\doibase https://doi.org/10.1016/0370-1573(95)00074-7} {\bibfield  {journal} {\bibinfo  {journal} {Phys. Rep.}\ }\textbf {\bibinfo {volume} {269}},\ \bibinfo {pages} {133--195} (\bibinfo {year} {1996})}\BibitemShut {NoStop}%
\bibitem [{\citenamefont {Chuna}\ \emph {et~al.}(2025)\citenamefont {Chuna}, \citenamefont {Barnfield}, \citenamefont {Dornheim}, \citenamefont {Friedlander},\ and\ \citenamefont {Hoheisel}}]{chuna2025dualformulationmaximumentropy}%
  \BibitemOpen
  \bibfield  {author} {\bibinfo {author} {\bibfnamefont {Thomas}\ \bibnamefont {Chuna}}, \bibinfo {author} {\bibfnamefont {Nicholas}\ \bibnamefont {Barnfield}}, \bibinfo {author} {\bibfnamefont {Tobias}\ \bibnamefont {Dornheim}}, \bibinfo {author} {\bibfnamefont {Michael~P.}\ \bibnamefont {Friedlander}}, \ and\ \bibinfo {author} {\bibfnamefont {Tim}\ \bibnamefont {Hoheisel}},\ }\href {https://arxiv.org/abs/2501.01869} {\enquote {\bibinfo {title} {Dual formulation of the maximum entropy method applied to analytic continuation of quantum {Monte Carlo} data},}\ } (\bibinfo {year} {2025}),\ \Eprint {http://arxiv.org/abs/2501.01869} {arXiv:2501.01869 [physics.comp-ph]} \BibitemShut {NoStop}%
\bibitem [{\citenamefont {Gradshteyn}\ and\ \citenamefont {Ryzhik}(2007)}]{GradshteynRyzhik}%
  \BibitemOpen
  \bibfield  {author} {\bibinfo {author} {\bibfnamefont {I.~S.}\ \bibnamefont {Gradshteyn}}\ and\ \bibinfo {author} {\bibfnamefont {I.~M.}\ \bibnamefont {Ryzhik}},\ }\href@noop {} {\emph {\bibinfo {title} {Table of Integrals, Series, and Products}}}\ (\bibinfo  {publisher} {Academic Press},\ \bibinfo {address} {New York},\ \bibinfo {year} {2007})\BibitemShut {NoStop}%
\bibitem [{\citenamefont {Tolias}\ and\ \citenamefont {Lucco~Castello}(2021)}]{Tolias_PoPbrief_2021}%
  \BibitemOpen
  \bibfield  {author} {\bibinfo {author} {\bibfnamefont {P.}~\bibnamefont {Tolias}}\ and\ \bibinfo {author} {\bibfnamefont {F.}~\bibnamefont {Lucco~Castello}},\ }\bibfield  {title} {\enquote {\bibinfo {title} {Description of longitudinal modes in moderately coupled {Yukawa} systems with the static local field correction},}\ }\href {\doibase 10.1063/5.0044871} {\bibfield  {journal} {\bibinfo  {journal} {Phys. Plasmas}\ }\textbf {\bibinfo {volume} {28}},\ \bibinfo {pages} {034502} (\bibinfo {year} {2021})}\BibitemShut {NoStop}%
\bibitem [{\citenamefont {Tolias}\ \emph {et~al.}(2025)\citenamefont {Tolias}, \citenamefont {Dornheim},\ and\ \citenamefont {Vorberger}}]{Tolias_CtPP2025}%
  \BibitemOpen
  \bibfield  {author} {\bibinfo {author} {\bibfnamefont {Panagiotis}\ \bibnamefont {Tolias}}, \bibinfo {author} {\bibfnamefont {Tobias}\ \bibnamefont {Dornheim}}, \ and\ \bibinfo {author} {\bibfnamefont {Jan}\ \bibnamefont {Vorberger}},\ }\bibfield  {title} {\enquote {\bibinfo {title} {On the density–density correlations of the non-interacting finite temperature electron gas},}\ }\href {\doibase https://doi.org/10.1002/ctpp.202400135} {\bibfield  {journal} {\bibinfo  {journal} {Contrib. Plasma Phys.}\ ,\ \bibinfo {pages} {e202400135}} (\bibinfo {year} {2025})}\BibitemShut {NoStop}%
\bibitem [{\citenamefont {Dornheim}\ \emph {et~al.}(2021{\natexlab{c}})\citenamefont {Dornheim}, \citenamefont {B\"ohme}, \citenamefont {Militzer},\ and\ \citenamefont {Vorberger}}]{Dornheim_PRB_nk_2021}%
  \BibitemOpen
  \bibfield  {author} {\bibinfo {author} {\bibfnamefont {Tobias}\ \bibnamefont {Dornheim}}, \bibinfo {author} {\bibfnamefont {Maximilian}\ \bibnamefont {B\"ohme}}, \bibinfo {author} {\bibfnamefont {Burkhard}\ \bibnamefont {Militzer}}, \ and\ \bibinfo {author} {\bibfnamefont {Jan}\ \bibnamefont {Vorberger}},\ }\bibfield  {title} {\enquote {\bibinfo {title} {Ab initio path integral monte carlo approach to the momentum distribution of the uniform electron gas at finite temperature without fixed nodes},}\ }\href {\doibase 10.1103/PhysRevB.103.205142} {\bibfield  {journal} {\bibinfo  {journal} {Phys. Rev. B}\ }\textbf {\bibinfo {volume} {103}},\ \bibinfo {pages} {205142} (\bibinfo {year} {2021}{\natexlab{c}})}\BibitemShut {NoStop}%
\bibitem [{\citenamefont {Boninsegni}\ \emph {et~al.}(2006{\natexlab{a}})\citenamefont {Boninsegni}, \citenamefont {Prokofev},\ and\ \citenamefont {Svistunov}}]{boninsegni1}%
  \BibitemOpen
  \bibfield  {author} {\bibinfo {author} {\bibfnamefont {M.}~\bibnamefont {Boninsegni}}, \bibinfo {author} {\bibfnamefont {N.~V.}\ \bibnamefont {Prokofev}}, \ and\ \bibinfo {author} {\bibfnamefont {B.~V.}\ \bibnamefont {Svistunov}},\ }\bibfield  {title} {\enquote {\bibinfo {title} {Worm algorithm and diagrammatic {M}onte {C}arlo: A new approach to continuous-space path integral {M}onte {C}arlo simulations},}\ }\href {https://journals.aps.org/pre/abstract/10.1103/PhysRevE.74.036701} {\bibfield  {journal} {\bibinfo  {journal} {Phys. Rev. E}\ }\textbf {\bibinfo {volume} {74}},\ \bibinfo {pages} {036701} (\bibinfo {year} {2006}{\natexlab{a}})}\BibitemShut {NoStop}%
\bibitem [{\citenamefont {Boninsegni}\ \emph {et~al.}(2006{\natexlab{b}})\citenamefont {Boninsegni}, \citenamefont {Prokofev},\ and\ \citenamefont {Svistunov}}]{boninsegni2}%
  \BibitemOpen
  \bibfield  {author} {\bibinfo {author} {\bibfnamefont {M.}~\bibnamefont {Boninsegni}}, \bibinfo {author} {\bibfnamefont {N.~V.}\ \bibnamefont {Prokofev}}, \ and\ \bibinfo {author} {\bibfnamefont {B.~V.}\ \bibnamefont {Svistunov}},\ }\bibfield  {title} {\enquote {\bibinfo {title} {Worm algorithm for continuous-space path integral {M}onte {C}arlo simulations},}\ }\href {https://journals.aps.org/prl/abstract/10.1103/PhysRevLett.96.070601} {\bibfield  {journal} {\bibinfo  {journal} {Phys. Rev. Lett}\ }\textbf {\bibinfo {volume} {96}},\ \bibinfo {pages} {070601} (\bibinfo {year} {2006}{\natexlab{b}})}\BibitemShut {NoStop}%
\bibitem [{\citenamefont {Dornheim}\ \emph {et~al.}(2024{\natexlab{d}})\citenamefont {Dornheim}, \citenamefont {Böhme},\ and\ \citenamefont {Schwalbe}}]{ISHTAR}%
  \BibitemOpen
  \bibfield  {author} {\bibinfo {author} {\bibfnamefont {Tobias}\ \bibnamefont {Dornheim}}, \bibinfo {author} {\bibfnamefont {Maximilian}\ \bibnamefont {Böhme}}, \ and\ \bibinfo {author} {\bibfnamefont {Sebastian}\ \bibnamefont {Schwalbe}},\ }\href {\doibase 10.5281/zenodo.10497098} {\enquote {\bibinfo {title} {{ISHTAR - Imaginary-time Stochastic High- performance Tool for Ab initio Research}},}\ } (\bibinfo {year} {2024}{\natexlab{d}})\BibitemShut {NoStop}%
\bibitem [{rep()}]{repo}%
  \BibitemOpen
  \href@noop {} {}\bibinfo {note} {A link to a repository containing all PIMC results will be made available upon publication.}\BibitemShut {Stop}%
\bibitem [{\citenamefont {Chiesa}\ \emph {et~al.}(2006)\citenamefont {Chiesa}, \citenamefont {Ceperley}, \citenamefont {Martin},\ and\ \citenamefont {Holzmann}}]{Chiesa_PRL_2006}%
  \BibitemOpen
  \bibfield  {author} {\bibinfo {author} {\bibfnamefont {Simone}\ \bibnamefont {Chiesa}}, \bibinfo {author} {\bibfnamefont {David~M.}\ \bibnamefont {Ceperley}}, \bibinfo {author} {\bibfnamefont {Richard~M.}\ \bibnamefont {Martin}}, \ and\ \bibinfo {author} {\bibfnamefont {Markus}\ \bibnamefont {Holzmann}},\ }\bibfield  {title} {\enquote {\bibinfo {title} {Finite-size error in many-body simulations with long-range interactions},}\ }\href {\doibase 10.1103/PhysRevLett.97.076404} {\bibfield  {journal} {\bibinfo  {journal} {Phys. Rev. Lett.}\ }\textbf {\bibinfo {volume} {97}},\ \bibinfo {pages} {076404} (\bibinfo {year} {2006})}\BibitemShut {NoStop}%
\bibitem [{\citenamefont {Drummond}\ \emph {et~al.}(2008)\citenamefont {Drummond}, \citenamefont {Needs}, \citenamefont {Sorouri},\ and\ \citenamefont {Foulkes}}]{Drummond_PRB_2008}%
  \BibitemOpen
  \bibfield  {author} {\bibinfo {author} {\bibfnamefont {N.~D.}\ \bibnamefont {Drummond}}, \bibinfo {author} {\bibfnamefont {R.~J.}\ \bibnamefont {Needs}}, \bibinfo {author} {\bibfnamefont {A.}~\bibnamefont {Sorouri}}, \ and\ \bibinfo {author} {\bibfnamefont {W.~M.~C.}\ \bibnamefont {Foulkes}},\ }\bibfield  {title} {\enquote {\bibinfo {title} {Finite-size errors in continuum quantum monte carlo calculations},}\ }\href {\doibase 10.1103/PhysRevB.78.125106} {\bibfield  {journal} {\bibinfo  {journal} {Phys. Rev. B}\ }\textbf {\bibinfo {volume} {78}},\ \bibinfo {pages} {125106} (\bibinfo {year} {2008})}\BibitemShut {NoStop}%
\bibitem [{\citenamefont {Dornheim}\ and\ \citenamefont {Vorberger}(2021)}]{Dornheim_JCP_2021}%
  \BibitemOpen
  \bibfield  {author} {\bibinfo {author} {\bibfnamefont {Tobias}\ \bibnamefont {Dornheim}}\ and\ \bibinfo {author} {\bibfnamefont {Jan}\ \bibnamefont {Vorberger}},\ }\bibfield  {title} {\enquote {\bibinfo {title} {Overcoming finite-size effects in electronic structure simulations at extreme conditions},}\ }\href {\doibase 10.1063/5.0045634} {\bibfield  {journal} {\bibinfo  {journal} {J. Chem. Phys.}\ }\textbf {\bibinfo {volume} {154}},\ \bibinfo {pages} {144103} (\bibinfo {year} {2021})}\BibitemShut {NoStop}%
\bibitem [{\citenamefont {Holzmann}\ \emph {et~al.}(2016)\citenamefont {Holzmann}, \citenamefont {Clay}, \citenamefont {Morales}, \citenamefont {Tubman}, \citenamefont {Ceperley},\ and\ \citenamefont {Pierleoni}}]{Holzmann_PRB_2016}%
  \BibitemOpen
  \bibfield  {author} {\bibinfo {author} {\bibfnamefont {Markus}\ \bibnamefont {Holzmann}}, \bibinfo {author} {\bibfnamefont {Raymond~C.}\ \bibnamefont {Clay}}, \bibinfo {author} {\bibfnamefont {Miguel~A.}\ \bibnamefont {Morales}}, \bibinfo {author} {\bibfnamefont {Norm~M.}\ \bibnamefont {Tubman}}, \bibinfo {author} {\bibfnamefont {David~M.}\ \bibnamefont {Ceperley}}, \ and\ \bibinfo {author} {\bibfnamefont {Carlo}\ \bibnamefont {Pierleoni}},\ }\bibfield  {title} {\enquote {\bibinfo {title} {Theory of finite size effects for electronic quantum monte carlo calculations of liquids and solids},}\ }\href {\doibase 10.1103/PhysRevB.94.035126} {\bibfield  {journal} {\bibinfo  {journal} {Phys. Rev. B}\ }\textbf {\bibinfo {volume} {94}},\ \bibinfo {pages} {035126} (\bibinfo {year} {2016})}\BibitemShut {NoStop}%
\bibitem [{\citenamefont {Dornheim}\ and\ \citenamefont {Vorberger}(2020)}]{Dornheim_PRE_2020}%
  \BibitemOpen
  \bibfield  {author} {\bibinfo {author} {\bibfnamefont {Tobias}\ \bibnamefont {Dornheim}}\ and\ \bibinfo {author} {\bibfnamefont {Jan}\ \bibnamefont {Vorberger}},\ }\bibfield  {title} {\enquote {\bibinfo {title} {Finite-size effects in the reconstruction of dynamic properties from ab initio path integral monte carlo simulations},}\ }\href {\doibase 10.1103/PhysRevE.102.063301} {\bibfield  {journal} {\bibinfo  {journal} {Phys. Rev. E}\ }\textbf {\bibinfo {volume} {102}},\ \bibinfo {pages} {063301} (\bibinfo {year} {2020})}\BibitemShut {NoStop}%
\bibitem [{\citenamefont {Dornheim}\ \emph {et~al.}(2020{\natexlab{c}})\citenamefont {Dornheim}, \citenamefont {Sjostrom}, \citenamefont {Tanaka},\ and\ \citenamefont {Vorberger}}]{dornheim_electron_liquid}%
  \BibitemOpen
  \bibfield  {author} {\bibinfo {author} {\bibfnamefont {Tobias}\ \bibnamefont {Dornheim}}, \bibinfo {author} {\bibfnamefont {Travis}\ \bibnamefont {Sjostrom}}, \bibinfo {author} {\bibfnamefont {Shigenori}\ \bibnamefont {Tanaka}}, \ and\ \bibinfo {author} {\bibfnamefont {Jan}\ \bibnamefont {Vorberger}},\ }\bibfield  {title} {\enquote {\bibinfo {title} {Strongly coupled electron liquid: Ab initio path integral monte carlo simulations and dielectric theories},}\ }\href {\doibase 10.1103/PhysRevB.101.045129} {\bibfield  {journal} {\bibinfo  {journal} {Phys. Rev. B}\ }\textbf {\bibinfo {volume} {101}},\ \bibinfo {pages} {045129} (\bibinfo {year} {2020}{\natexlab{c}})}\BibitemShut {NoStop}%
\bibitem [{\citenamefont {Takada}(2016)}]{Takada_PRB_2016}%
  \BibitemOpen
  \bibfield  {author} {\bibinfo {author} {\bibfnamefont {Yasutami}\ \bibnamefont {Takada}},\ }\bibfield  {title} {\enquote {\bibinfo {title} {Emergence of an excitonic collective mode in the dilute electron gas},}\ }\href {\doibase 10.1103/PhysRevB.94.245106} {\bibfield  {journal} {\bibinfo  {journal} {Phys. Rev. B}\ }\textbf {\bibinfo {volume} {94}},\ \bibinfo {pages} {245106} (\bibinfo {year} {2016})}\BibitemShut {NoStop}%
\bibitem [{\citenamefont {Bowen}\ \emph {et~al.}(1994)\citenamefont {Bowen}, \citenamefont {Sugiyama},\ and\ \citenamefont {Alder}}]{bowen2}%
  \BibitemOpen
  \bibfield  {author} {\bibinfo {author} {\bibfnamefont {C.}~\bibnamefont {Bowen}}, \bibinfo {author} {\bibfnamefont {G.}~\bibnamefont {Sugiyama}}, \ and\ \bibinfo {author} {\bibfnamefont {B.~J.}\ \bibnamefont {Alder}},\ }\bibfield  {title} {\enquote {\bibinfo {title} {Static dielectric response of the electron gas},}\ }\href {http://link.aps.org/doi/10.1103/PhysRevB.50.14838} {\bibfield  {journal} {\bibinfo  {journal} {Phys. Rev. B}\ }\textbf {\bibinfo {volume} {50}},\ \bibinfo {pages} {14838} (\bibinfo {year} {1994})}\BibitemShut {NoStop}%
\bibitem [{\citenamefont {Anderson}(2007)}]{anderson2007quantum}%
  \BibitemOpen
  \bibfield  {author} {\bibinfo {author} {\bibfnamefont {J.B.}\ \bibnamefont {Anderson}},\ }\href {https://books.google.de/books?id=\_QUSDAAAQBAJ} {\emph {\bibinfo {title} {Quantum Monte Carlo: Origins, Development, Applications}}}\ (\bibinfo  {publisher} {Oxford University Press, USA},\ \bibinfo {year} {2007})\BibitemShut {NoStop}%
\bibitem [{\citenamefont {Ceperley}(1991)}]{Ceperley1991}%
  \BibitemOpen
  \bibfield  {author} {\bibinfo {author} {\bibfnamefont {D.~M.}\ \bibnamefont {Ceperley}},\ }\bibfield  {title} {\enquote {\bibinfo {title} {Fermion nodes},}\ }\href {\doibase 10.1007/BF01030009} {\bibfield  {journal} {\bibinfo  {journal} {Journal of Statistical Physics}\ }\textbf {\bibinfo {volume} {63}},\ \bibinfo {pages} {1237--1267} (\bibinfo {year} {1991})}\BibitemShut {NoStop}%
\bibitem [{\citenamefont {Moroni}\ \emph {et~al.}(1992)\citenamefont {Moroni}, \citenamefont {Ceperley},\ and\ \citenamefont {Senatore}}]{moroni}%
  \BibitemOpen
  \bibfield  {author} {\bibinfo {author} {\bibfnamefont {S.}~\bibnamefont {Moroni}}, \bibinfo {author} {\bibfnamefont {D.~M.}\ \bibnamefont {Ceperley}}, \ and\ \bibinfo {author} {\bibfnamefont {G.}~\bibnamefont {Senatore}},\ }\bibfield  {title} {\enquote {\bibinfo {title} {Static response from quantum {M}onte {C}arlo calculations},}\ }\href {https://journals.aps.org/prl/abstract/10.1103/PhysRevLett.69.1837} {\bibfield  {journal} {\bibinfo  {journal} {Phys. Rev. Lett}\ }\textbf {\bibinfo {volume} {69}},\ \bibinfo {pages} {1837} (\bibinfo {year} {1992})}\BibitemShut {NoStop}%
\bibitem [{\citenamefont {Moroni}\ \emph {et~al.}(1995)\citenamefont {Moroni}, \citenamefont {Ceperley},\ and\ \citenamefont {Senatore}}]{moroni2}%
  \BibitemOpen
  \bibfield  {author} {\bibinfo {author} {\bibfnamefont {S.}~\bibnamefont {Moroni}}, \bibinfo {author} {\bibfnamefont {D.~M.}\ \bibnamefont {Ceperley}}, \ and\ \bibinfo {author} {\bibfnamefont {G.}~\bibnamefont {Senatore}},\ }\bibfield  {title} {\enquote {\bibinfo {title} {Static response and local field factor of the electron gas},}\ }\href {http://link.aps.org/doi/10.1103/PhysRevLett.75.689} {\bibfield  {journal} {\bibinfo  {journal} {Phys. Rev. Lett}\ }\textbf {\bibinfo {volume} {75}},\ \bibinfo {pages} {689} (\bibinfo {year} {1995})}\BibitemShut {NoStop}%
\bibitem [{\citenamefont {Dornheim}\ \emph {et~al.}(2017{\natexlab{b}})\citenamefont {Dornheim}, \citenamefont {Groth}, \citenamefont {Vorberger},\ and\ \citenamefont {Bonitz}}]{dornheim_pre}%
  \BibitemOpen
  \bibfield  {author} {\bibinfo {author} {\bibfnamefont {T.}~\bibnamefont {Dornheim}}, \bibinfo {author} {\bibfnamefont {S.}~\bibnamefont {Groth}}, \bibinfo {author} {\bibfnamefont {J.}~\bibnamefont {Vorberger}}, \ and\ \bibinfo {author} {\bibfnamefont {M.}~\bibnamefont {Bonitz}},\ }\bibfield  {title} {\enquote {\bibinfo {title} {Permutation blocking path integral {M}onte {C}arlo approach to the static density response of the warm dense electron gas},}\ }\href {https://journals.aps.org/pre/abstract/10.1103/PhysRevE.96.023203} {\bibfield  {journal} {\bibinfo  {journal} {Phys. Rev. E}\ }\textbf {\bibinfo {volume} {96}},\ \bibinfo {pages} {023203} (\bibinfo {year} {2017}{\natexlab{b}})}\BibitemShut {NoStop}%
\bibitem [{\citenamefont {Groth}\ \emph {et~al.}(2017)\citenamefont {Groth}, \citenamefont {Dornheim},\ and\ \citenamefont {Bonitz}}]{groth_jcp}%
  \BibitemOpen
  \bibfield  {author} {\bibinfo {author} {\bibfnamefont {S.}~\bibnamefont {Groth}}, \bibinfo {author} {\bibfnamefont {T.}~\bibnamefont {Dornheim}}, \ and\ \bibinfo {author} {\bibfnamefont {M.}~\bibnamefont {Bonitz}},\ }\bibfield  {title} {\enquote {\bibinfo {title} {Configuration path integral {M}onte {C}arlo approach to the static density response of the warm dense electron gas},}\ }\href {https://aip.scitation.org/doi/abs/10.1063/1.4999907} {\bibfield  {journal} {\bibinfo  {journal} {J. Chem. Phys}\ }\textbf {\bibinfo {volume} {147}},\ \bibinfo {pages} {164108} (\bibinfo {year} {2017})}\BibitemShut {NoStop}%
\bibitem [{\citenamefont {Moldabekov}\ \emph {et~al.}(2023)\citenamefont {Moldabekov}, \citenamefont {B{\"o}hme}, \citenamefont {Vorberger}, \citenamefont {Blaschke},\ and\ \citenamefont {Dornheim}}]{Moldabekov_JCTC_2023}%
  \BibitemOpen
  \bibfield  {author} {\bibinfo {author} {\bibfnamefont {Zhandos}\ \bibnamefont {Moldabekov}}, \bibinfo {author} {\bibfnamefont {Maximilian}\ \bibnamefont {B{\"o}hme}}, \bibinfo {author} {\bibfnamefont {Jan}\ \bibnamefont {Vorberger}}, \bibinfo {author} {\bibfnamefont {David}\ \bibnamefont {Blaschke}}, \ and\ \bibinfo {author} {\bibfnamefont {Tobias}\ \bibnamefont {Dornheim}},\ }\bibfield  {title} {\enquote {\bibinfo {title} {Ab initio static exchange--correlation kernel across jacob's ladder without functional derivatives},}\ }\href {\doibase 10.1021/acs.jctc.2c01180} {\bibfield  {journal} {\bibinfo  {journal} {J. Chem. Theory Comput.}\ }\textbf {\bibinfo {volume} {19}},\ \bibinfo {pages} {1286--1299} (\bibinfo {year} {2023})}\BibitemShut {NoStop}%
\bibitem [{\citenamefont {Moldabekov}\ \emph {et~al.}(2025)\citenamefont {Moldabekov}, \citenamefont {Vorberger},\ and\ \citenamefont {Dornheim}}]{moldabekov2025density}%
  \BibitemOpen
  \bibfield  {author} {\bibinfo {author} {\bibfnamefont {Z.}~\bibnamefont {Moldabekov}}, \bibinfo {author} {\bibfnamefont {J.}~\bibnamefont {Vorberger}}, \ and\ \bibinfo {author} {\bibfnamefont {T.}~\bibnamefont {Dornheim}},\ }\bibfield  {title} {\enquote {\bibinfo {title} {From density response to energy functionals and back: An ab initio perspective on matter under extreme conditions},}\ }\href {\doibase 10.1016/j.ppnp.2024.104144} {\bibfield  {journal} {\bibinfo  {journal} {Progress in Particle and Nuclear Physics}\ }\textbf {\bibinfo {volume} {140}},\ \bibinfo {pages} {104144} (\bibinfo {year} {2025})}\BibitemShut {NoStop}%
\bibitem [{\citenamefont {Militzer}\ and\ \citenamefont {Ceperley}(2001)}]{Militzer_PRE_2001}%
  \BibitemOpen
  \bibfield  {author} {\bibinfo {author} {\bibfnamefont {B.}~\bibnamefont {Militzer}}\ and\ \bibinfo {author} {\bibfnamefont {D.~M.}\ \bibnamefont {Ceperley}},\ }\bibfield  {title} {\enquote {\bibinfo {title} {Path integral monte carlo simulation of the low-density hydrogen plasma},}\ }\href {\doibase 10.1103/PhysRevE.63.066404} {\bibfield  {journal} {\bibinfo  {journal} {Phys. Rev. E}\ }\textbf {\bibinfo {volume} {63}},\ \bibinfo {pages} {066404} (\bibinfo {year} {2001})}\BibitemShut {NoStop}%
\bibitem [{\citenamefont {Brown}\ \emph {et~al.}(2013)\citenamefont {Brown}, \citenamefont {Clark}, \citenamefont {DuBois},\ and\ \citenamefont {Ceperley}}]{Brown_PRL_2013}%
  \BibitemOpen
  \bibfield  {author} {\bibinfo {author} {\bibfnamefont {Ethan~W.}\ \bibnamefont {Brown}}, \bibinfo {author} {\bibfnamefont {Bryan~K.}\ \bibnamefont {Clark}}, \bibinfo {author} {\bibfnamefont {Jonathan~L.}\ \bibnamefont {DuBois}}, \ and\ \bibinfo {author} {\bibfnamefont {David~M.}\ \bibnamefont {Ceperley}},\ }\bibfield  {title} {\enquote {\bibinfo {title} {Path-integral monte carlo simulation of the warm dense homogeneous electron gas},}\ }\href {\doibase 10.1103/PhysRevLett.110.146405} {\bibfield  {journal} {\bibinfo  {journal} {Phys. Rev. Lett.}\ }\textbf {\bibinfo {volume} {110}},\ \bibinfo {pages} {146405} (\bibinfo {year} {2013})}\BibitemShut {NoStop}%
\bibitem [{\citenamefont {Fraser}\ \emph {et~al.}(1996)\citenamefont {Fraser}, \citenamefont {Foulkes}, \citenamefont {Rajagopal}, \citenamefont {Needs}, \citenamefont {Kenny},\ and\ \citenamefont {Williamson}}]{Fraser_PRB_1996}%
  \BibitemOpen
  \bibfield  {author} {\bibinfo {author} {\bibfnamefont {Louisa~M.}\ \bibnamefont {Fraser}}, \bibinfo {author} {\bibfnamefont {W.~M.~C.}\ \bibnamefont {Foulkes}}, \bibinfo {author} {\bibfnamefont {G.}~\bibnamefont {Rajagopal}}, \bibinfo {author} {\bibfnamefont {R.~J.}\ \bibnamefont {Needs}}, \bibinfo {author} {\bibfnamefont {S.~D.}\ \bibnamefont {Kenny}}, \ and\ \bibinfo {author} {\bibfnamefont {A.~J.}\ \bibnamefont {Williamson}},\ }\bibfield  {title} {\enquote {\bibinfo {title} {Finite-size effects and coulomb interactions in quantum monte carlo calculations for homogeneous systems with periodic boundary conditions},}\ }\href {\doibase 10.1103/PhysRevB.53.1814} {\bibfield  {journal} {\bibinfo  {journal} {Phys. Rev. B}\ }\textbf {\bibinfo {volume} {53}},\ \bibinfo {pages} {1814--1832} (\bibinfo {year} {1996})}\BibitemShut {NoStop}%
\bibitem [{\citenamefont {Dornheim}\ \emph {et~al.}(2023{\natexlab{e}})\citenamefont {Dornheim}, \citenamefont {Wicaksono}, \citenamefont {Suarez-Cardona}, \citenamefont {Tolias}, \citenamefont {B\"ohme}, \citenamefont {Moldabekov}, \citenamefont {Hecht},\ and\ \citenamefont {Vorberger}}]{Dornheim_moments_2023}%
  \BibitemOpen
  \bibfield  {author} {\bibinfo {author} {\bibfnamefont {Tobias}\ \bibnamefont {Dornheim}}, \bibinfo {author} {\bibfnamefont {Damar~C.}\ \bibnamefont {Wicaksono}}, \bibinfo {author} {\bibfnamefont {Juan~E.}\ \bibnamefont {Suarez-Cardona}}, \bibinfo {author} {\bibfnamefont {Panagiotis}\ \bibnamefont {Tolias}}, \bibinfo {author} {\bibfnamefont {Maximilian~P.}\ \bibnamefont {B\"ohme}}, \bibinfo {author} {\bibfnamefont {Zhandos~A.}\ \bibnamefont {Moldabekov}}, \bibinfo {author} {\bibfnamefont {Michael}\ \bibnamefont {Hecht}}, \ and\ \bibinfo {author} {\bibfnamefont {Jan}\ \bibnamefont {Vorberger}},\ }\bibfield  {title} {\enquote {\bibinfo {title} {Extraction of the frequency moments of spectral densities from imaginary-time correlation function data},}\ }\href {\doibase 10.1103/PhysRevB.107.155148} {\bibfield  {journal} {\bibinfo  {journal} {Phys. Rev. B}\ }\textbf {\bibinfo {volume} {107}},\ \bibinfo {pages} {155148} (\bibinfo {year} {2023}{\natexlab{e}})}\BibitemShut {NoStop}%
\bibitem [{\citenamefont {Tanaka}\ and\ \citenamefont {Ichimaru}(1986)}]{stls}%
  \BibitemOpen
  \bibfield  {author} {\bibinfo {author} {\bibfnamefont {S.}~\bibnamefont {Tanaka}}\ and\ \bibinfo {author} {\bibfnamefont {S.}~\bibnamefont {Ichimaru}},\ }\bibfield  {title} {\enquote {\bibinfo {title} {Thermodynamics and correlational properties of finite-temperature electron liquids in the {S}ingwi-{T}osi-{Land}-{S}j\"olander approximation},}\ }\href {http://journals.jps.jp/doi/abs/10.1143/JPSJ.55.2278} {\bibfield  {journal} {\bibinfo  {journal} {J. Phys. Soc. Jpn}\ }\textbf {\bibinfo {volume} {55}},\ \bibinfo {pages} {2278--2289} (\bibinfo {year} {1986})}\BibitemShut {NoStop}%
\bibitem [{\citenamefont {Tolias}\ \emph {et~al.}(2024)\citenamefont {Tolias}, \citenamefont {Kalkavouras},\ and\ \citenamefont {Dornheim}}]{Tolias_JCP_2024}%
  \BibitemOpen
  \bibfield  {author} {\bibinfo {author} {\bibfnamefont {Panagiotis}\ \bibnamefont {Tolias}}, \bibinfo {author} {\bibfnamefont {Fotios}\ \bibnamefont {Kalkavouras}}, \ and\ \bibinfo {author} {\bibfnamefont {Tobias}\ \bibnamefont {Dornheim}},\ }\bibfield  {title} {\enquote {\bibinfo {title} {{Fourier–Matsubara series expansion for imaginary–time correlation functions}},}\ }\href {\doibase 10.1063/5.0211814} {\bibfield  {journal} {\bibinfo  {journal} {J. Chem. Phys.}\ }\textbf {\bibinfo {volume} {160}},\ \bibinfo {pages} {181102} (\bibinfo {year} {2024})}\BibitemShut {NoStop}%
\bibitem [{\citenamefont {Glenzer}\ and\ \citenamefont {Redmer}(2009)}]{siegfried_review}%
  \BibitemOpen
  \bibfield  {author} {\bibinfo {author} {\bibfnamefont {S.~H.}\ \bibnamefont {Glenzer}}\ and\ \bibinfo {author} {\bibfnamefont {R.}~\bibnamefont {Redmer}},\ }\bibfield  {title} {\enquote {\bibinfo {title} {X-ray thomson scattering in high energy density plasmas},}\ }\href {https://journals.aps.org/rmp/abstract/10.1103/RevModPhys.81.1625} {\bibfield  {journal} {\bibinfo  {journal} {Rev. Mod. Phys}\ }\textbf {\bibinfo {volume} {81}},\ \bibinfo {pages} {1625} (\bibinfo {year} {2009})}\BibitemShut {NoStop}%
\bibitem [{\citenamefont {Gawne}\ \emph {et~al.}(2024)\citenamefont {Gawne}, \citenamefont {Moldabekov}, \citenamefont {Humphries}, \citenamefont {Appel}, \citenamefont {Baehtz}, \citenamefont {Bouffetier}, \citenamefont {Brambrink}, \citenamefont {Cangi}, \citenamefont {G\"ode}, \citenamefont {Kon\^opkov\'a}, \citenamefont {Makita}, \citenamefont {Mishchenko}, \citenamefont {Nakatsutsumi}, \citenamefont {Ramakrishna}, \citenamefont {Randolph}, \citenamefont {Schwalbe}, \citenamefont {Vorberger}, \citenamefont {Wollenweber}, \citenamefont {Zastrau}, \citenamefont {Dornheim},\ and\ \citenamefont {Preston}}]{Gawne_PRB_2024}%
  \BibitemOpen
  \bibfield  {author} {\bibinfo {author} {\bibfnamefont {Thomas}\ \bibnamefont {Gawne}}, \bibinfo {author} {\bibfnamefont {Zhandos~A.}\ \bibnamefont {Moldabekov}}, \bibinfo {author} {\bibfnamefont {Oliver~S.}\ \bibnamefont {Humphries}}, \bibinfo {author} {\bibfnamefont {Karen}\ \bibnamefont {Appel}}, \bibinfo {author} {\bibfnamefont {Carsten}\ \bibnamefont {Baehtz}}, \bibinfo {author} {\bibfnamefont {Victorien}\ \bibnamefont {Bouffetier}}, \bibinfo {author} {\bibfnamefont {Erik}\ \bibnamefont {Brambrink}}, \bibinfo {author} {\bibfnamefont {Attila}\ \bibnamefont {Cangi}}, \bibinfo {author} {\bibfnamefont {Sebastian}\ \bibnamefont {G\"ode}}, \bibinfo {author} {\bibfnamefont {Zuzana}\ \bibnamefont {Kon\^opkov\'a}}, \bibinfo {author} {\bibfnamefont {Mikako}\ \bibnamefont {Makita}}, \bibinfo {author} {\bibfnamefont {Mikhail}\ \bibnamefont {Mishchenko}}, \bibinfo {author} {\bibfnamefont {Motoaki}\ \bibnamefont {Nakatsutsumi}}, \bibinfo {author} {\bibfnamefont {Kushal}\ \bibnamefont {Ramakrishna}}, \bibinfo {author}
  {\bibfnamefont {Lisa}\ \bibnamefont {Randolph}}, \bibinfo {author} {\bibfnamefont {Sebastian}\ \bibnamefont {Schwalbe}}, \bibinfo {author} {\bibfnamefont {Jan}\ \bibnamefont {Vorberger}}, \bibinfo {author} {\bibfnamefont {Lennart}\ \bibnamefont {Wollenweber}}, \bibinfo {author} {\bibfnamefont {Ulf}\ \bibnamefont {Zastrau}}, \bibinfo {author} {\bibfnamefont {Tobias}\ \bibnamefont {Dornheim}}, \ and\ \bibinfo {author} {\bibfnamefont {Thomas~R.}\ \bibnamefont {Preston}},\ }\bibfield  {title} {\enquote {\bibinfo {title} {Ultrahigh resolution x-ray thomson scattering measurements at the european x-ray free electron laser},}\ }\href {\doibase 10.1103/PhysRevB.109.L241112} {\bibfield  {journal} {\bibinfo  {journal} {Phys. Rev. B}\ }\textbf {\bibinfo {volume} {109}},\ \bibinfo {pages} {L241112} (\bibinfo {year} {2024})}\BibitemShut {NoStop}%
\bibitem [{\citenamefont {Bellenbaum}\ \emph {et~al.}(2025)\citenamefont {Bellenbaum}, \citenamefont {Bachmann}, \citenamefont {Kraus}, \citenamefont {Gawne}, \citenamefont {Böhme}, \citenamefont {Döppner}, \citenamefont {Fletcher}, \citenamefont {MacDonald}, \citenamefont {Moldabekov}, \citenamefont {Preston}, \citenamefont {Vorberger},\ and\ \citenamefont {Dornheim}}]{Bellenbaum_APL_2025}%
  \BibitemOpen
  \bibfield  {author} {\bibinfo {author} {\bibfnamefont {H.~M.}\ \bibnamefont {Bellenbaum}}, \bibinfo {author} {\bibfnamefont {B.}~\bibnamefont {Bachmann}}, \bibinfo {author} {\bibfnamefont {D.}~\bibnamefont {Kraus}}, \bibinfo {author} {\bibfnamefont {Th.}\ \bibnamefont {Gawne}}, \bibinfo {author} {\bibfnamefont {M.~P.}\ \bibnamefont {Böhme}}, \bibinfo {author} {\bibfnamefont {T.}~\bibnamefont {Döppner}}, \bibinfo {author} {\bibfnamefont {L.~B.}\ \bibnamefont {Fletcher}}, \bibinfo {author} {\bibfnamefont {M.~J.}\ \bibnamefont {MacDonald}}, \bibinfo {author} {\bibfnamefont {Zh.~A.}\ \bibnamefont {Moldabekov}}, \bibinfo {author} {\bibfnamefont {T.~R.}\ \bibnamefont {Preston}}, \bibinfo {author} {\bibfnamefont {J.}~\bibnamefont {Vorberger}}, \ and\ \bibinfo {author} {\bibfnamefont {T.}~\bibnamefont {Dornheim}},\ }\bibfield  {title} {\enquote {\bibinfo {title} {Toward model-free temperature diagnostics of warm dense matter from multiple scattering angles},}\ }\href {\doibase 10.1063/5.0248230} {\bibfield
  {journal} {\bibinfo  {journal} {Applied Physics Letters}\ }\textbf {\bibinfo {volume} {126}},\ \bibinfo {pages} {044104} (\bibinfo {year} {2025})}\BibitemShut {NoStop}%
\bibitem [{\citenamefont {Huotari}\ \emph {et~al.}(2010)\citenamefont {Huotari}, \citenamefont {Soininen}, \citenamefont {Pylkk\"anen}, \citenamefont {H\"am\"al\"ainen}, \citenamefont {Issolah}, \citenamefont {Titov}, \citenamefont {McMinis}, \citenamefont {Kim}, \citenamefont {Esler}, \citenamefont {Ceperley}, \citenamefont {Holzmann},\ and\ \citenamefont {Olevano}}]{Huotari}%
  \BibitemOpen
  \bibfield  {author} {\bibinfo {author} {\bibfnamefont {Simo}\ \bibnamefont {Huotari}}, \bibinfo {author} {\bibfnamefont {J.~Aleksi}\ \bibnamefont {Soininen}}, \bibinfo {author} {\bibfnamefont {Tuomas}\ \bibnamefont {Pylkk\"anen}}, \bibinfo {author} {\bibfnamefont {Keijo}\ \bibnamefont {H\"am\"al\"ainen}}, \bibinfo {author} {\bibfnamefont {Arezki}\ \bibnamefont {Issolah}}, \bibinfo {author} {\bibfnamefont {Andrey}\ \bibnamefont {Titov}}, \bibinfo {author} {\bibfnamefont {Jeremy}\ \bibnamefont {McMinis}}, \bibinfo {author} {\bibfnamefont {Jeongnim}\ \bibnamefont {Kim}}, \bibinfo {author} {\bibfnamefont {Ken}\ \bibnamefont {Esler}}, \bibinfo {author} {\bibfnamefont {David~M.}\ \bibnamefont {Ceperley}}, \bibinfo {author} {\bibfnamefont {Markus}\ \bibnamefont {Holzmann}}, \ and\ \bibinfo {author} {\bibfnamefont {Valerio}\ \bibnamefont {Olevano}},\ }\bibfield  {title} {\enquote {\bibinfo {title} {Momentum distribution and renormalization factor in sodium and the electron gas},}\ }\href {\doibase
  10.1103/PhysRevLett.105.086403} {\bibfield  {journal} {\bibinfo  {journal} {Phys. Rev. Lett.}\ }\textbf {\bibinfo {volume} {105}},\ \bibinfo {pages} {086403} (\bibinfo {year} {2010})}\BibitemShut {NoStop}%
\bibitem [{\citenamefont {Fletcher}\ \emph {et~al.}(2022)\citenamefont {Fletcher}, \citenamefont {Vorberger}, \citenamefont {Schumaker}, \citenamefont {Ruyer}, \citenamefont {Goede}, \citenamefont {Galtier}, \citenamefont {Zastrau}, \citenamefont {Alves}, \citenamefont {Baalrud}, \citenamefont {Baggott}, \citenamefont {Barbrel}, \citenamefont {Chen}, \citenamefont {Döppner}, \citenamefont {Gauthier}, \citenamefont {Granados}, \citenamefont {Kim}, \citenamefont {Kraus}, \citenamefont {Lee}, \citenamefont {MacDonald}, \citenamefont {Mishra}, \citenamefont {Pelka}, \citenamefont {Ravasio}, \citenamefont {Roedel}, \citenamefont {Fry}, \citenamefont {Redmer}, \citenamefont {Fiuza}, \citenamefont {Gericke},\ and\ \citenamefont {Glenzer}}]{Fletcher_Frontiers_2022}%
  \BibitemOpen
  \bibfield  {author} {\bibinfo {author} {\bibfnamefont {L.~B.}\ \bibnamefont {Fletcher}}, \bibinfo {author} {\bibfnamefont {J.}~\bibnamefont {Vorberger}}, \bibinfo {author} {\bibfnamefont {W.}~\bibnamefont {Schumaker}}, \bibinfo {author} {\bibfnamefont {C.}~\bibnamefont {Ruyer}}, \bibinfo {author} {\bibfnamefont {S.}~\bibnamefont {Goede}}, \bibinfo {author} {\bibfnamefont {E.}~\bibnamefont {Galtier}}, \bibinfo {author} {\bibfnamefont {U.}~\bibnamefont {Zastrau}}, \bibinfo {author} {\bibfnamefont {E.~P.}\ \bibnamefont {Alves}}, \bibinfo {author} {\bibfnamefont {S.~D.}\ \bibnamefont {Baalrud}}, \bibinfo {author} {\bibfnamefont {R.~A.}\ \bibnamefont {Baggott}}, \bibinfo {author} {\bibfnamefont {B.}~\bibnamefont {Barbrel}}, \bibinfo {author} {\bibfnamefont {Z.}~\bibnamefont {Chen}}, \bibinfo {author} {\bibfnamefont {T.}~\bibnamefont {Döppner}}, \bibinfo {author} {\bibfnamefont {M.}~\bibnamefont {Gauthier}}, \bibinfo {author} {\bibfnamefont {E.}~\bibnamefont {Granados}}, \bibinfo {author} {\bibfnamefont {J.~B.}\
  \bibnamefont {Kim}}, \bibinfo {author} {\bibfnamefont {D.}~\bibnamefont {Kraus}}, \bibinfo {author} {\bibfnamefont {H.~J.}\ \bibnamefont {Lee}}, \bibinfo {author} {\bibfnamefont {M.~J.}\ \bibnamefont {MacDonald}}, \bibinfo {author} {\bibfnamefont {R.}~\bibnamefont {Mishra}}, \bibinfo {author} {\bibfnamefont {A.}~\bibnamefont {Pelka}}, \bibinfo {author} {\bibfnamefont {A.}~\bibnamefont {Ravasio}}, \bibinfo {author} {\bibfnamefont {C.}~\bibnamefont {Roedel}}, \bibinfo {author} {\bibfnamefont {A.~R.}\ \bibnamefont {Fry}}, \bibinfo {author} {\bibfnamefont {R.}~\bibnamefont {Redmer}}, \bibinfo {author} {\bibfnamefont {F.}~\bibnamefont {Fiuza}}, \bibinfo {author} {\bibfnamefont {D.~O.}\ \bibnamefont {Gericke}}, \ and\ \bibinfo {author} {\bibfnamefont {S.~H.}\ \bibnamefont {Glenzer}},\ }\bibfield  {title} {\enquote {\bibinfo {title} {Electron-ion temperature relaxation in warm dense hydrogen observed with picosecond resolved x-ray scattering},}\ }\href {\doibase 10.3389/fphy.2022.838524} {\bibfield  {journal}
  {\bibinfo  {journal} {Frontiers in Physics}\ }\textbf {\bibinfo {volume} {10}} (\bibinfo {year} {2022}),\ 10.3389/fphy.2022.838524}\BibitemShut {NoStop}%
\bibitem [{\citenamefont {Zastrau}\ \emph {et~al.}(2014)\citenamefont {Zastrau}, \citenamefont {Sperling}, \citenamefont {Harmand}, \citenamefont {Becker}, \citenamefont {Bornath}, \citenamefont {Bredow}, \citenamefont {Dziarzhytski}, \citenamefont {Fennel}, \citenamefont {Fletcher}, \citenamefont {F\"orster}, \citenamefont {G\"ode}, \citenamefont {Gregori}, \citenamefont {Hilbert}, \citenamefont {Hochhaus}, \citenamefont {Holst}, \citenamefont {Laarmann}, \citenamefont {Lee}, \citenamefont {Ma}, \citenamefont {Mithen}, \citenamefont {Mitzner}, \citenamefont {Murphy}, \citenamefont {Nakatsutsumi}, \citenamefont {Neumayer}, \citenamefont {Przystawik}, \citenamefont {Roling}, \citenamefont {Schulz}, \citenamefont {Siemer}, \citenamefont {Skruszewicz}, \citenamefont {Tiggesb\"aumker}, \citenamefont {Toleikis}, \citenamefont {Tschentscher}, \citenamefont {White}, \citenamefont {W\"ostmann}, \citenamefont {Zacharias}, \citenamefont {D\"oppner}, \citenamefont {Glenzer},\ and\ \citenamefont {Redmer}}]{Zastrau}%
  \BibitemOpen
  \bibfield  {author} {\bibinfo {author} {\bibfnamefont {U.}~\bibnamefont {Zastrau}}, \bibinfo {author} {\bibfnamefont {P.}~\bibnamefont {Sperling}}, \bibinfo {author} {\bibfnamefont {M.}~\bibnamefont {Harmand}}, \bibinfo {author} {\bibfnamefont {A.}~\bibnamefont {Becker}}, \bibinfo {author} {\bibfnamefont {T.}~\bibnamefont {Bornath}}, \bibinfo {author} {\bibfnamefont {R.}~\bibnamefont {Bredow}}, \bibinfo {author} {\bibfnamefont {S.}~\bibnamefont {Dziarzhytski}}, \bibinfo {author} {\bibfnamefont {T.}~\bibnamefont {Fennel}}, \bibinfo {author} {\bibfnamefont {L.~B.}\ \bibnamefont {Fletcher}}, \bibinfo {author} {\bibfnamefont {E.}~\bibnamefont {F\"orster}}, \bibinfo {author} {\bibfnamefont {S.}~\bibnamefont {G\"ode}}, \bibinfo {author} {\bibfnamefont {G.}~\bibnamefont {Gregori}}, \bibinfo {author} {\bibfnamefont {V.}~\bibnamefont {Hilbert}}, \bibinfo {author} {\bibfnamefont {D.}~\bibnamefont {Hochhaus}}, \bibinfo {author} {\bibfnamefont {B.}~\bibnamefont {Holst}}, \bibinfo {author} {\bibfnamefont
  {T.}~\bibnamefont {Laarmann}}, \bibinfo {author} {\bibfnamefont {H.~J.}\ \bibnamefont {Lee}}, \bibinfo {author} {\bibfnamefont {T.}~\bibnamefont {Ma}}, \bibinfo {author} {\bibfnamefont {J.~P.}\ \bibnamefont {Mithen}}, \bibinfo {author} {\bibfnamefont {R.}~\bibnamefont {Mitzner}}, \bibinfo {author} {\bibfnamefont {C.~D.}\ \bibnamefont {Murphy}}, \bibinfo {author} {\bibfnamefont {M.}~\bibnamefont {Nakatsutsumi}}, \bibinfo {author} {\bibfnamefont {P.}~\bibnamefont {Neumayer}}, \bibinfo {author} {\bibfnamefont {A.}~\bibnamefont {Przystawik}}, \bibinfo {author} {\bibfnamefont {S.}~\bibnamefont {Roling}}, \bibinfo {author} {\bibfnamefont {M.}~\bibnamefont {Schulz}}, \bibinfo {author} {\bibfnamefont {B.}~\bibnamefont {Siemer}}, \bibinfo {author} {\bibfnamefont {S.}~\bibnamefont {Skruszewicz}}, \bibinfo {author} {\bibfnamefont {J.}~\bibnamefont {Tiggesb\"aumker}}, \bibinfo {author} {\bibfnamefont {S.}~\bibnamefont {Toleikis}}, \bibinfo {author} {\bibfnamefont {T.}~\bibnamefont {Tschentscher}}, \bibinfo {author}
  {\bibfnamefont {T.}~\bibnamefont {White}}, \bibinfo {author} {\bibfnamefont {M.}~\bibnamefont {W\"ostmann}}, \bibinfo {author} {\bibfnamefont {H.}~\bibnamefont {Zacharias}}, \bibinfo {author} {\bibfnamefont {T.}~\bibnamefont {D\"oppner}}, \bibinfo {author} {\bibfnamefont {S.~H.}\ \bibnamefont {Glenzer}}, \ and\ \bibinfo {author} {\bibfnamefont {R.}~\bibnamefont {Redmer}},\ }\bibfield  {title} {\enquote {\bibinfo {title} {Resolving ultrafast heating of dense cryogenic hydrogen},}\ }\href {https://journals.aps.org/prl/abstract/10.1103/PhysRevLett.112.105002} {\bibfield  {journal} {\bibinfo  {journal} {Phys. Rev. Lett}\ }\textbf {\bibinfo {volume} {112}},\ \bibinfo {pages} {105002} (\bibinfo {year} {2014})}\BibitemShut {NoStop}%
\bibitem [{\citenamefont {Hamann}\ \emph {et~al.}(2023)\citenamefont {Hamann}, \citenamefont {Kordts}, \citenamefont {Filinov}, \citenamefont {Bonitz}, \citenamefont {Dornheim},\ and\ \citenamefont {Vorberger}}]{Hamann_PRR_2023}%
  \BibitemOpen
  \bibfield  {author} {\bibinfo {author} {\bibfnamefont {Paul}\ \bibnamefont {Hamann}}, \bibinfo {author} {\bibfnamefont {Linda}\ \bibnamefont {Kordts}}, \bibinfo {author} {\bibfnamefont {Alexey}\ \bibnamefont {Filinov}}, \bibinfo {author} {\bibfnamefont {Michael}\ \bibnamefont {Bonitz}}, \bibinfo {author} {\bibfnamefont {Tobias}\ \bibnamefont {Dornheim}}, \ and\ \bibinfo {author} {\bibfnamefont {Jan}\ \bibnamefont {Vorberger}},\ }\bibfield  {title} {\enquote {\bibinfo {title} {Prediction of a roton-type feature in warm dense hydrogen},}\ }\href {\doibase 10.1103/PhysRevResearch.5.033039} {\bibfield  {journal} {\bibinfo  {journal} {Phys. Rev. Res.}\ }\textbf {\bibinfo {volume} {5}},\ \bibinfo {pages} {033039} (\bibinfo {year} {2023})}\BibitemShut {NoStop}%
\end{thebibliography}%
\end{document}